%% file: paper_blazars.tex
\documentclass{aa}  

\usepackage{graphicx}
\usepackage{txfonts}
\usepackage{lscape}
\usepackage{booktabs}
\usepackage{threeparttablex}
\usepackage{caption}
\usepackage{wasysym}
\usepackage{arydshln}
\usepackage[bookmarks=false,colorlinks=true,linkcolor=black,citecolor=black,filecolor=black,urlcolor=cyan]{hyperref}

\begin{document}

   \title{X-ray absorption towards high-redshift sources: probing the intergalactic medium with blazars}

   \author{R. Arcodia
          \inst{1,2,3}
          \and
          S. Campana\inst{2}
          \and
          R. Salvaterra\inst{4}
          \and
          G. Ghisellini\inst{2}
          }

   \institute{Department of Physics G. Occhialini, University of Milano-Bicocca, Piazza della Scienza 3, I-20126 Milano, Italy\\
              \email{arcodia@mpe.mpg.de}
         \and
             INAF-Osservatorio Astronomico di Brera, Via Bianchi 46, I-23807 Merate (LC), Italy
         \and
	         Max-Planck-Institut f\"ur extraterrestrische Physik (MPE), Giessenbachstrasse 1, D-85748 Garching bei M\"unchen, Germany
         \and
             INAF, IASF Milano, Via E. Bassini 15, I-20133 Milano, Italy
             }

   \date{Received ; accepted }

 
  \abstract{
        The role played by the intergalactic medium (IGM) in the X-ray absorption towards high-redshift sources has recently drawn more attention in spectral analysis studies. Here, we study the X-ray absorption towards 15 flat-spectrum radio quasars at $z>2$, relying on high counting statistic ($\gtrsim10\,000$ photons) provided by \emph{XMM-Newton}, with additional \emph{NuSTAR} (and simultaneous \emph{Swift-XRT}) observations when available. Blazars can be confidently considered to have negligible X-ray absorption along the line of sight within the host galaxy, likely swept by the kpc-scale relativistic jet. This makes our sources ideal for testing the absorption component along the IGM. 
        Our new approach is to revisit the origin of the soft X-ray spectral hardening observed in high-$z$ blazars in terms of X-ray absorption occurring along the IGM, with the help of a low-$z$ sample used as comparison. We verify that the presence of absorption in excess of the Galactic value is the preferred explanation to explain the observed hardening, while intrinsic energy breaks, predicted by blazars' emission models, can easily occur out of the observing energy band in most sources. First, we perform an indirect analysis comparing the inferred amount of absorption in excess of the Galactic value with a simulated IGM absorption contribution, that increases with redshift and includes both a minimum component from diffuse IGM metals, and the additional contribution of discrete denser intervening regions. Then, we directly investigate the warm-hot IGM with a spectral model on the best candidates of our sample, obtaining an average IGM density of $n_0=1.01^{+0.53}_{-0.72}\times10^{-7}$\,cm$^{-3}$ and temperature of $\log(T/\text{K})=6.45^{+0.51}_{-2.12}$. A more dedicated study is currently beyond reach, but our results can be used as a stepping stone for future more accurate analysis, involving \emph{Athena}. 
        }

   \titlerunning{}
   \authorrunning{R. Arcodia et al.} 
   \maketitle

%

\section{Introduction}
X-ray spectral analysis involving extragalactic sources is currently not able to detect simple absorption spectral features (lines, edges\dots), hence typically only the total amount of absorbing matter (in $N_H$, cm$^{-2}$) can be inferred. Of this, X-ray absorption occurring within our Galaxy usually accounts for a known fraction \citep[e.g.][]{Dickey90:DL_HI,Kalberla:LAB,Willingale13:GalacticH2}. An absorption component in addition to the Galactic value (hereafter simply called \emph{excess absorption}), if needed, was often uniquely attributed to the galaxy hosting the X-ray source, while the absorption produced by intergalactic intervening matter (IGM) was rarely included. As a matter of fact, most of the cosmic matter resides among galaxies within the IGM \citep[see][for an extensive and recent review]{McQuinn16:WHIMreview}. Here, we are interested in the low-redshift IGM ($z\leq2$), focusing on the missing "metal fog" that is predicted to lie in a hot phase at $\sim(10^5-10^7)\,$K, composing the so-called warm-hot IGM \citep[WHIM, see][and references therein]{Cen99:WHIM,Cen06:WHIMsimul,Dave01:WHIM,Bregman07:IGMreview,Shull12:WHIM}. Hence, when fitting a low-energy X-ray spectrum requires excess absorption, both the host and the IGM component should be considered. Besides, while the former varies among different types of sources and should be physically motivated depending on their environment, an IGM absorption component is unaffected by the type of source emitting behind.

A soft X-ray spectral hardening has been typically observed towards distant sources, i.e. high-$z$ Active Galactic Nuclei \citep[AGN,][for a recent review]{Padovani17:AGNreview} and Gamma-ray Bursts \citep[GRBs,][for a recent review]{Schady17:GRBsreview}. In principle, it is unclear whether the observed spectrum is congruent to the emitted one or some excess absorption is occurring. In GRBs the observed X-ray hardening was promptly attributed to excess absorption intrinsic to the host \citep[e.g.][]{Owens98:GRBxhardening,Galama01:GRBxhardening,Stratta04:highNHGRB,DeLuca05:XMMgrbs,Campana06:GRBabs,Campana10:nhzswift,Campana12:nhxcomplete,Arcodia16:GRBsNhz}, since their environment is known to be dense \citep{Fruchter:environment,Woosley:SN}. By contrast, in distant quasars also different origins, e.g. spectral breaks intrinsic to the emission, were considered as alternative to the excess absorption scenario \citep[see][]{Elvis94:absquasar,Cappi97:Xspectraquasar,Fiore98:ROSATquasar,Reeves00:ASCAquasar,Fabian01:7Cwarmabsorber,Fabian01:0525warmabs,Worsley04:7Cflatt,Worsley04:0525warm,Page05:XMMquasar,Yuan06:RLQshighz,Grupe06:XMMhighzquasar,Sambruna07:SwiftRLQ,Saez11:RLQshighz}. Explaining the observed hardening with excess absorption uniquely attributed to the host galaxy resulted in an increasing trend of the intrinsic column densities (hereafter $N_H(z)$, with $z=z_{source}$) with redshift. For all extragalactic sources, this so-called $N_H(z)-z$ relation typically showed at low-$z$ both non- and highly-absorbed sources (from columns slightly above the Galactic value to $N_H(z)\geq10^{23}$ cm$^{-2}$), while at high-$z$, surprisingly, only heavily-absorbed sources were observed. Apparently, this is against the idea of an environment less polluted by metals within galaxies of the younger Universe. This contrast can be resolved interpreting the excess absorption values in the $N_H(z)-z$ relation not only as intrinsic to the source, but also due to intervening IGM matter. 

The idea of an IGM absorbing component common to all sources emerged through the years, first as a simple suggestive hypothesis \citep[e.g. see][]{Fabian01:7Cwarmabsorber}. A more quantitative approach was adopted only recently in a series of papers, in which both a diffuse IGM and additional discrete intervening systems were considered towards quasars and GRBs \citep{Behar:baryons,Campana12:nhxcomplete,Starling:evoluzIGM,Eitan:2013}. This scenario was later confirmed with dedicated cosmological simulations by \citet{Campana:missing}, who matched the observed $N_H(z)-z$ relation with a simulated absorption contribution occurring along the IGM, that obviously increases with redshift. This would explain the lack of unabsorbed high-$z$ sources without invoking complicated scenarios occurring within distant host galaxies.

Here, we report a study of X-ray absorption towards high-redshift blazars \citep[][for recent reviews]{Madejski:BLAZreview,Foschini17:blazarsreview}. They consist in jetted-AGN in which the relativistic jet is pointing towards us. Then, it is reasonable to assume that any host absorber was likely swept by the kpc-scale jet. This assumption makes blazars, in principle, the ideal sources to test the IGM absorption scenario. Their spectral energy distribution (SED) is characterised by two broad humps (in $\nu F_{\nu}$), tracing the beamed emission of the relativistic jet, that dominates over the typical AGN emission at almost all frequencies. The two humps are thought to be related to synchrotron and inverse Compton (IC) processes, at low and high frequencies, respectively. The photons emitted by the former mechanisms can be used as seed by the latter via synchrotron self-Compton (SSC), but in most powerful blazars electrons responsible for the IC emission are thought to interact with photons external to the jet (External Compton, EC), the most accredited being produced by the broad line region (BLR) or by the dusty torus \citep[][and references therein]{Ghisellini10:physicalPROP}. These most powerful blazars, called flat-spectrum radio quasars (FSRQs), are of interest in this work, the other sub-class being BL Lacertae objects \citep[BL Lacs,][]{Urry95:unified_scheme,Ghisellini11:transition_0521sed}.

Most of the objects analysed in this work were already studied in the last two decades \citep[e.g.][]{Elvis94:absquasar,Elvis94:0014ASCA,Cappi97:Xspectraquasar,Fiore98:ROSATquasar,Reeves97:ASCAquasars,Reeves00:ASCAquasar,Reeves01:0537XMM,Fabian01:7Cwarmabsorber,Fabian01:0525warmabs,Worsley04:7Cflatt,Worsley04:0525warm,Worsley06:7Cflatt,Page05:XMMquasar,Yuan00:flatt1026,Yuan05:abs1026,Yuan06:RLQshighz,Grupe04:1026,Grupe06:XMMhighzquasar,Tavecchio00:BEPPOSAXquasar,Tavecchio07:RBS315Suzaku-intrinsicbreak,Sambruna07:SwiftRLQ,Eitan:2013,Tagliaferri15:NuSTAR-2149and4C,Paliya15:4Cbroadband,Paliya16:broadband,Sbarrato16:NUSTARblazars}. In these works, if the observed soft X-ray spectral hardening was attributed to excess absorption, only an absorber intrinsic to the host galaxy or few discrete intervening absorbers, namely Damped or sub-Damped Lyman-$\alpha$ absorbers \citep[DLAs or subDLAs][]{Wolfe86:DLAs,Wolfe05:DLAs}, were investigated. Nonetheless, in blazars the former contribution is negligible and typically in contrast with the optical-UV observations, and the latter is insufficient \citep[e.g. see the discussions in][]{Elvis94:absquasar,Cappi97:Xspectraquasar,Fabian01:7Cwarmabsorber,Fabian01:0525warmabs,Worsley04:7Cflatt,Worsley04:0525warm,Page05:XMMquasar}. Hence, an alternative explanation involving intrinsic energy breaks started to be preferred to account for the observed hardening in blazars' X-ray spectra \citep[e.g.][]{Tavecchio07:RBS315Suzaku-intrinsicbreak}.

Our new approach is to consider the observed X-ray spectrum of distant blazars to be absorbed, in excess of the Galactic component, uniquely by the WHIM plus additional intervening systems along the IGM line of sight, if known. Intrinsic spectral breaks, predicted by blazars' emission models \citep[see, e.g.,][and references therein]{Sikora94:blazarsmodels,Sikora97:blazarsmodels,Sikora09:blazarsmodels,Tavecchio07:RBS315Suzaku-intrinsicbreak,Tavecchio08:BLRspectrum,Ghisellini09:CanonicalBlazars-7Csed,Ghisellini15:blazarsMODELS}, are also considered, although they can easily occur out of the observing band. 

In Section~\ref{sec:samples} we outline the criteria with which we built our samples. In Section~\ref{sec:observations} (and Appendix~\ref{sec:appA}) we report the details of the filtering and processing of our sources. In Section~\ref{sec:timing} we check for possible flux variations in the processed observations. In Section~\ref{sec:analysis} we describe the models adopted and our fitting methods, along with spectral results. In Section~\ref{sec:discussion}, we discuss how our results fit in the $N_H(z)-z$ relation along with the current literature, testing the IGM absorption contribution in an indirect way. Then, we also directly test the WHIM absorption component with a spectral model. The coexistence between the IGM excess absorption scenario and blazars' emission models is investigated in details in Appendix~\ref{sec:appB}. Conclusions are drawn is Section~\ref{sec:conclusions}.

\section{Samples}
\label{sec:samples}
The IGM absorption contribution is thought to become dominant at $z\gtrsim2$ \citep{Starling:evoluzIGM,Campana:missing}, thus our first criterion was to select $z\geq2$ blazars. Moreover, high signal-to-noise X-ray spectra are necessary to properly assess the presence of a curved spectrum in distant extragalactic sources with a fine-tuned analysis. This purpose could be fulfilled with the \emph{XMM-Newton} satellite \citep{Jansen:XMM}. The second criterion was then to include in the analysis all blazars for which the three EPIC cameras \citep{Struder:pn,Turner:MOS} jointly recorded more than $\sim10\,000$ photons\footnote{This criterion was adopted a priori, but at a later time it turned out to be a fair limit to ensure high signal-to-noise spectra.}.

We used several published catalogues of blazars and other publications to build up our samples. We first obtained a list of $z\geq2$ blazars cross-checking the BAT70 catalogue \citep{Baumgartner:BAT70} and the "List of LAT AGN" catalogue\footnote{see \href{http://www.asdc.asi.it/fermiagn/}{http://www.asdc.asi.it/fermiagn/}}. The latter consists in a list of all the AGN published by both the LAT team in several catalogue, i.e. 1LAC \citep{Abdo:1LAC}, 2LAC \citep{Ackermann:2LAC}, 3LAC \citep{Ackermann:3LAC}, 1FHL \citep{Ackermann:1FHL}, 2FHL \citep{Ackermann:2FHL}, ATels\footnote{\href{http://www-glast.stanford.edu/cgi-bin/pub\_rapid}{http://www-glast.stanford.edu/cgi-bin/pub\_rapid}} and papers from scientists external to the LAT collaboration. In addition, we cross-checked this provisional list with the sample of X-ray selected quasars of \citet{Eitan:2013}.

Starting from this huge parent sample, we selected all blazars with at least an \emph{XMM-Newton} observation and then discarded all faint objects. In order to securely classify the remaining sources as FSRQs, we relied on several blazar catalogues, SIMBAD and other works available in the literature. We also analysed their SED with the 
ASDC "SED Builder"\footnote{\href{https://tools.asdc.asi.it/}{https://tools.asdc.asi.it/}}.

We call \emph{Silver Sample} the final catalogue containing 15 \emph{XMM-Newton} FSRQs selected according to the above criteria, ranging from redshift $2.07$ to $4.72$ (see Table~\ref{tab:Silver_Sample}). Among them, we searched for \emph{NuSTAR} \citep{Harrison13:NuSTAR} and simultaneous \emph{Swift-XRT} \citep{Burrows05:swiftXRT} observations, when possible, with the aim to provide a superior broadband analysis of blazars' X-ray spectral curvature. They were available for six objects of the Silver sample, highlighted in bold in Table~\ref{tab:Silver_Sample}, that form the \emph{Golden Sample}.

\begin{table}[tb]
	\caption{List of $z>2$ \emph{XMM-Newton} FSRQs of the \emph{Silver Sample}. In bold, Blazars of the \emph{Golden Sample} (i.e. with additional \emph{NuSTAR} observations).}
	\label{tab:Silver_Sample}
	\centering
	\begin{tabular}{cccc}
		\toprule
		\multicolumn{1}{c}{Name} &
		\multicolumn{1}{c}{$z$} &
		\multicolumn{1}{c}{RA} &
		\multicolumn{1}{c}{dec} \\
		\midrule
		\textbf{7C 1428+4218} & 4.715 & 14 30 23.74 & +42 04 36.49 \\
		QSO J0525-3343 & 4.413 & 05 25 06.2 & -33 43 05 \\
		QSO B1026-084 & 4.276 & 10 28 38.79 & -08 44 38.44 \\ 
		\textbf{QSO B0014+810} & 3.366 & 00 17 08.48 & +81 35 08.14 \\
		PKS 2126-158 & 3.268 & 21 29 12.18 & -15 38 41.02 \\
		QSO B0537-286 & 3.104 & 05 39 54.28 & -28 39 55.90\\
		QSO B0438-43 & 2.852 & 04 40 17.17 & -43 33 08.62 \\
		\textbf{RBS 315} & 2.69 & 02 25 04.67 & +18 46 48.77	\\
		QSO J2354-1513 & 2.675 & 23 54 30.20 & -15 13 11.16 \\
		\textbf{PBC J1656.2-3303} & 2.4	& 16 56 16.78 & -33 02 12.7 \\
		QSO J0555+3948 & 2.363 & 05 55 30.81 & +39 48 49.16 \\
		\textbf{PKS 2149-306} & 2.345 & 21 51 55.52 & -30 27 53.63 \\
		QSO B0237-2322 & 2.225 & 02 40 08.18 & -23 09 15.78 \\
		\textbf{4C 71.07} & 2.172 & 08 41 24.4 & +70 53 42 \\
		PKS 0528+134 & 2.07	& 05 30 56.42 & +13 31 55.15 \\
		\bottomrule
	\end{tabular}
\end{table}

We also built a low-redshift sample to perform useful comparisons. We restricted the selection to sources in which a significant amount of excess IGM absorption is not expected, thus we opted for the $0-0.5$ redshift range. We selected low-redshift blazars following the same criteria and methods outlined for the high-$z$ Silver Sample, drawing a list of candidates from the same catalogues, mostly from the "List of LAT AGN". Being interested in comparing the same region of the SED, short of the redshift scaling, we looked for FSRQs but considered also low-energy peaked BL Lac objects (LBLs), that show a low synchrotron peak frequency ($<10^{14}\,$Hz). After discarding the faint objects and all the unsuited BL Lacs, we ended up with five FSRQs\footnote{The classification of blazar PKS 0521-365 is less certain. It was classified as a misaligned blazar \citep{D'ammando15:0521misaligned}, not corresponding probably to the most canonical FSRQ-type, although its SED resembles their characteristics \citep{Ghisellini11:transition_0521sed}.}, one NLS1\footnote{$\gamma$-ray emitting NLS1s are thought to be FSRQs at an early stage of their evolution or rejuvenated by a recent merger \citep[see][for a recent review]{Foschini17:blazarsreview}. The difference is in the jet power, likely due to a lower mass of the central black hole since the environment is photon-rich as in FSRQs. Consequently, we expect the X-ray emission to be similar. This possibility was confirmed for the specific case of PKS 2004-447 \citep{Paliya13:PKS2004NLS1,Kreikenbohm16:PKS2004NLS1}.} and two LBLs (see Table~\ref{tab:Lowzsample}).

\begin{figure}[tb]
	\centering
	\includegraphics[width=\columnwidth]{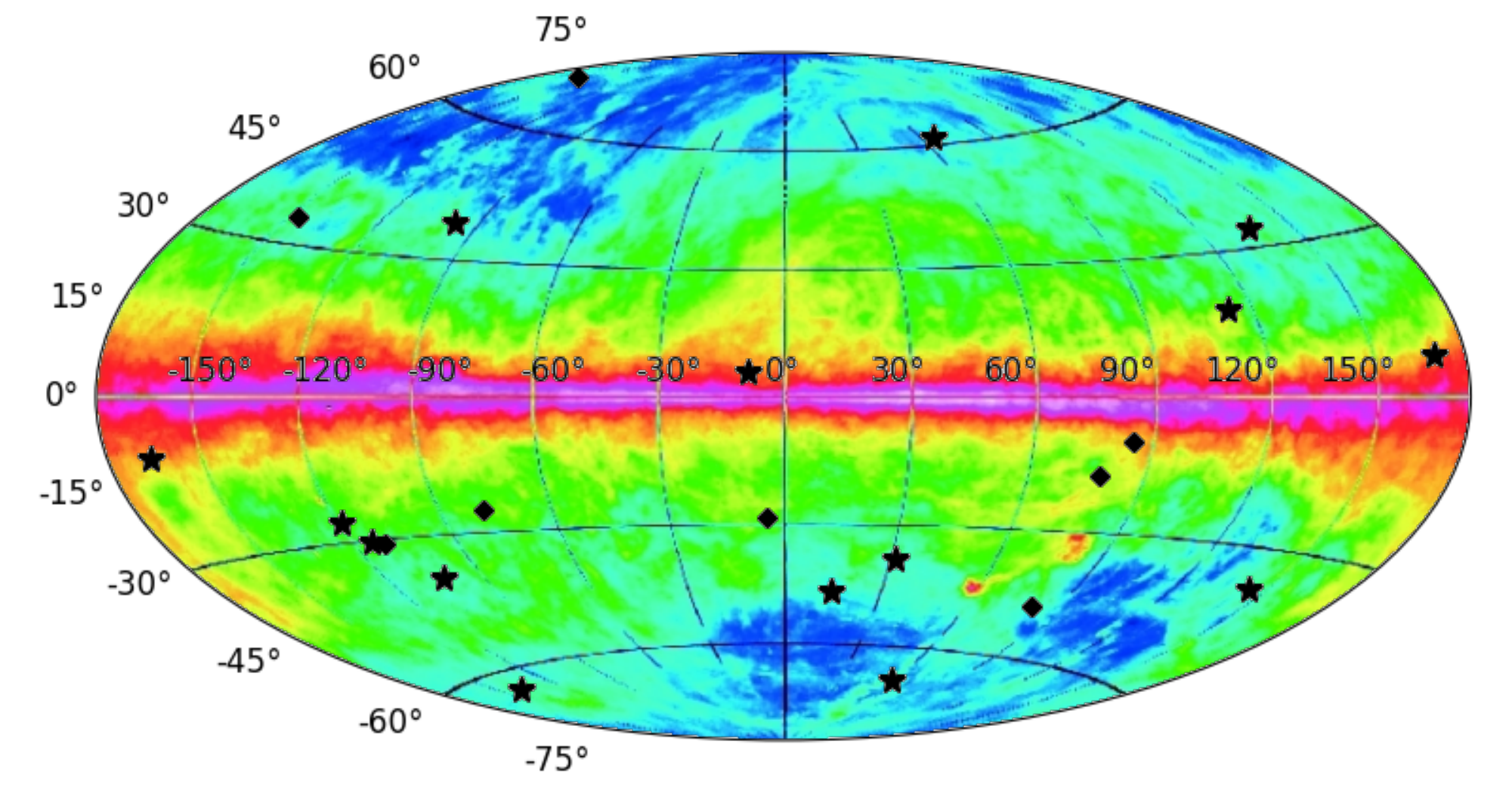}
	\caption{Distribution of blazars in the sky in Galactic coordinates (Aitoff projection), superimposed to the LAB absorption map, representing the distribution of Galactic column densities \citep[see][for further details]{Kalberla:LAB}. \emph{Black} stars represent the high-$z$ \emph{Silver Sample}. As \emph{black} diamonds, blazars of the low-$z$ sample.}
	\label{fig:Aitoff_LAB}
\end{figure}

In Figure~\ref{fig:Aitoff_LAB}, the position in the sky of both high- and low-$z$ blazars is shown superimposed to the Leiden Argentine Bonn (LAB) absorption map \citep{Kalberla:LAB}, that represents Galactic column densities yielded by the $H_I$ integrated emission.

\begin{table}[tb]
	\tiny
	\caption{List of low-redshift \emph{XMM-Newton} blazars. Objects for which additional \emph{NuSTAR} observations were analysed are highlighted in bold.}
	\label{tab:Lowzsample}
	\centering
		\begin{tabular}{ccccc}
			\toprule
			\multicolumn{1}{c}{Name} &
			\multicolumn{1}{c}{$z$} &
			\multicolumn{1}{c}{Class.} &
			\multicolumn{1}{c}{RA} &
			\multicolumn{1}{c}{dec}	\\
			\midrule
			TXS 2331+073 & 0.401 & FSRQ & 23 34 12.83 & +07 36 27.55 \\
			4C +31.63 & 0.295 & FSRQ & 22 03 14.97 & -31 45 38.26  \\
			B2 1128+31 & 0.29 & FSRQ & 10 28 38.79 & -08 44 38.44  \\ 
			\textbf{PKS 2004-447} & 0.24 & NLS1 & 20 07 55.18 & -44 34 44.28 \\
			PMN J0623-6436 & 0.129 & FSRQ & 06 23 07.70 & -64 36 20.72\\
			PKS 0521-365 & 0.055 & FSRQ? & 05 22 57.98 & -36 27 30.85 \\
			OJ 287 & 0.306 & BLL & 08 54 48.87 & +20 06 30.64  \\
			BL Lacertae & 0.069 & BLL & 22 02 43.29 & +42 16 39.98 \\
			\bottomrule
		\end{tabular}
\end{table}

\section{Observations}
\label{sec:observations}
Here we report the tools used for the processing, screening and analysis of \emph{XMM-Newton} data, along with the procedure adopted for \emph{NuSTAR} and \emph{Swift-XRT} data. Details on the processed observation(s) for each of ours high-$z$ blazars are provided in Appendix~\ref{sec:appA}. 
\subsection{XMM-Newton}
For the processing, screening, and analysis of the data from the EPIC MOS1, MOS2 \citep{Turner:MOS} and pn \citep{Struder:pn} cameras, standard tools have been used (XMM SAS v. 15.0.0 and HEAsoft v. 6.20). Observation Data Files (ODFs) were downloaded and regularly processed according to the SAS Data Analysis Threads\footnote{\href{https://www.cosmos.esa.int/web/xmm-newton/sas-threads}{https://www.cosmos.esa.int/web/xmm-newton/sas-threads}}. The event file of each observation was filtered from \emph{Flaring Particle Background} (FPB): a good time interval (GTI) was created accepting only times when the background count rate of single pixel events ("PATTERN==0") with high energies ($\geq10\,$keV for EPIC-MOS and $10-12\,$keV for EPIC-pn) was less than a chosen threshold (e.g. the default choice is $<0.35\,$c$\,$s$^{-1}$ for MOS1 and MOS2, $<0.4$\,c$\,$s$^{-1}$ for pn). 

The source spectrum was first extracted from a circular region. Background was extracted from a nearby region with the same radius for EPIC-MOS cameras, whilst for EPIC-pn it was extracted from a region at the same distance to the readout node (RAWY position) as the source region\footnote{for further details, see \href{https://www.cosmos.esa.int/web/xmm-newton/sas-thread-pn-spectrum}{XMM-SOC-CAL-TN-0018}}. When extracting the source and background EPIC-pn spectrum with the SAS \texttt{evselect} task, the strings "FLAG==0" and "PATTERN<=4" (i.e. up to double-pixel events) were included in the selection expression, while for EPIC-MOS we included the string "PATTERN<=12" (i.e. up to quadruple-pixel events). The "FLAG==0" string omits parts of the detector area like border pixels or columns with higher offset.

Any possible pile-up effect on each spectrum was then checked with the SAS task \texttt{epatplot}. The plot allows us to compare the observed versus the expected pattern distribution within a source extraction region. If both agree, pile-up is not considered to be present for the observation. In some cases, also the (more approximate) tool \texttt{WebPIMMS} was used for consistency. In some sources (see Appendix~\ref{sec:appA}) pile-up was present and the circular source region was corrected excising a core with increasing radius up to the best agreement between the expected and observed pattern distribution in the \texttt{epatplot}.

For all sources, \emph{XMM-Newton} spectra were rebinned, so that each energy bin contained a minimum of 20 counts. Moreover, the SAS task \texttt{oversample=3} was adopted to ensure that no group was narrower than 1/3 of the FWHM resolution\footnote{see the \href{https://xmm-tools.cosmos.esa.int/external/sas/current/doc/specgroup.pdf}{\texttt{specgroup}} documentation}. 

\subsection{NuSTAR}
Throughout this work, the \emph{NuSTAR} Focal Plane Module A (FPMA) and B (FPMB) data were processed with \texttt{NuSTARDAS} v1.7.1, jointly developed by the ASI Science Data Center (ASDC, Italy) and the California Institute of Technology (Caltech, USA). Event files were calibrated and cleaned using the \texttt{nupipeline} task (v0.4.6). After the selection of the source (and background) region, spectra were obtained with the \texttt{nuproducts} task (v0.3.0), in the energy range $3-79\,$keV. Since \emph{NuSTAR} has a triggered readout, it does not suffer from pile-up effects \citep{Harrison13:NuSTAR}. Throughout this work every \emph{NuSTAR} spectrum was binned to ensure a minimum of 20 counts per bin.
\subsubsection{Swift-XRT}
We processed \emph{Swift-XRT} data through the UK \emph{Swift} Science Data Centre (UKSSDC) XRT tool\footnote{\href{http://www.swift.ac.uk/user_objects/}{http://www.swift.ac.uk/user\_objects/}}, designed to build XRT products \citep{Evans09:XRTspectra}. Spectra were all extracted in Photon Counting mode and the analysis was carried out in the $0.3-10\,$keV energy range. Spectra were then rebinned with a minimum of 20 counts, through the \texttt{group min 20} command within the \texttt{grppha} tool. 
\section{Variability analysis}
\label{sec:timing}
Due to the spectral variability commonly observed in blazars \citep[e.g.][]{Marscher85:extragalsou_variab,Wagner95:IDVBLlac,Ulrich97:AGNvariab} we checked for possible flux variations extracting X-ray light curves for every processed \emph{XMM-Newton} observation. Source and background regions were the same selected for the extraction of the spectra (see Appendix~\ref{sec:appA}). 

After the extraction, light curves were corrected for various effects (vignetting, bad pixels, PSF variation and quantum efficiency, dead time and GTIs) at once with the task \texttt{epiclccorr}. A time bin-size of 500\,s was adopted. The exposure time of the observations set the x-axis, the holes in the data representing the time-regions filtered from FPB. No significant flux variations were observed within the single observation of any source, hence spectral results (see Section~\ref{sec:analysis}) are to be considered free from intra-observation variability.
\section{Spectral analysis}
\label{sec:analysis}

\subsection{Rationale}
\emph{XMM-Newton} data of the three EPIC cameras were jointly fitted (in the $0.2-10\,$keV energy range for the EPIC-pn detector and $0.3-10\,$keV for EPIC-MOS) in Silver sample's blazars, with a floating constant representing the cross-normalization parameter among the different cameras, fixed at 1 for EPIC-pn \citep[see][]{Madsen17:IACHECcrosscalibr}. If several observations were present, the different states of the source were fitted with untied parameters (i.e. photon indexes, normalizations, curvature terms were left free to vary among the different observations). X-ray absorption terms were always tied together.

In case of additional \emph{NuSTAR} observations of the source, a broadband $0.2-79\,$keV fit was performed. Due to the high variability typically observed in blazars, non-simultaneous \emph{XMM-Newton} and \emph{NuSTAR} observations are expected to describe different states of the object, thus we used varying photon indexes, normalizations, curvature terms and spectral breaks. In Golden sample's blazars the simultaneous \emph{Swift-XRT} and \emph{NuSTAR} observations were then fitted keeping the same source parameters, jointly with \emph{XMM-Newton} data fitted using different parameters. The absorption column densities were held fixed between \emph{XMM-Newton} and \emph{Swift-XRT}+\emph{NuSTAR}. Similarly to the adopted procedure for the EPIC cameras, inter-calibration constants were left free to vary for FPMB and \emph{Swift-XRT} with respect to FPMA, fixed at 1 \citep[see][]{Madsen17:IACHECcrosscalibr}.

No significant background contaminations were found in our data, as the observed background-to-source ratio was typically around or below $1\%$. Even in QSO B1026-084, the source with the lowest number of photons ($\sim10\,000$), the ratio reached $\sim10\%$ only above $\sim8\,$keV, and for EPIC-MOS cameras only. Moreover, the impact of the current relative uncertainties on the \emph{XMM-Newton} effective area calibration on our fitted parameters was minimal. We acknowledge the use of the \texttt{CORRAREA} correction\footnote{see\,\href{http://xmm2.esac.esa.int/docs/documents/CAL-SRN-0321-1-2.pdf}{http://xmm2.esac.esa.int/docs/documents/CAL-SRN-0321-1-2.pdf}} for this verification.

\begin{table}[tb]
	\small
	\caption{List of DLA candidates from the literature (see text) included in the spectral analysis.}
	\label{tab:DLAs}
	\centering
	\begin{tabular}{cccc}
		\toprule
		\multicolumn{1}{c}{Source} &
		\multicolumn{1}{c}{$z_{source}$} &		
		\multicolumn{1}{c}{$z_{abs}$} &
		\multicolumn{1}{c}{$N_{HI}$} \\ 
		&
		&
		&
		(unity of $10^{19}\,\text{cm}^{-2}$) \\
		\midrule
		QSO B1026-084 & 4.276 & 3.420 & 12.50  \\ 
		&	&  4.050 & 5.01  \\ 
		PKS 2126-158 & 3.268 & 2.638 & 1.78  \\
		& & 2.769 & 1.58  \\ 		
		QSO B0537-286 & 3.104 & 2.975 & 20.00 \\
		QSO B0438-43 & 2.852 & 2.347 & 60.30  \\
		QSO B0237-2322 & 2.225 & 1.636 & 1.58 \\
		&  & 1.673 & 6.03  \\ 
		\bottomrule		
	\end{tabular}
\end{table}

One or more DLA or sub-DLA systems were detected in the literature towards QSO B0237-2322, QSO B0537-287, QSO B0438-43, QSO B1026-084 and PKS 2126-158 \citep{Peroux2001:optspec66quasar,Ellison01:CORALS-DLA,Fathivavsari13:LOSof0237,Quiret16:subDLA_DLA,Lehner16:LLSs}. The systems were included in the analysis and are shown in Table~\ref{tab:DLAs}. In any case their contribution to the overall curvature is minor.

\subsection{Simple power-law fits}
\label{sec:PL}

\begin{figure*}[tb]
	\centering
	\includegraphics[height=6cm,angle=-90]{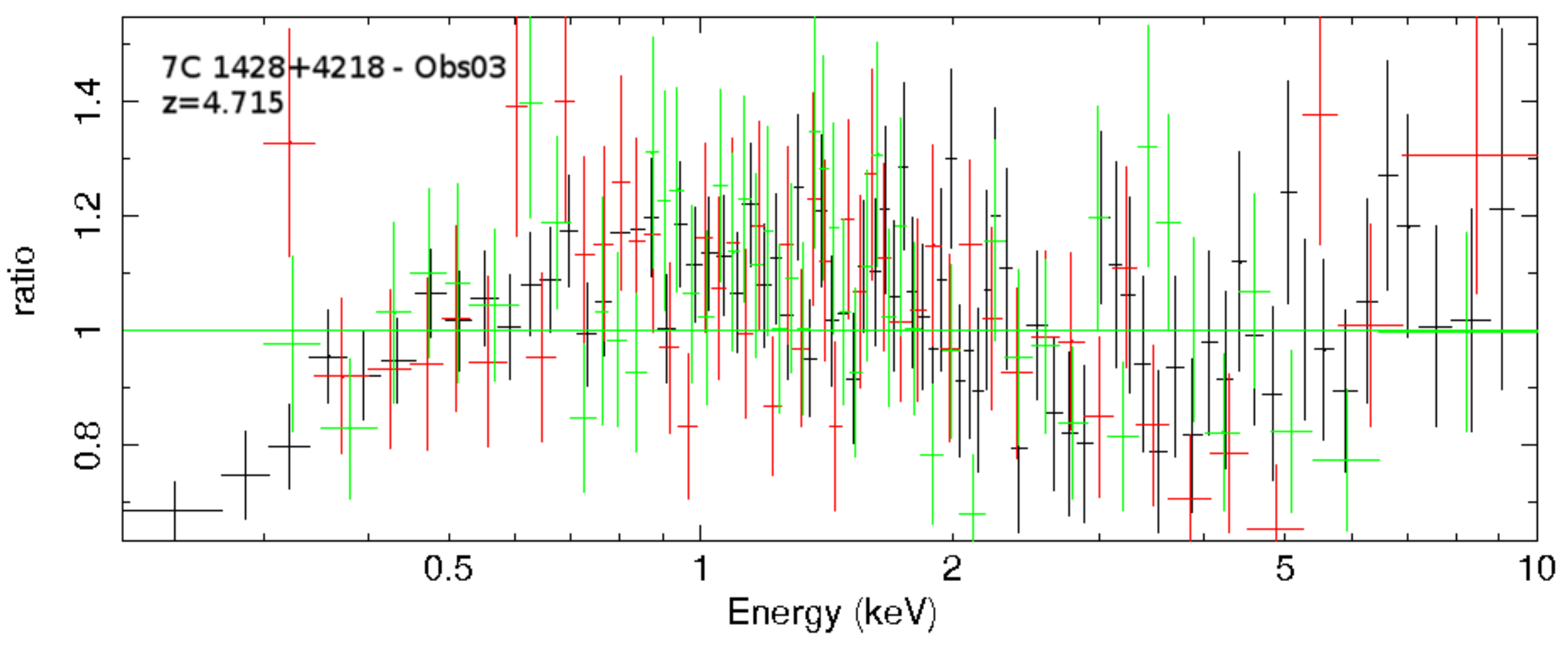}%
	\includegraphics[height=6cm,angle=-90]{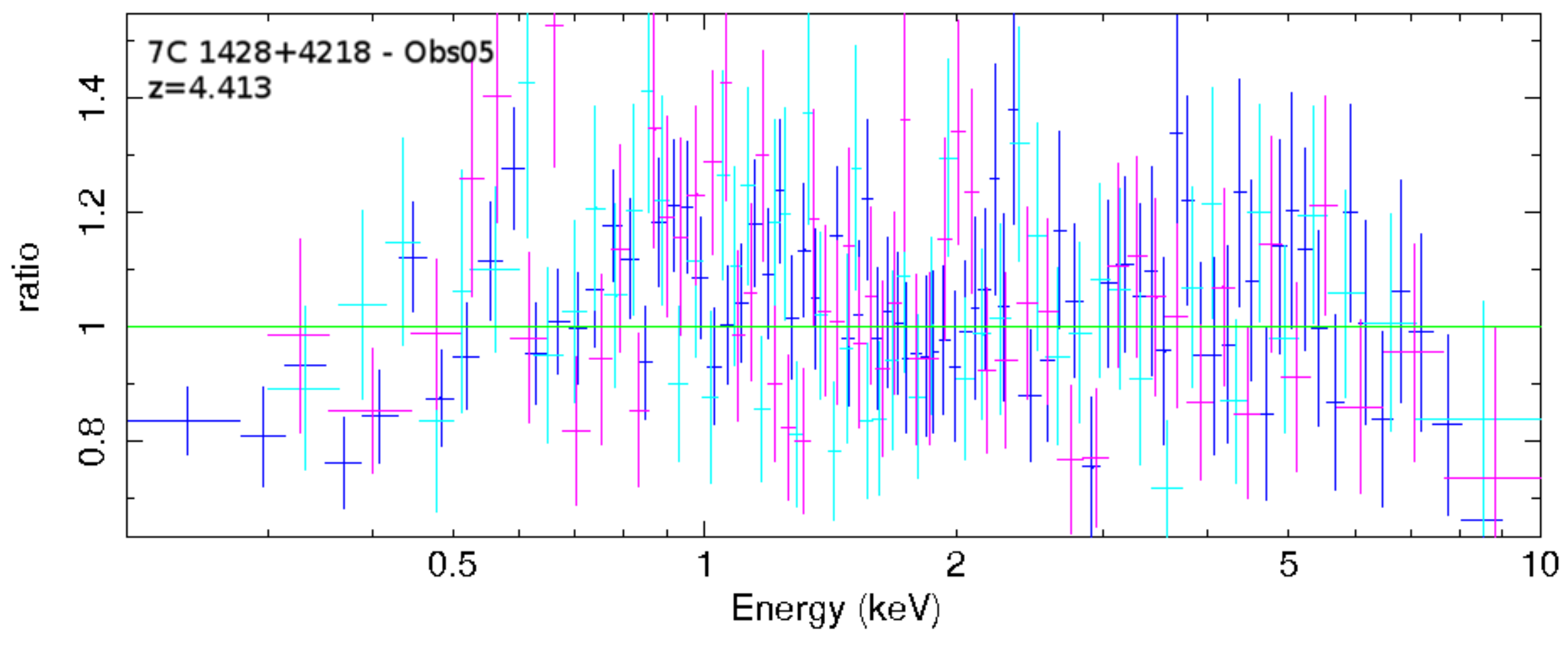}%
	\includegraphics[height=6cm,angle=-90]{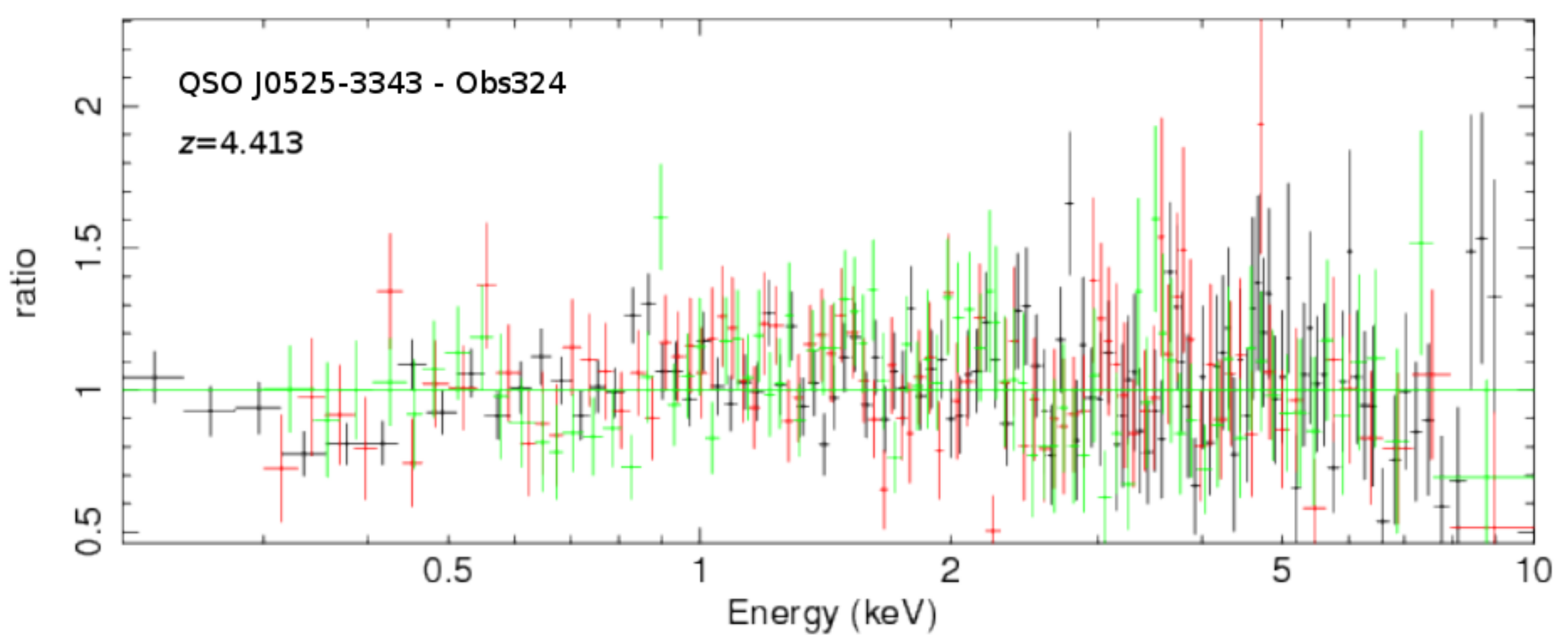}%
	\\
	\includegraphics[height=6cm,angle=-90]{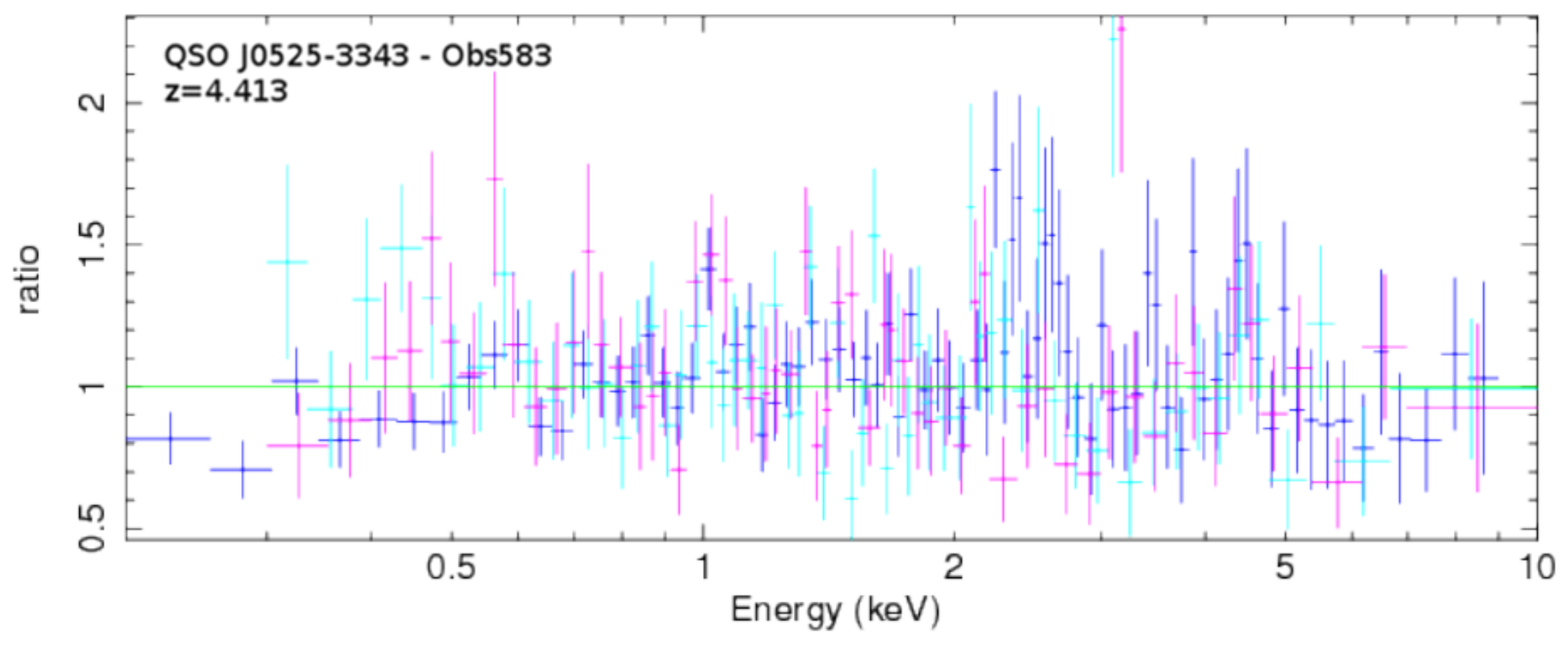}%
	\includegraphics[height=6cm,angle=-90]{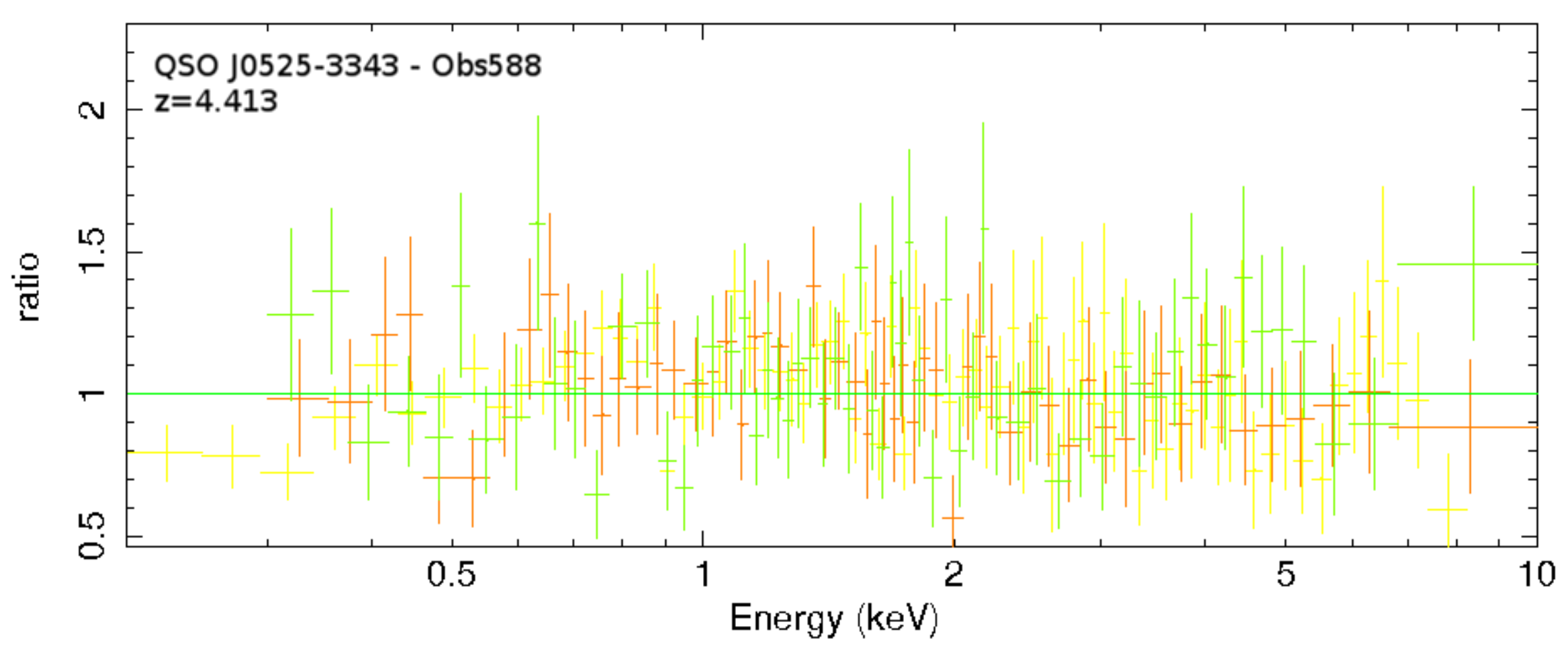}%
	\includegraphics[height=6cm,angle=-90]{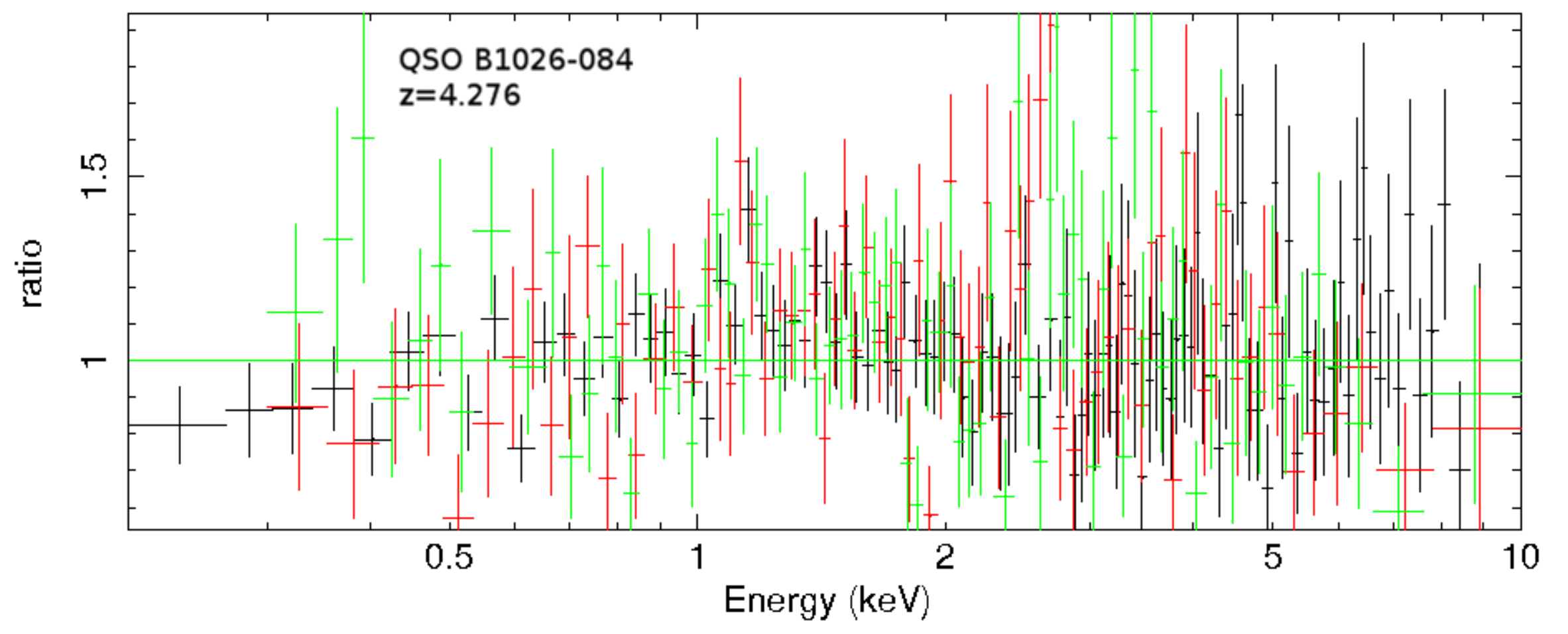}%
	\\
	\includegraphics[height=6cm,angle=-90]{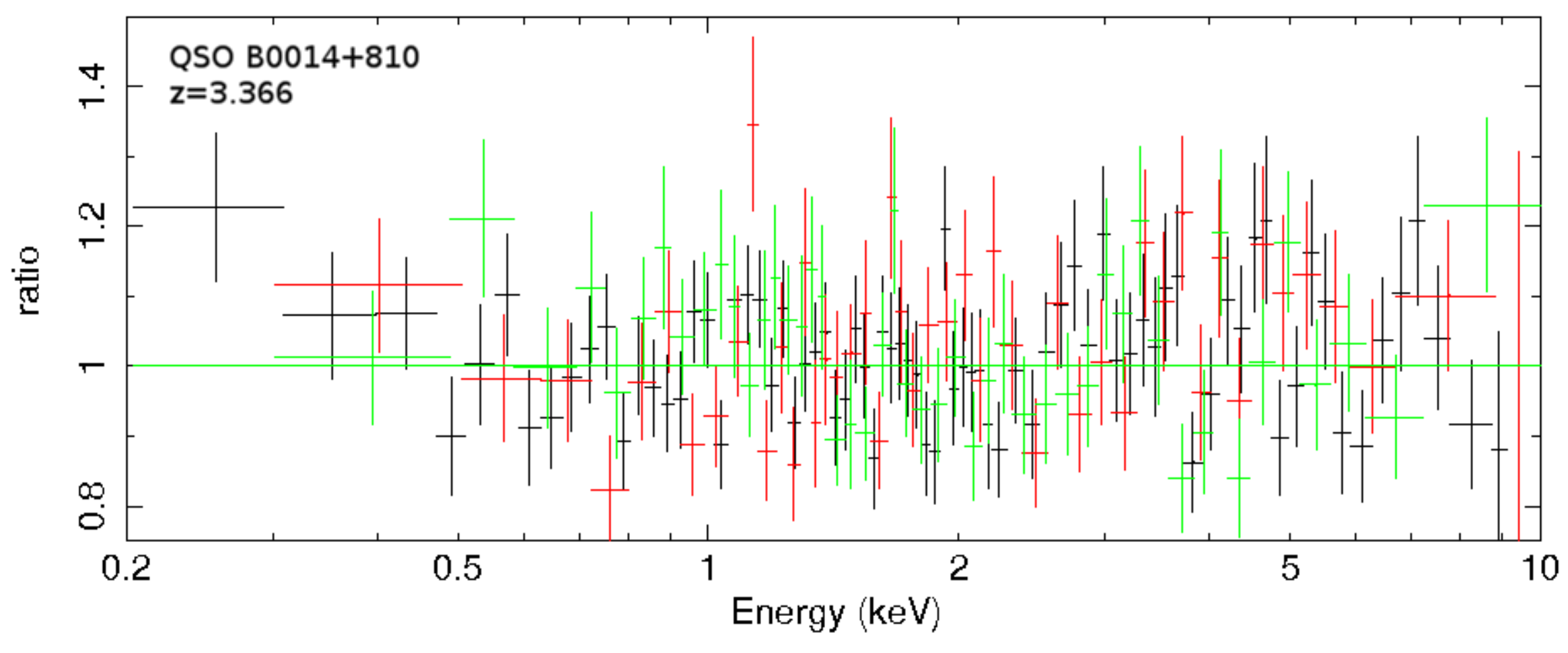}%
	\includegraphics[height=6cm,angle=-90]{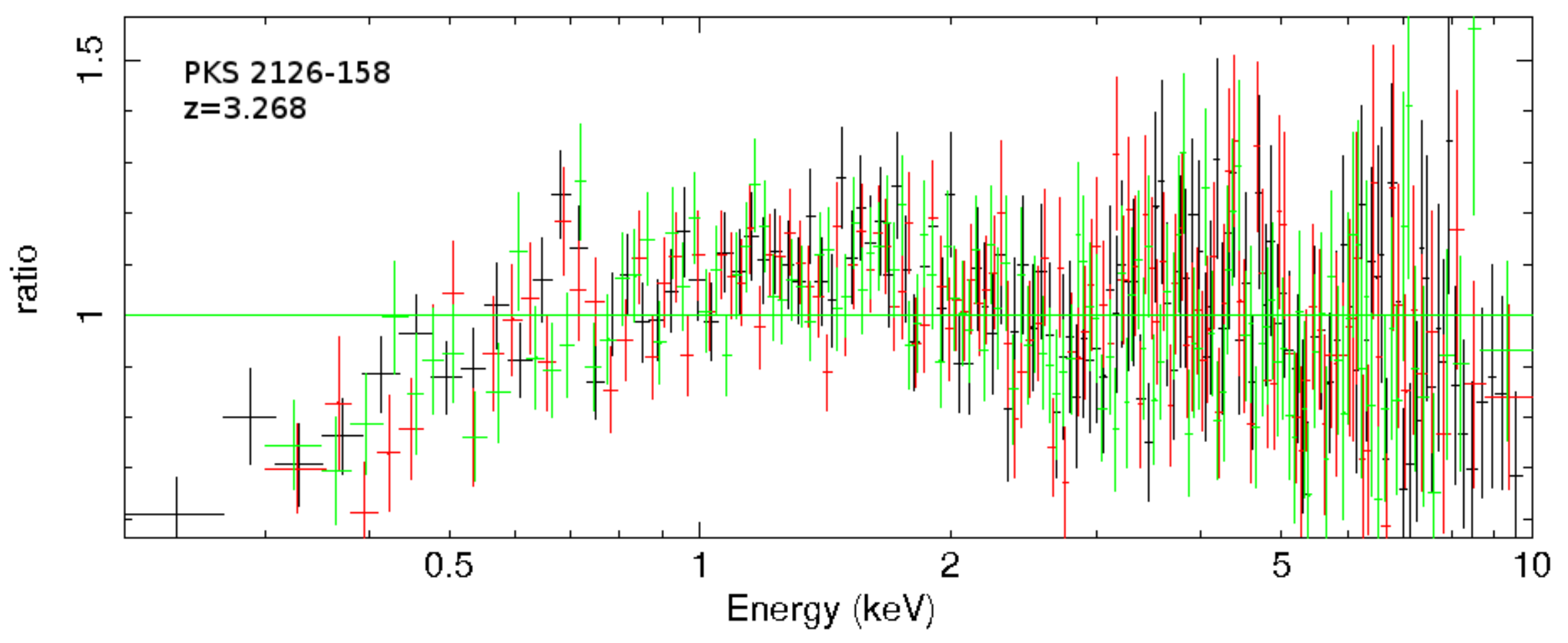}%
	\includegraphics[height=6cm,angle=-90]{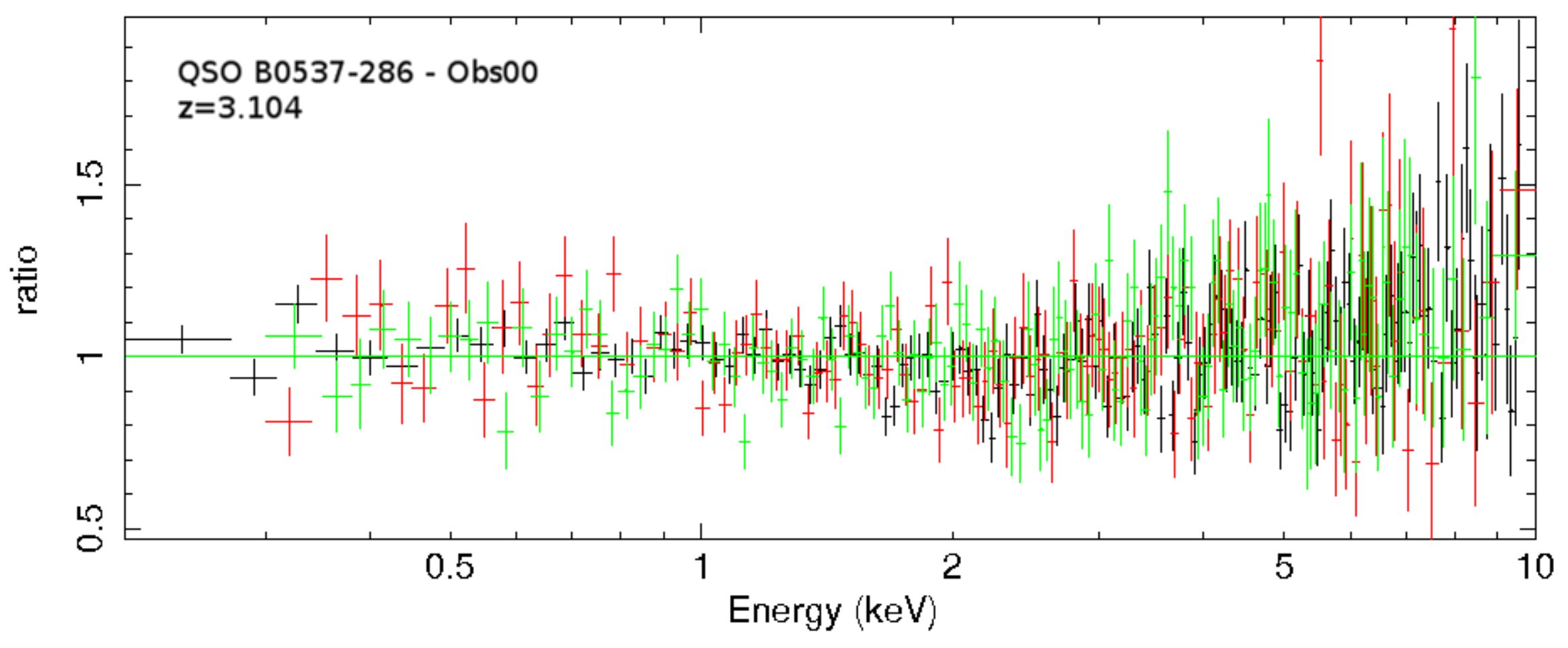}%
	\\
	\includegraphics[height=6cm,angle=-90]{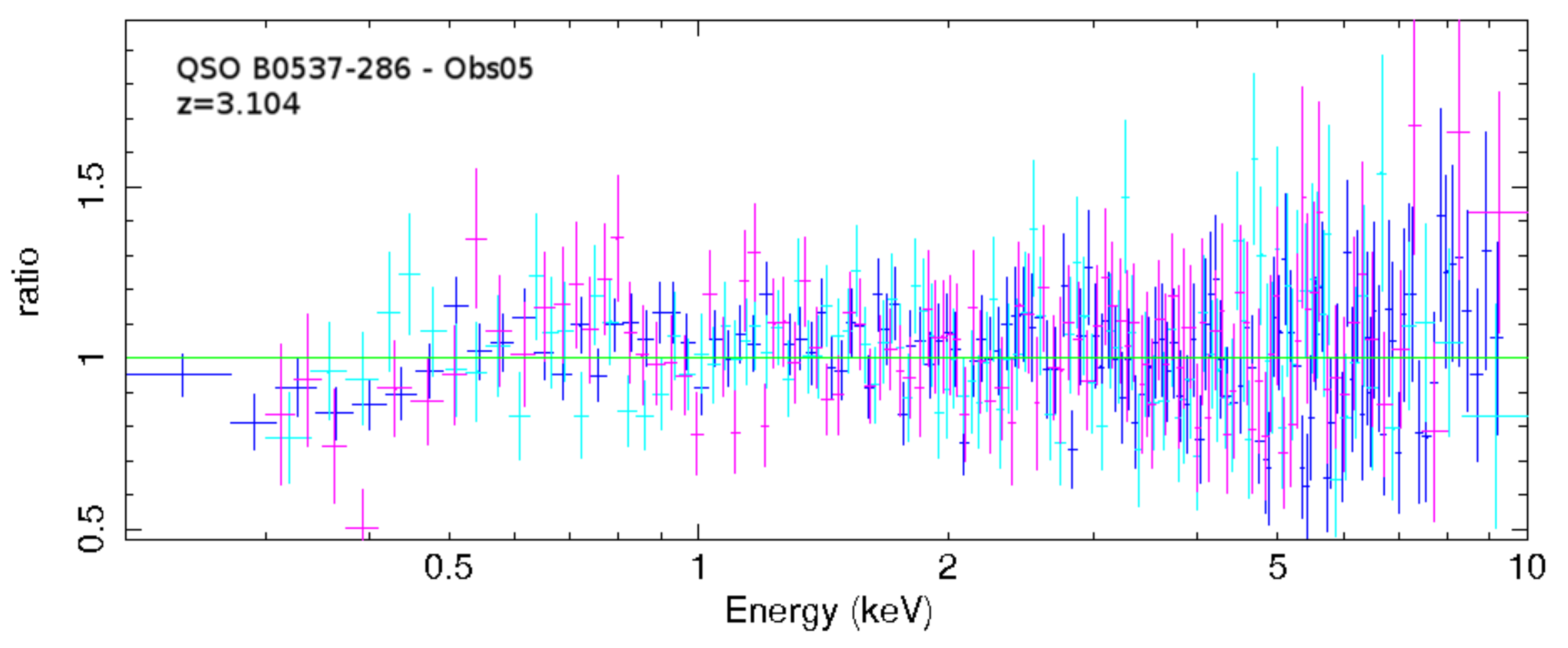}%
	\includegraphics[height=6cm,angle=-90]{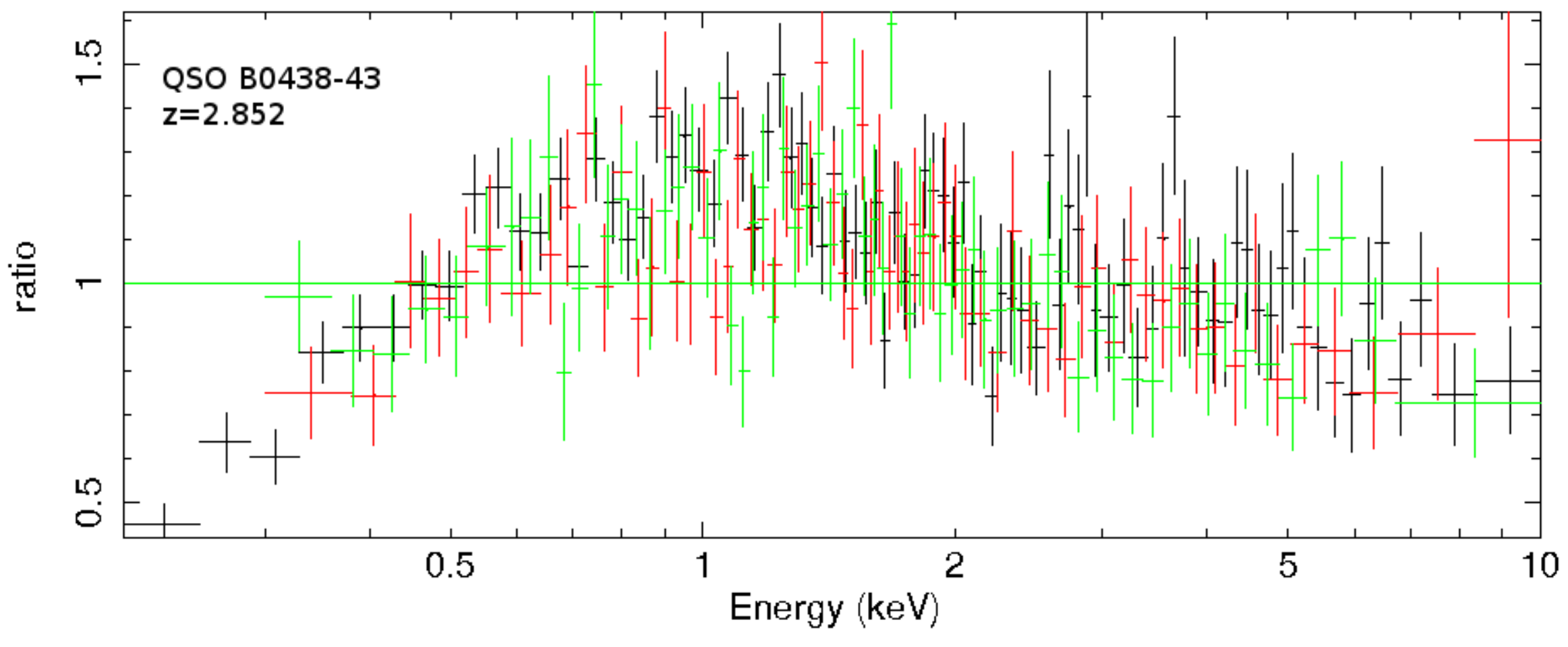}%
	\includegraphics[height=6cm,angle=-90]{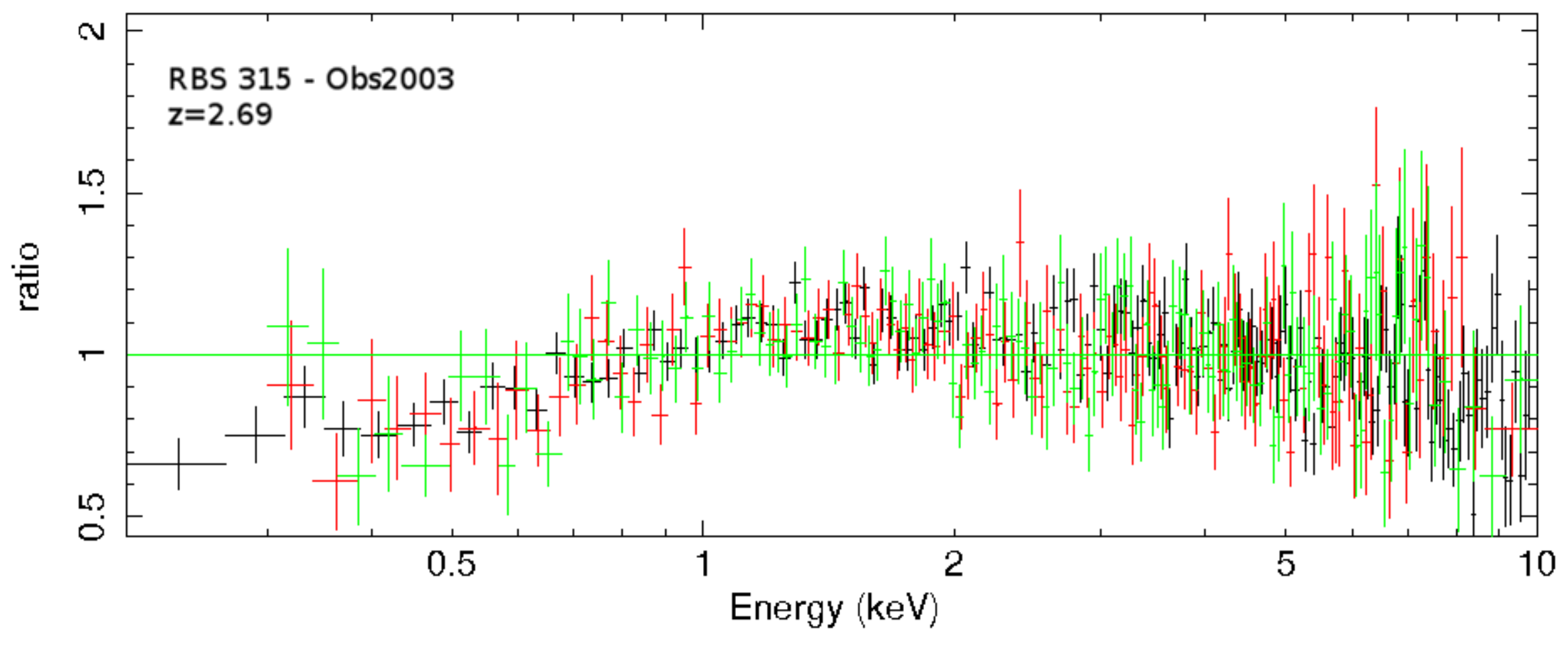}%
	\\
	\includegraphics[height=6cm,angle=-90]{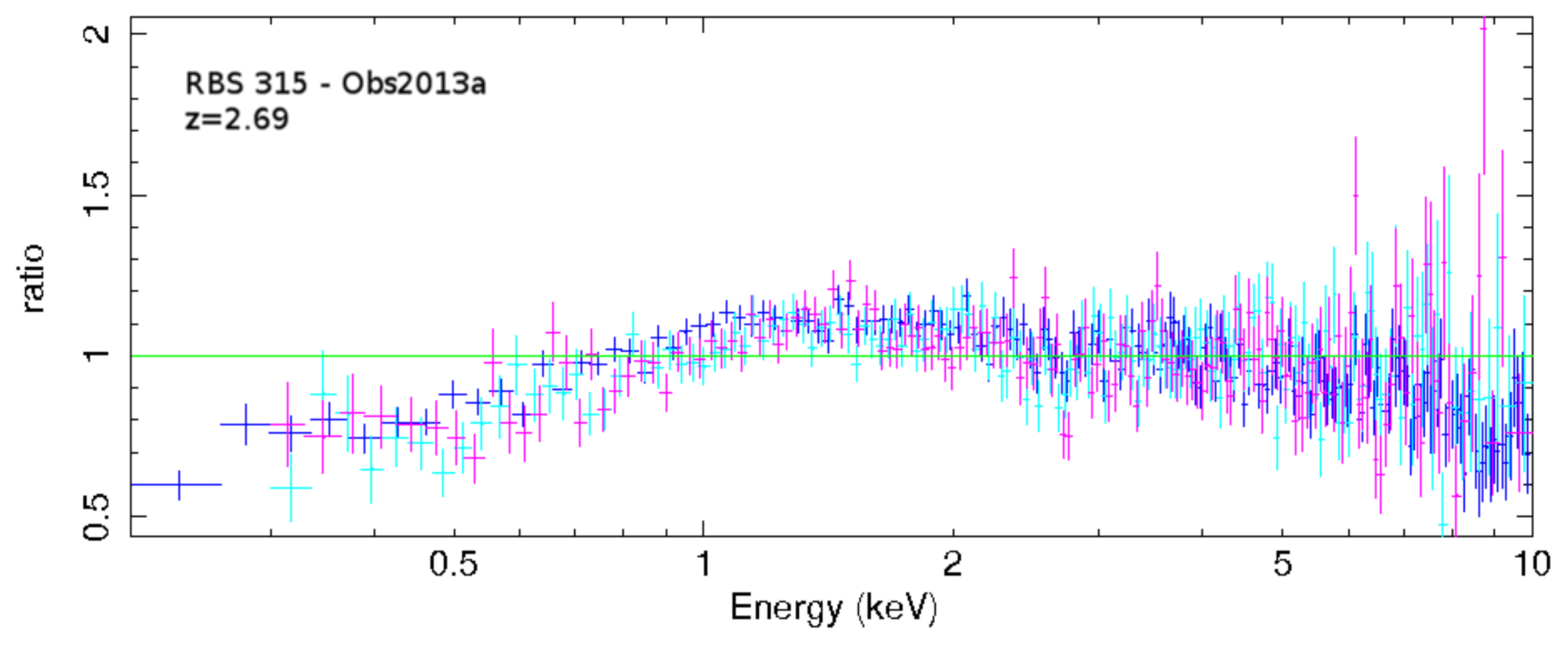}%
	\includegraphics[height=6cm,angle=-90]{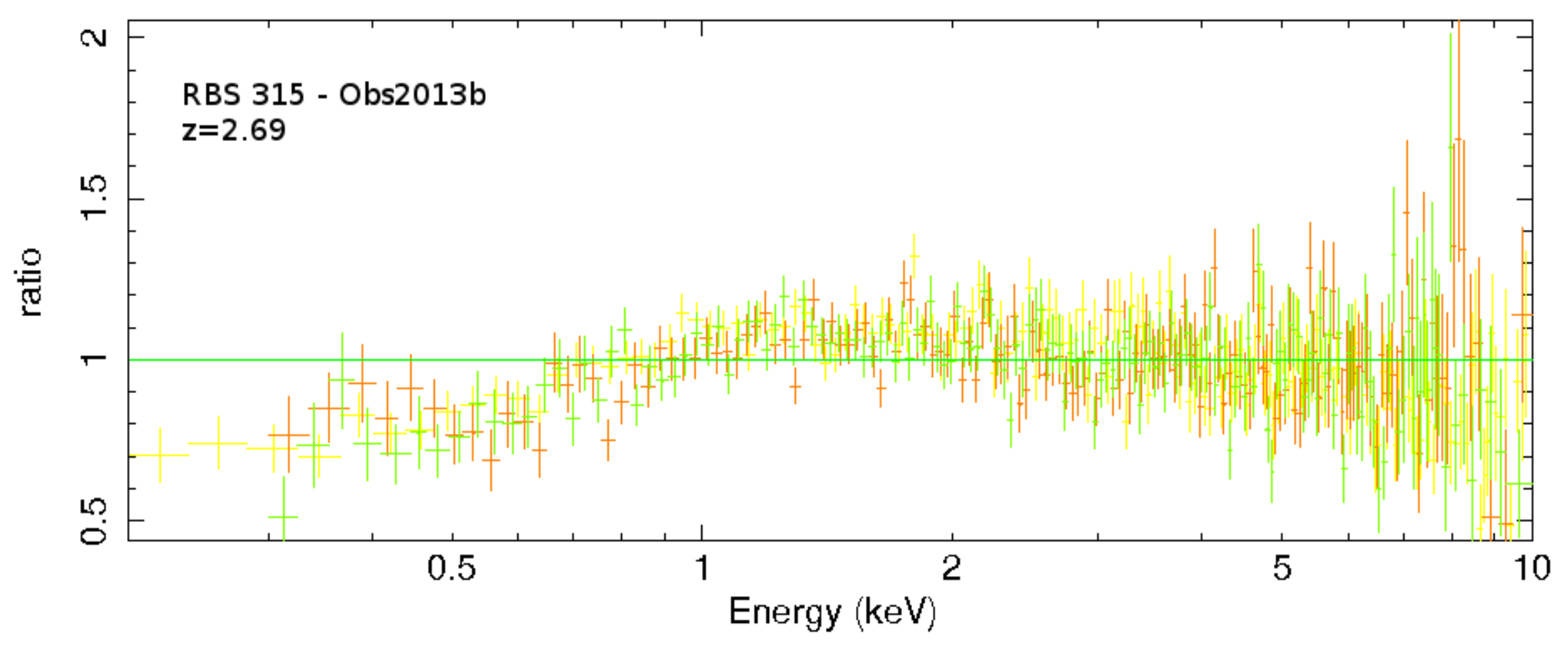}%
	\includegraphics[height=6cm,angle=-90]{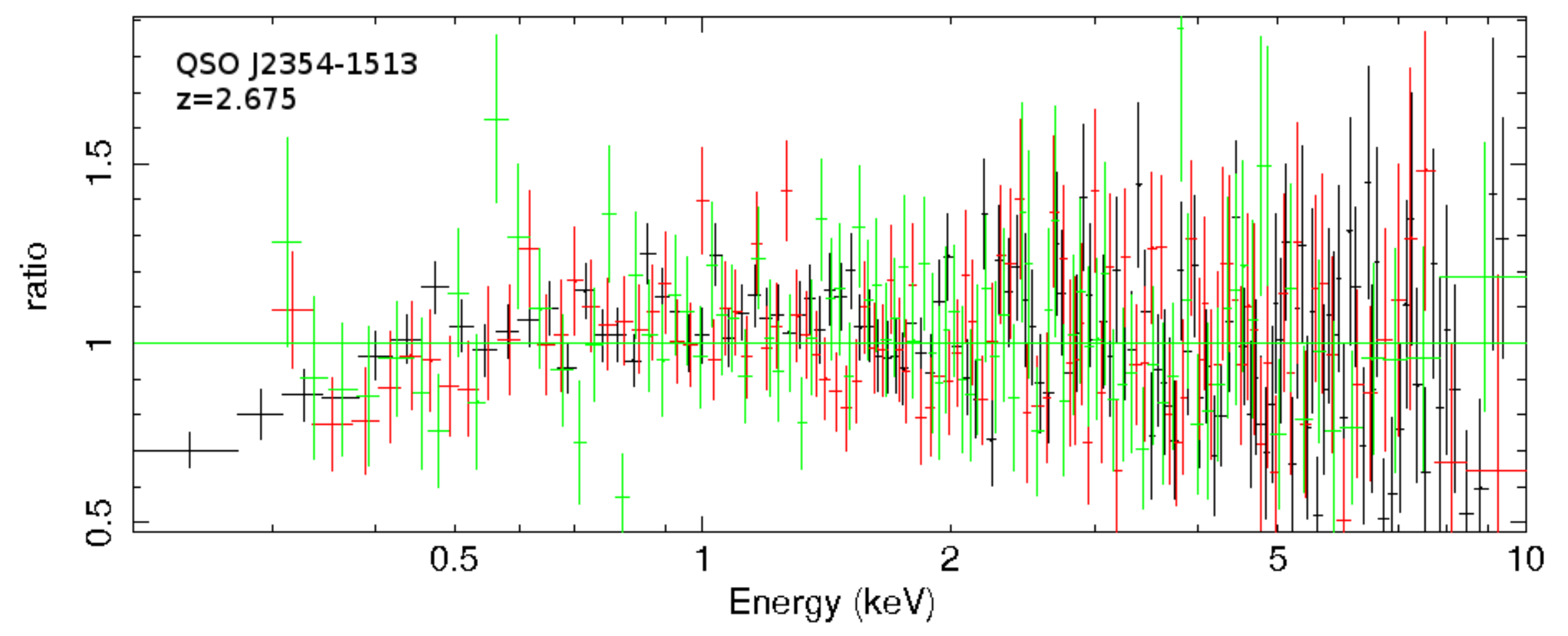}%
	\\
	\includegraphics[height=6cm,angle=-90]{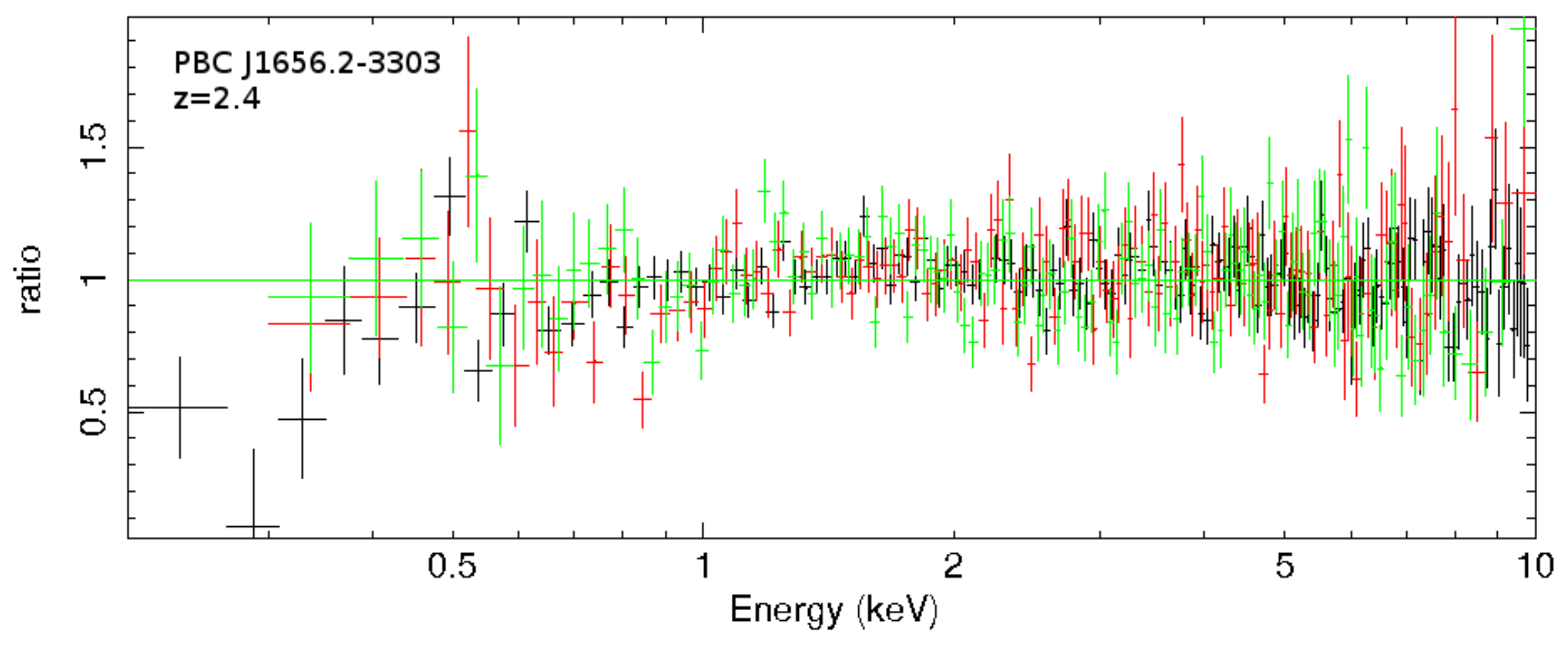}%
	\includegraphics[height=6cm,angle=-90]{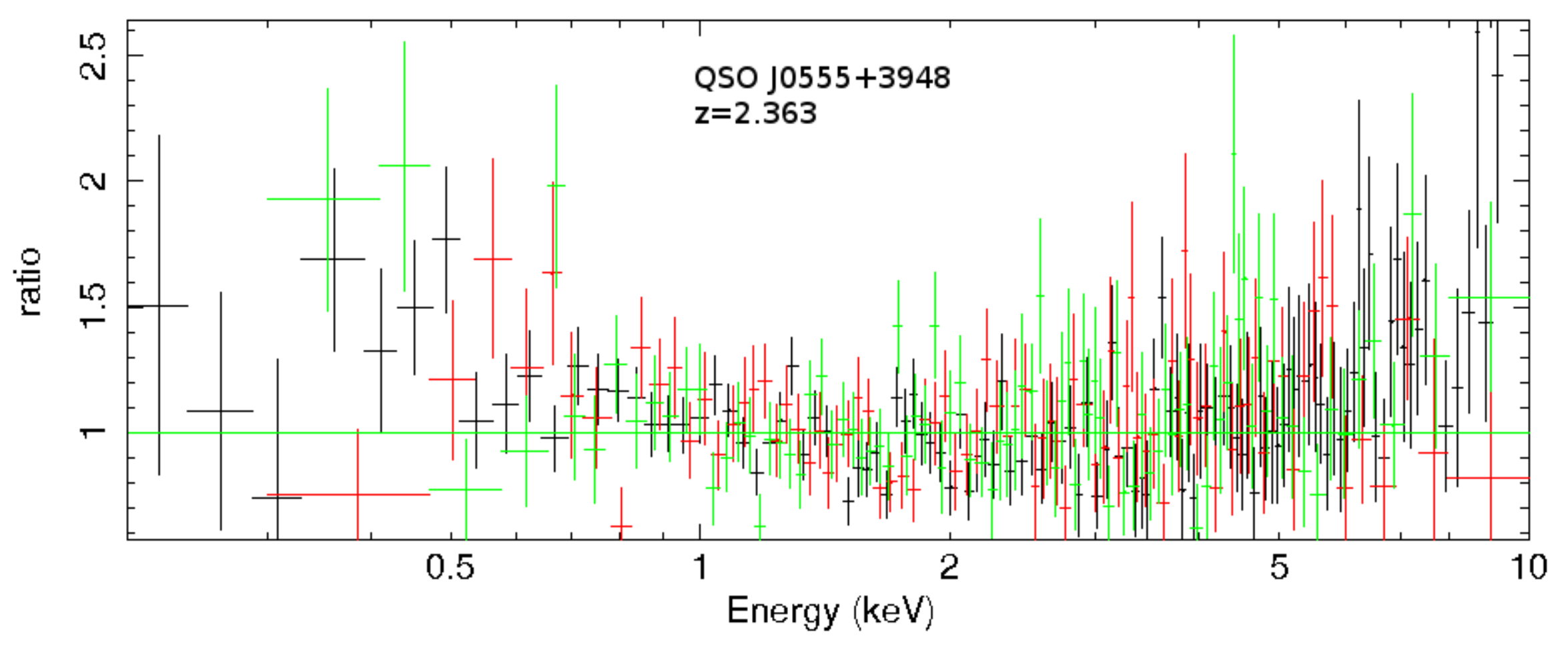}%
	\includegraphics[height=6cm,angle=-90]{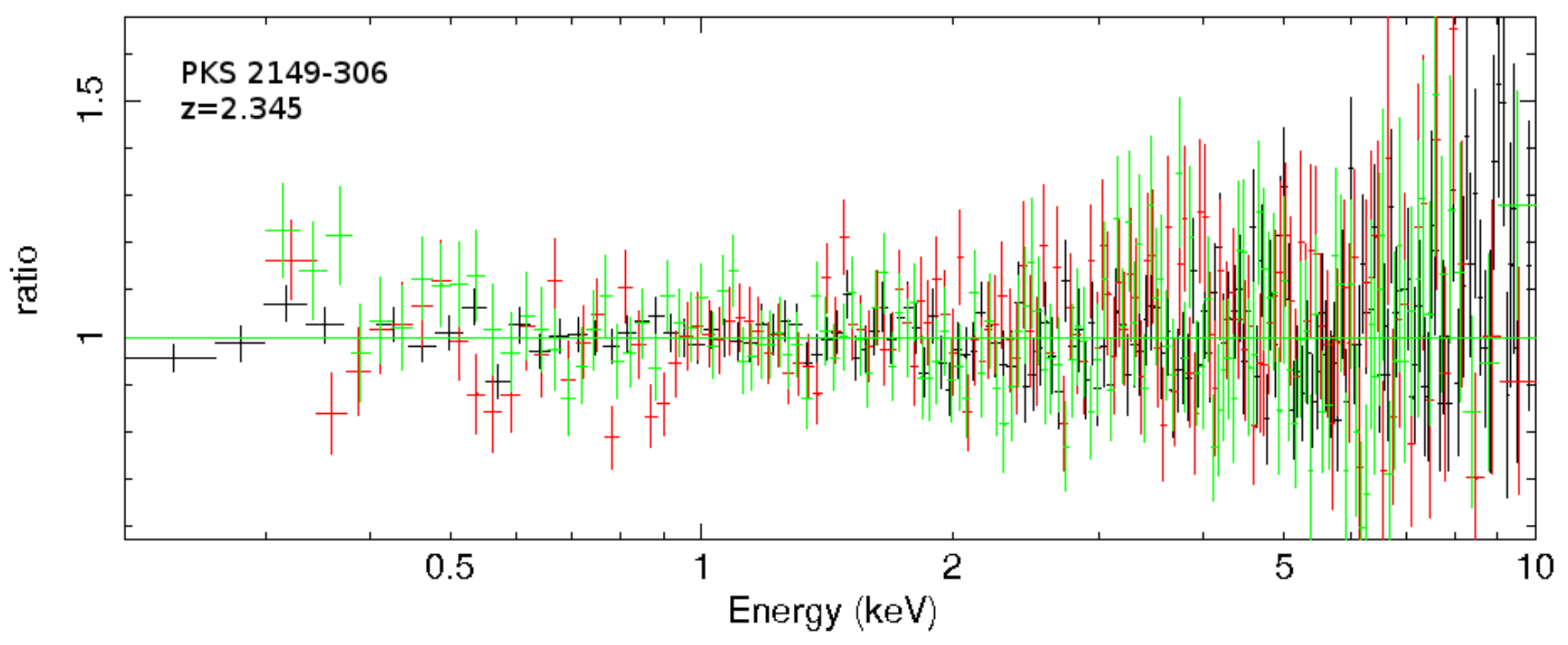}%
	\\
	\includegraphics[height=6cm,angle=-90]{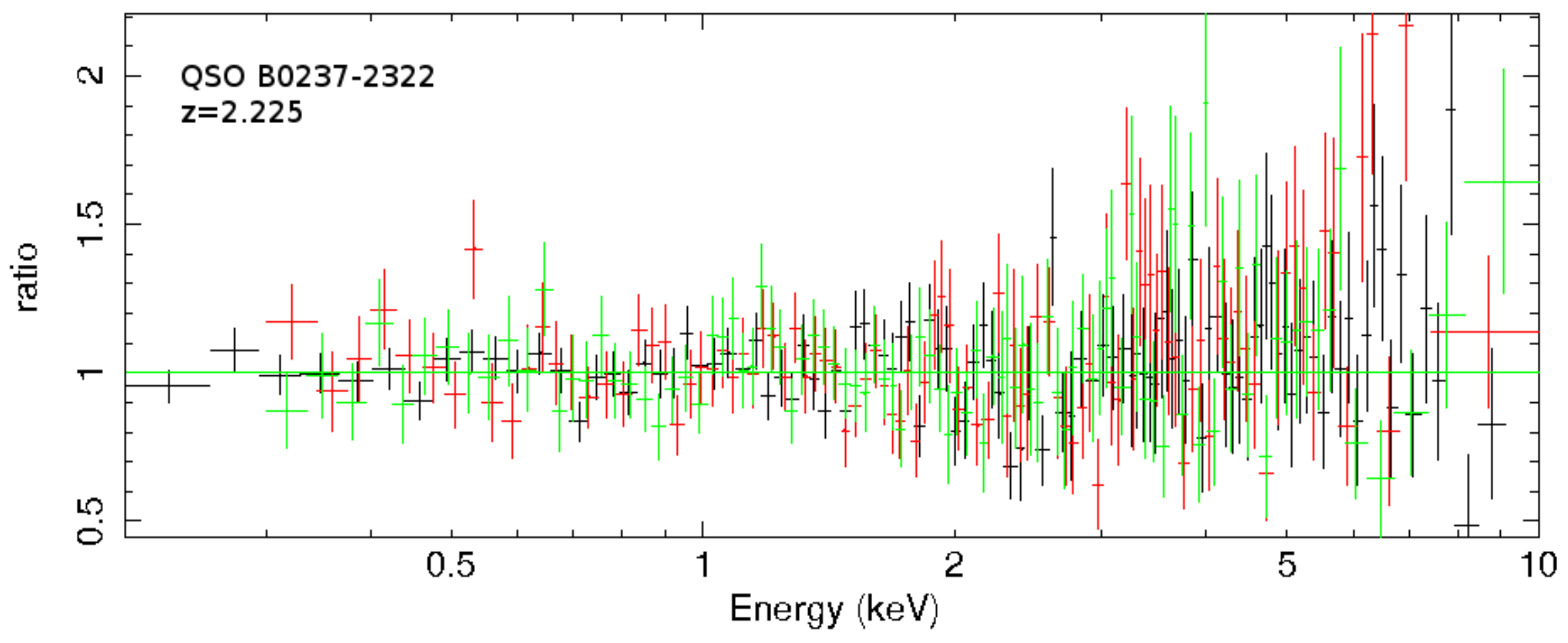}%
	\includegraphics[height=6cm,angle=-90]{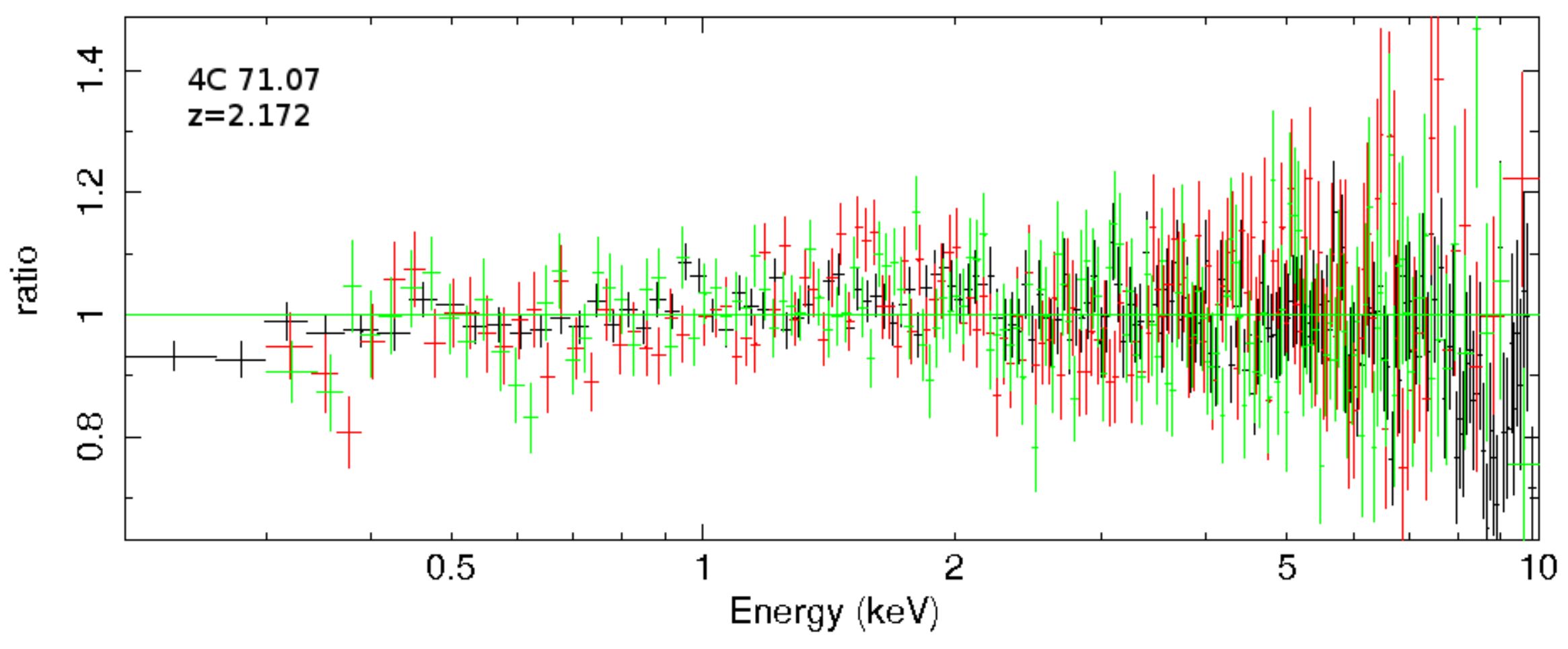}%
	\includegraphics[height=6cm,angle=-90]{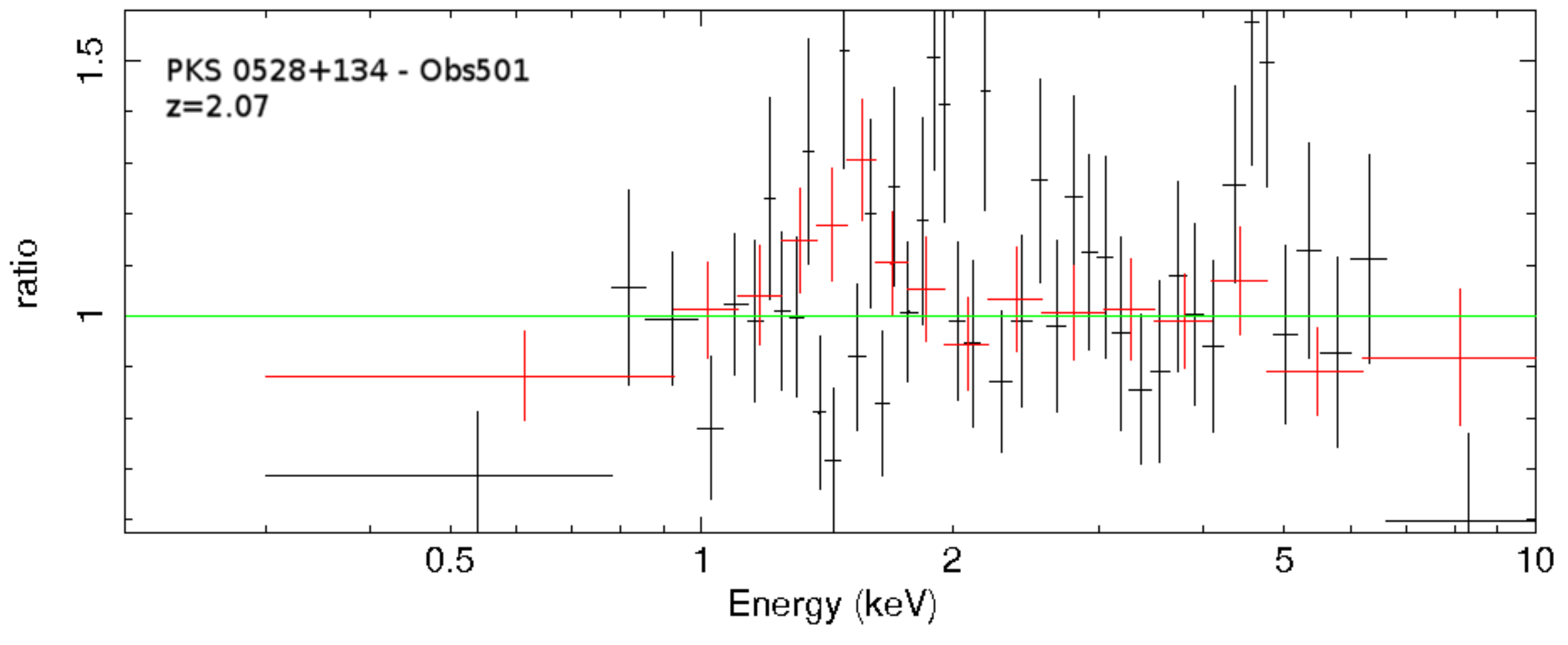}%
	\\
	\includegraphics[height=6cm,angle=-90]{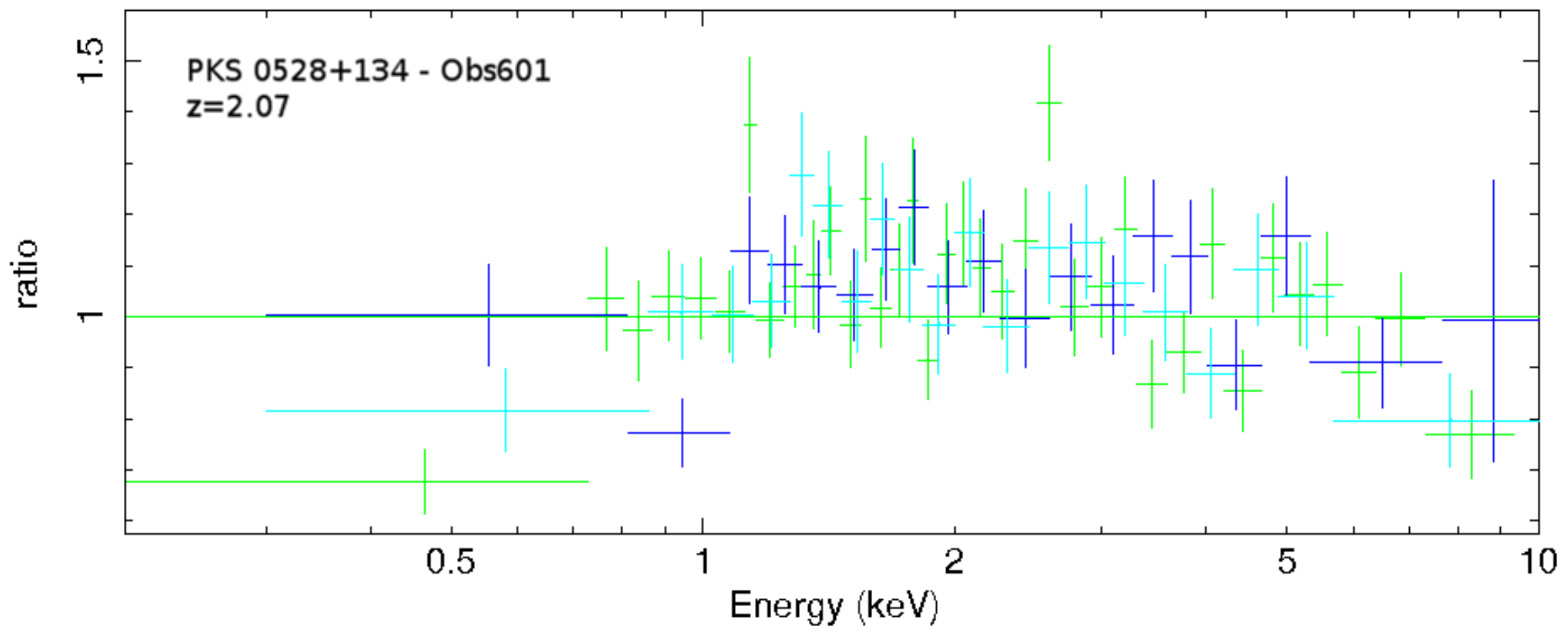}%
	\includegraphics[height=6cm,angle=-90]{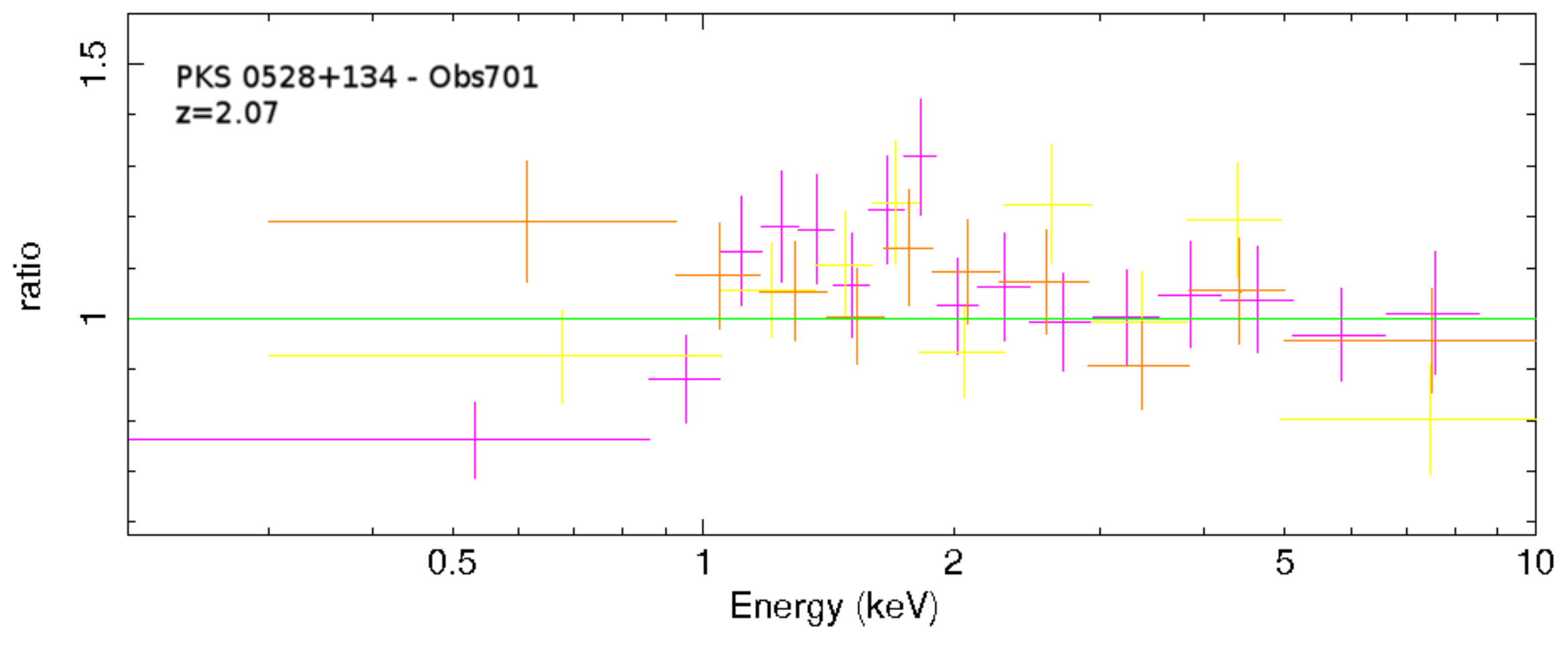}%
	\caption{Data/Model ratio of a simple PL model for \emph{XMM-Newton} observations of Silver-sample blazars. EPIC-pn, MOS1 and MOS2 data are displayed for each observation in different colours, with pn data extending down to 0.2\,keV.}
	\label{fig:simplePLrat}
\end{figure*}

Blazars' emission can be approximated by simple power-laws in limited energy ranges, e.g. within the rise (in $\nu F_{\nu}$) of the IC hump. This is the SED region that we are likely observing in $z>2$ FSRQs. Then, we first modelled the observed spectra using a power-law continuum with fixed Galactic column density \citep{Willingale13:GalacticH2}. This "null" model, hereafter PL, is described by:
\begin{equation*} N(E)=Ke^{-N_H\sigma(E)}E^{-\Gamma}
\end{equation*}
in which the photon flux $N(E)$ [photons s$^{-1}$ cm$^{-2}$ keV$^{-1}$] is modelled with a power-law with photon index $\Gamma$ (\texttt{powerlaw} within \texttt{XSPEC}) and an exponential cut-off caused by a column density of absorbing matter (in unity of $10^{22}$ atoms/cm$^{-2}$) interacting with an energy-dependent cross-section $\sigma(E)$. $K$ is the normalization at 1 keV. This Tuebingen-Boulder ISM absorption model (\texttt{tbabs} within \texttt{XSPEC}) is actually the improved version \texttt{tbnew}\footnote{\href{http://pulsar.sternwarte.uni-erlangen.de/wilms/research/tbabs/}{http://pulsar.sternwarte.uni-erlangen.de/wilms/research/tbabs/}}, automatically included within \texttt{XSPEC} 12.9.1. Cross sections from \citet{Verner96:ISMcrosssec} and abundances from \citet{Wilms00:ISMabs} are used.

PL fit results of \emph{XMM-Newton} data are shown in Table~\ref{tab:spectral_analysis} for each source. Related data-model ratios are reported in Figure~\ref{fig:simplePLrat}. They both suggest that a significant additional curvature is required in almost all objects, except for QSO B0014+810, PKS 2149-306 and QSO B0237-2322. The aggregate reduced chi-square is $\chi^2_{\nu,tot}=1.414$ (10675/7552), suggesting that a more complex modelling overall is needed.

Adding \emph{NuSTAR} (with simultaneous \emph{Swift-XRT}) data to the analysis allowed us to extend the observing bandwidth up to 79 keV in the six blazars belonging to the Golden sample. The results are reported in Table~\ref{tab:spectral_analysis} for the single source. The total reduced chi-square for the PL model is $\chi^2_{\nu,tot}=1.275$ (12150/9529) and confirms that some additional curvature is suggested, to a greater or lesser extent, in all the high-$z$ blazars.

\subsection{Intrinsic curvature fits}
\label{sec:CURV}

The curvature in addition to the PL model could be due to spectral breaks intrinsic to the emission. Such features are predicted by blazars' emission models \citep[e.g.][]{Sikora09:blazarsmodels,Tavecchio07:RBS315Suzaku-intrinsicbreak,Ghisellini09:CanonicalBlazars-7Csed,Ghisellini15:blazarsMODELS} and details will be discussed in Appendix~\ref{sec:appB}.

The power-law continuum can be improved with a broken power-law (BKN) or with a log-parabola (LGP), still with a fixed Galactic absorption value. The broken power-law model (\texttt{bknpower} within \texttt{XSPEC}) simply consists in two different power-laws separated by a break at $E_b$ (in keV):
\begin{equation*} N(E)=\begin{cases}
		K E^{-\Gamma_1} & \text{  if  } E\leq E_b
		\\
		K E_b^{\Gamma_2-\Gamma_1} E^{-\Gamma_2} & \text{  if  } E>E_b
	\end{cases}
\end{equation*} where $\Gamma_1$ and $\Gamma_2$ are the low- and high-energy photon index, respectively. 

The log-parabolic model \citep[][\texttt{logpar} within \texttt{XSPEC}]{Massaro04:logpar_blazars,Massaro06:logpar_blazars} is given by the following equation: \begin{equation*}N(E)=K\left(\frac{E}{E_1}\right)^{-a-b\log(E/E_1)}\end{equation*}where $E_1$ is the fixed pivot energy (typically $1\,$keV in soft X-ray fits), $a$ is the slope at $E_1$ and $b$ the curvature term. In both BKN and LGP models, the photon flux is absorbed by a Galactic column density represented by the same exponential cut-off of the PL equation.

Results obtained with both models are shown in Table~\ref{tab:spectral_analysis} for each source, along with the F-test \citep{Protassov02:ftest} $p$-value computed with respect to the null PL model, that represents a clear improvement in most cases. In order to compare the overall improvement, we then calculated the total reduced chi-square for both BKN and LGP model, obtaining $\chi^2_{\nu,tot}=1.035$ ($\chi^2_{tot}/dof_{tot}=7758/7498$) and $1.094$ (8228/7521), respectively. The F-test yielded a telling $p$-value $<10^{-200}$ in both cases. When broadband data are fitted for Golden sample's sources, the narrowband conclusions are confirmed (see Table~\ref{tab:spectral_analysis} for individual results). The overall reduced chi-squares are $1.027$ (9728/9473) and $1.043$ (9899/9491), for BKN and LGP model, respectively. The related F-test $p$-values are again $<10^{-200}$.

\subsection{Excess absorption fits}
\label{sec:EX}

The PL model can also be improved adding absorption in excess of the Galactic value, to account for the additional curvature required. We already stressed the concept that in blazars any excess absorber should be considered intervening, since no intrinsic absorption likely occurs due to the presence of a relativistic jet sweeping the local environment up to kpc-scales. However, using a cold absorber intrinsic to the host galaxy (\texttt{ztbabs} within \texttt{XSPEC}) is the easiest and fastest way to investigate the presence of additional absorbers in excess of the Galactic value.

Individual fit results for all blazars of the Silver sample are reported in Table~\ref{tab:spectral_analysis}. In general, excess absorption always improved the simple PL fit. The majority of sources yielded a detection of a significant column density, while un upper limit was obtained for QSO B0014+810, QSO B0537-286, PBC J1656.2-3303, QSO J0555+3948, PKS 2149-306, QSO B0237-2322. The total reduced chi-square was computed, i.e. $\chi^2_{\nu, tot}=1.048$ (7880/7520). The F-test $p$-value with respect to the PL model is $<10^{-200}$.

When \emph{NuSTAR} (with simultaneous \emph{Swift-XRT}) data are added to the analysis, the fitted column densities are fully consistent, within the errors, with the results of the "narrow-band" \emph{XMM-Newton} fits. The overall reduced chi-square is $1.081$ (10265/9497), yielding a $p$-value of $<10^{-200}$.

Note that in this model (PL+EX), the Galactic value was left free to vary between $\pm15\%$ boundaries of the tabulated value (see Section~\ref{sec:lowzanalysis} for a motivation). However, this choice did not favour the detection of excess absorption within spectral fits, since the Galactic value was fitted towards the lower boundary allowed by the $\pm15\%$ errors only in 5 blazars out of 15. On the contrary, in 8 sources it was fitted towards the upper boundary, thus disfavouring any extra-absorber.

\subsection{Intrinsic curvature + excess absorption fits}
\label{sec:CURV+EX}

Poor PL fits were adequately improved with both excess absorption \emph{or} an intrinsic spectral break. We can not discern which model among BKN, LGP and PL+EX is better using the F-test, nor looking at the residuals, as these models are nearly statistically undistinguishable for the single source. A more complex modelling could be hardly introduced by these arguments. Nonetheless, we fitted \emph{XMM-Newton} spectra with both models simultaneously, assuming a priori that radiation coming from every $z>2$ source could be partly absorbed along the IGM and that, in addition, for some of the sources an intrinsic energy break could have occurred within the observed energy band. Then, a posteriori we verified the inclusion of this model in the analysis with more thorough statistical tools and arguments (see Section~\ref{sec:bestfitmodel}).

This LGP+EX model includes \texttt{ztbabs} and \texttt{logpar} (the LGP was chosen as reference for modelling a curved continuum, see Sections~\ref{sec:bestfitmodel} and~\ref{sec:lowzanalysis}). Note that also in the LGP+EX fits the Galactic value was left free to vary between $\pm15\%$ boundaries. Results are reported in Table~\ref{tab:spectral_analysis}, along with the F-test $p$-value computed with respect to the PL+EX model, for each individual source. The total reduced chi-square for the LGP+EX model, i.e. $\chi^2_{\nu,tot}=1.028$ (7705/7495), is a clear improvement of the PL+EX model (F-test $p$-value$\sim 10^{-23}$), but also of the LGP ($p$-value$\sim 10^{-87}$). 

Moreover, despite the presence of some degeneracies, we were able to draw general conclusions. The excess absorption component was always fitted, with column density values compatibles with the PL+EX scenario, while continuum curvature terms were consistent with a power-law in 11 out of 15 blazars. Only in few cases both terms appeared to be required by the data, e.g. in QSO B0537-286, RBS 315, QSO J0555+3948 and 4C 71.07. These sources were fundamental, since they proved that when excess absorption is present \emph{and} some intrinsic curvature is within the observed band, they both can be fitted. 

Also when \emph{NuSTAR} (with simultaneous \emph{Swift-XRT}) data were added to the analysis in the 6 blazars of the Golden sample, they were, as a general rule, better modelled with a LGP+EX (see Table~\ref{tab:spectral_analysis} for individual results). Exceptions were 7C 1428+4218 and QSO B0014+810, in which a curved continuum was not statistically required. The overall chi square is $1.018$ (9654/9480), with $p$-values of $\sim 10^{-44}$ and $\sim 10^{-113}$ with respect to LGP and PL+EX models, respectively.

\subsection{Best-fit model}
\label{sec:bestfitmodel}

Using the F-test, we were only able to tell that every suggested alternative model (namely BKN, LGP, PL+EX and LGP+EX) was a clear refinement with respect to a simple PL model, with no information on the relative quality between these models. We now want to infer the overall best-fit model, balancing the quality of the fit (given by the chi-square statistic) with the complexity of the model (the number of parameters involved), taking always into account the physics behind it.

The ideal statistics for this purpose is represented by the Akaike Information Criterion \citep[AIC,][]{Akaike74:AIC}, since it can be used to compare non-nested models as well. The AIC has been widely applied to astrophysical problems \citep[e.g.][]{Liddle04:AIC,Liddle07:AIC,Tan12:AIC}, defined as:\begin{equation*}AIC=-2\ln L_{max}+2k\end{equation*}where $L_{max}$ is the maximum likelihood that can be achieved by the model and $k$ is the number of parameters of the model. The second term is a penalty for models that yield better fits but with many more parameters. With the assumption of Gaussian-distributed errors, the equation further reduces to:\begin{equation}\label{eq:AIC}AIC=\chi^2+2k\end{equation}where $\chi^2$ is yielded by the spectral fits for each model. Hence, the model with the smallest AIC value is determined to be the "best", although a confidence level needs to be associated for distinguishing the best among several models. Given two models $A$ and $B$, $A$ is ranked to be better than $B$ if \begin{equation*}\lvert\Delta_{A,B}\rvert=\lvert AIC(A)-AIC(B)\rvert >\Delta_{threshold}\end{equation*}where $\Delta_{threshold}$ is conventionally 5 (10) for a "strong" ("decisive") evidence against the model with higher criterion value \citep[see][and references therein]{Liddle07:AIC}.

We computed the AIC for \emph{XMM-Newton} results of each Silver-sample blazar (see Table~\ref{tab:AIC}), confirming the ambiguity outlined in the previous sections, as well as in other works. For almost each individual source, the model with the lowest AIC (among BKN, LGP, PL+EX and LGP+EX) had at least another model within a $\Delta_{threshold}=10$. In Table~\ref{tab:AIC} we highlighted in bold the lowest AIC and in italics any additional model within a $\Delta_{threshold}=10$. This states that as long as the single source is analysed, the suggested models are mostly statistically undistinguishable.

\begin{table}[tb]
	\tiny
	\caption{AIC values for each blazar of the Silver sample, computed with \emph{XMM-Newton} spectral fits from Eq.~\ref{eq:AIC}. BKN, LGP, PL+EX, LGP+EX stand for the different models used, the reader is referred to the top of this Section for a description. For each blazar, we highlighted in bold the lowest AIC and in italics any additional model within a $\Delta_{threshold}=10$.}
	\label{tab:AIC}
	\centering
	\begin{tabular}{ccccc}
		\toprule
		\multicolumn{1}{c}{Source} &
		\multicolumn{4}{c}{AIC} \\
		&
		\multicolumn{1}{c}{BKN} &
		\multicolumn{1}{c}{LGP} &
		\multicolumn{1}{c}{PL+EX} & 
		\multicolumn{1}{c}{LGP+EX} \\
		\midrule
		7C 1428+4218 	& \emph{491} & 545 & \textbf{489} & \textbf{489}\\
		QSO J0525-3343 	& \textbf{704} & \emph{713}  & \textbf{704} &  \emph{708}\\
		QSO B1026-084 	& \textbf{288} & \emph{295} & \emph{296} & \emph{296} \\ 
		QSO B0014+810 	& \textbf{393} & \emph{398} & \emph{398} & \emph{400} \\
		PKS 2126-158 	& 413 & 446 & \textbf{399} & \emph{400} \\
		QSO B0537-286 	& \textbf{803} & 829 & 874 & \emph{810} \\
		QSO B0438-43 	& 314 & 422 & \emph{288} & \textbf{287} \\
		RBS 315 		& 1538 	& 1690 & 1555 & \textbf{1405} \\
		QSO J2354-1513	& \emph{348} & 370 & \textbf{344} & \emph{346} \\
		PBC J1656.2-3303& \textbf{485}& \emph{495} & \emph{490} & \emph{492} \\
		QSO J0555+3948 	& \emph{303} & \textbf{295} & 310 & \emph{300} \\
		PKS 2149-306 	& \textbf{458} & \emph{460} & \emph{461} & \emph{459} \\
		QSO B0237-2322 	& \textbf{299} & \textbf{299} &  \emph{302} & \emph{302} \\
		4C 71.07 		& \emph{557} &  \textbf{555} & 575 & \emph{559}	 \\
		PKS 0528+134 	& \emph{646}& 653 & \textbf{636} & \emph{638} \\
		\midrule
		Total		 	& 8040 & 8464 & 8118 & \textbf{7993} \\		
		\bottomrule
	\end{tabular}
\end{table}

Then, we computed the total AIC value for each model, inserting in Eq.~\ref{eq:AIC} the total chi-square values and the sum of the parameters. The values correspond to a $\chi^2_{\nu,tot}=7758$, 8228, 7880, 7705 with 141, 118, 119 and 144 parameters involved, for BKN, LGP, PL+EX and LGP+EX, respectively. The total AIC is reported in the last row of Table~\ref{tab:AIC}. Results indicate that on the strength of an overall analysis on the whole sample, the best-fit model is indeed LGP+EX. Hence, the coexistence of excess absorption and intrinsic curvature is the preferred explanation for high-$z$ blazars, from physical and statistical motivations.

We would like to highlight that the LGP+EX model is basically equivalent to the PL+EX model for 11 sources, in which the fitted curvature term was consistent with zero (see Section~\ref{sec:CURV+EX}). Hence, the better AIC value of LGP+EX is probably driven by the good description of the PL+EX for these 11 sources, with the additional optimal description of LGP+EX for the remaining 4, namely QSO B0537-286, RBS 315, QSO J0555+3948 and 4C 71.07.

Among the intrinsic curvature models, a BKN seems to be significantly preferred with respect to a LGP. This is clear in Table~\ref{tab:AIC}, but it was also evident in Section~\ref{sec:CURV} looking at the $\chi^2_{\nu,tot}$ values. However, note that several BKN fits yielded excellent results with unlikely parameters, e.g. low-energy photon indexes consistent with zero or negative values, and energy-breaks close to one of the \emph{XMM-Newton} energy-band limit (see Table~\ref{tab:spectral_analysis}). In blazars QSO B0014+810, PKS 2149-306 and QSO B0237-2322 these non-physical parameters were consistent with having good results also in the simple PL fits, but in other objects (e.g. 7C 1428+4218, QSO J0525-3343, QSO B0438-43 and QSO J2354-1513) some additional curvature was indeed required by the data, hence physical BKN parameters were expected. 

We suggest that in the scenario (strengthened by the AIC overall results) in which both excess absorption and intrinsic curvature are present, when the analysis is limited to the sole intrinsic curvature term (i.e. with a BKN or LGP model) it could be possible that BKN sharp-break parameters yield very good results favoured by absorption features (e.g. edges). On the other hand, a LGP would yield worse results, since it simply discerns a curved from a non-curved continuum. An existing excess absorption feature would be likely better mimicked by the BKN model, rather than a LGP. To better understand this ambiguity, we used the low-$z$ blazar sample as comparison for determining the reference model for a curved continuum, as in close objects even the excess absorption along the IGM would be negligible. It turned out that at low-$z$ better fits were obtained with a LGP rather than a BKN (see Section~\ref{sec:lowzanalysis} for details).

\subsection{The low-$z$ sample}
\label{sec:lowzanalysis}

We analysed \emph{XMM-Newton} spectra of six FSRQs and two LBLs below $z=0.5$ (see Section~\ref{sec:samples}) and individual results are reported in Table~\ref{tab:lowz_XMM}. Overall, the simple PL model resulted in a poor $\chi^2_{\nu,tot}$ of 1.39 (with 5748 dof). Intrinsic curvature models improved the fits, yielding $\chi^2_{\nu,tot}=1.09$ (5501/5037) and 1.11 (5549/5024) for the LGP and BKN models, that correspond to F-test $p$-values of $\sim10^{-124}$ and $\sim10^{-115}$, respectively. The two $\chi^2_{\nu,tot}$ are similar, hence we computed the total AIC value (from Eq.~\ref{eq:AIC}). In low-$z$ blazars, in which even the IGM absorption contribution is not expected due to their proximity, the AIC statistics would allow us to assess the preferred model for a curved continuum. The total number of parameters is 66 and 79 for LGP and BKN, that yield $AIC=5633$ and 5707, respectively. The $\Delta_{threshold}$ is significantly greater than 10, indicating a "decisive" evidence in favour of LGP against the BKN model. Hence, on the strength of an overall analysis, a LGP is the reference for a curved continuum. Then, we also performed LGP+EX fits to provide upper limits for the $N_H(z)-z$ relation. 

All spectra showed a concave curvature (see Table~\ref{tab:lowz_XMM}). This can be explained by the appearance of the SSC component. Note that, while at high redshift we were selecting the most powerful sources, that show an almost "naked" EC component (namely without the SSC contribution or the X-ray corona component), at low redshift also weaker blazars could be easily observed. Two low-$z$ FSRQs, namely TXS 2331+073 and 4C 31.63, were analysed with the Very Long Baseline Array (VLBA) during the MOJAVE program\footnote{\href{http://www.physics.purdue.edu/MOJAVE/}{http://www.physics.purdue.edu/MOJAVE/}}. Relatively low apparent velocities ($\beta_{app}$) were reported, i.e. up to $5.35\pm0.74$ \citep{Lister13:MOJAVE} and up to $8.3\pm0.1$ \citep{Homan15:MOJAVE} for TXS 2331+073 and 4C 31.63, respectively. This indicates a moderate beaming, and since the EC component is more dependent than SSC from the beaming factor, we expect that in these sources the SSC can contribute. A concave spectrum in low-$z$ blazars can be also produced by the upturn from the steep high-energy tail of the synchrotron emission and the flatter low-energy rise of the IC hump \citep[see, e.g.,][]{Gaur17:lowzBLAZCONCAVE}.

\subsubsection{Comparison with high-$z$ results}
\label{sec:intr_br_only}

In Section~\ref{sec:bestfitmodel} we obtained with an AIC test that the best-fit model for our high-$z$ FSRQs is the LGP+EX. Here, benefiting from the low-$z$ sample, we further disfavour the pure BKN scenario, in which no excess absorption is required.

Intrinsic spectral breaks predicted by blazars' models are convex (see Appendix~\ref{sec:appB} for details). If the spectral hardening observed in high-$z$ blazars is uniquely attributed to energy breaks intrinsic to the emission, their absence within the observing band in the low-$z$ sample (we even reached 79\,keV with \emph{NuSTAR} in PKS 2004-447) is striking. In fact, any spectral break observed around $1-2\,$keV at $z=3$ could be, in principle, observed around $4-8\,$ keV in the same sources at low-$z$. Furthermore, the SSC component cannot be invoked for covering the putative breaks at $4-8\,$ keV, since in our low-$z$ FSRQs it appears below $\sim3\,$keV (the fitted breaks are concave and within $\sim1-3\,$keV, see Table~\ref{tab:lowz_XMM}). On the other hand, low-$z$ blazars are consistent with the excess absorption scenario, since they show only a marginal IGM excess absorption contribution, in agreement with their proximity (see Section~\ref{sec:discussion}).

\subsubsection{Errors on the Galactic value}
\label{sec:NhGal_errors}

\begin{figure}[tb]
	\centering
	\includegraphics[width=0.9\columnwidth]{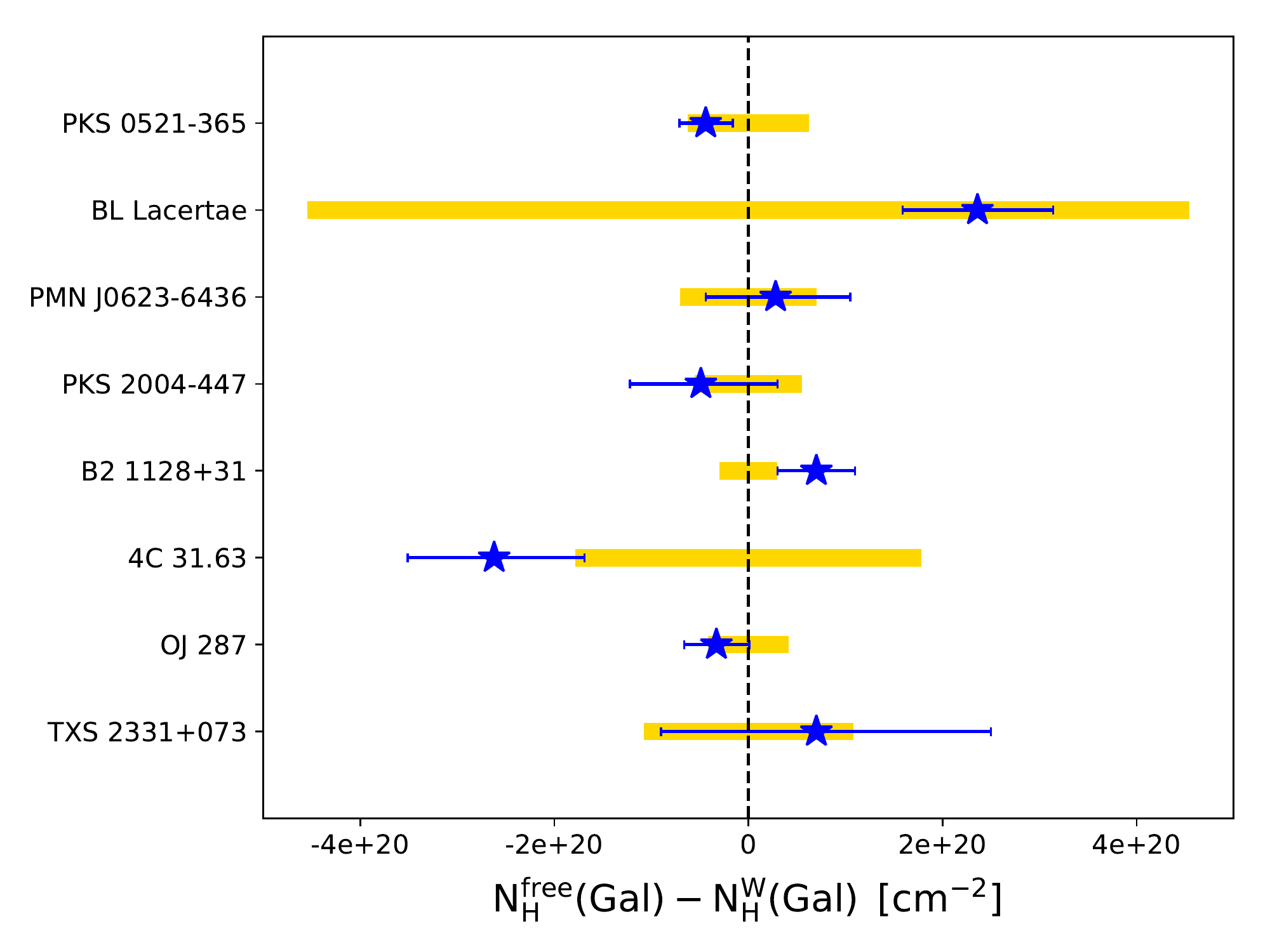}%
	\caption{The difference between the fitted Galactic column densities, $N_H^{free}(Gal)$ and the tabulated values \citep[$N_H^{W}(Gal)$,][]{Willingale13:GalacticH2} is displayed, for each blazar of the low-redshift sample. \emph{Yellow} regions represent their $\pm15\%$ boundaries.}
	\label{fig:NhGal_boundaries}
\end{figure}

Here, we also investigate with low-$z$ blazars the accuracy of the tabulated Galactic column densities. An error should always be added, given the many uncertainties in the determination of Galactic column densities from radio surveys\footnote{i.e. due to scale, stray radiation, noise, baseline errors, RFIs\dots see \href{https://www.astro.uni-bonn.de/hisurvey/AllSky_profiles/index.php}{https://www.astro.uni-bonn.de/hisurvey/AllSky\_profiles/index.php}}, plus the averaging over a conical region, e.g. with a 1-deg radius \citep{Kalberla:LAB}, around the input position of the source. Hence, an error should be always expected, also in values provided by \citet{Willingale13:GalacticH2}, that basically added the molecular hydrogen contribution to the LAB absorption map \citep{Kalberla:LAB}.

We first explored the literature and found that it is quite common to add an arbitrary error to the Galactic value \citep{Elvis86:NhGalerrors,Elvis89:21beamNH}. The issue was to adopt a boundary without biasing our analysis, as a wide range of values have been adopted through the years, e.g. a $\pm20\%$ \citep{Watson07:GRBenvironment,Campana16:narrowlinesGRB} or even a $\pm30\%$ \citep{Cappi97:Xspectraquasar}.

We opted for a $\pm15\%$ error on our Galactic values, to be verified a posteriori with our low-$z$ blazars. Without excess absorption in play, the fitted Galactic values, $N_H^{free}(Gal)$, should have settled nearby the tabulated value provided by \citet{Willingale13:GalacticH2}, $N_H^{W}(Gal)$. The fitted Galactic values, along with their errors, were compared\footnote{We compared with a difference between $N_H^{W}(Gal)$ and $N_H^{free}(Gal)$ and not with a ratio, since dividing the difference for $N_H^{W}(Gal)$ or $N_H^{free}(Gal)$ would have equally led to problems. In the former case, we would have been dividing $N_H^{free}(Gal)$ errors by $N_H^{W}(Gal)$, thus obtaining incorrect percentage errors; in the latter, we would have been dividing the difference for $N_H^{free}(Gal)$, that was unrelated to the $\pm15\%$ boundaries.} to tabulated values and then to their $\pm15\%$ boundaries (see Figure~\ref{fig:NhGal_boundaries}). The result confirmed our choice, since the new fitted values were not always compatible with Willingale's values (vertical dashed line in Figure~\ref{fig:NhGal_boundaries}), but they were indeed consistent within the errors with its $\pm15\%$ boundaries (yellow region in Figure~\ref{fig:NhGal_boundaries}). Note also that a $\pm15\%$ boundary is among the lowest adopted by the literature.

\section{Discussion}
\label{sec:discussion}

In Section~\ref{sec:analysis} we obtained for high-$z$ blazars that the best-fit model is LGP+EX. An excess absorption component, modelled as intrinsic for simplicity, was always fitted, and this component in blazars should be attributed to the IGM. Here, we first test the role of IGM X-ray absorption indirectly with the $N_H(z)-z$ relation (Section~\ref{sec:nhzrelation}), then directly with a spectral model for a WHIM (Section~\ref{sec:igmabs}). In a few sources, there was evidence of a spectral break within the observed band, in addition to the fitted excess absorption. The coexistence between the excess absorption and the presence/absence of intrinsic spectral breaks will be thoroughly treated for each source in Appendix~\ref{sec:appB}.
\subsection{The $N_H(z)-z$ relation}
\label{sec:nhzrelation}
\begin{figure*}[tb]
	\centering
	\includegraphics[width=1.25\columnwidth]{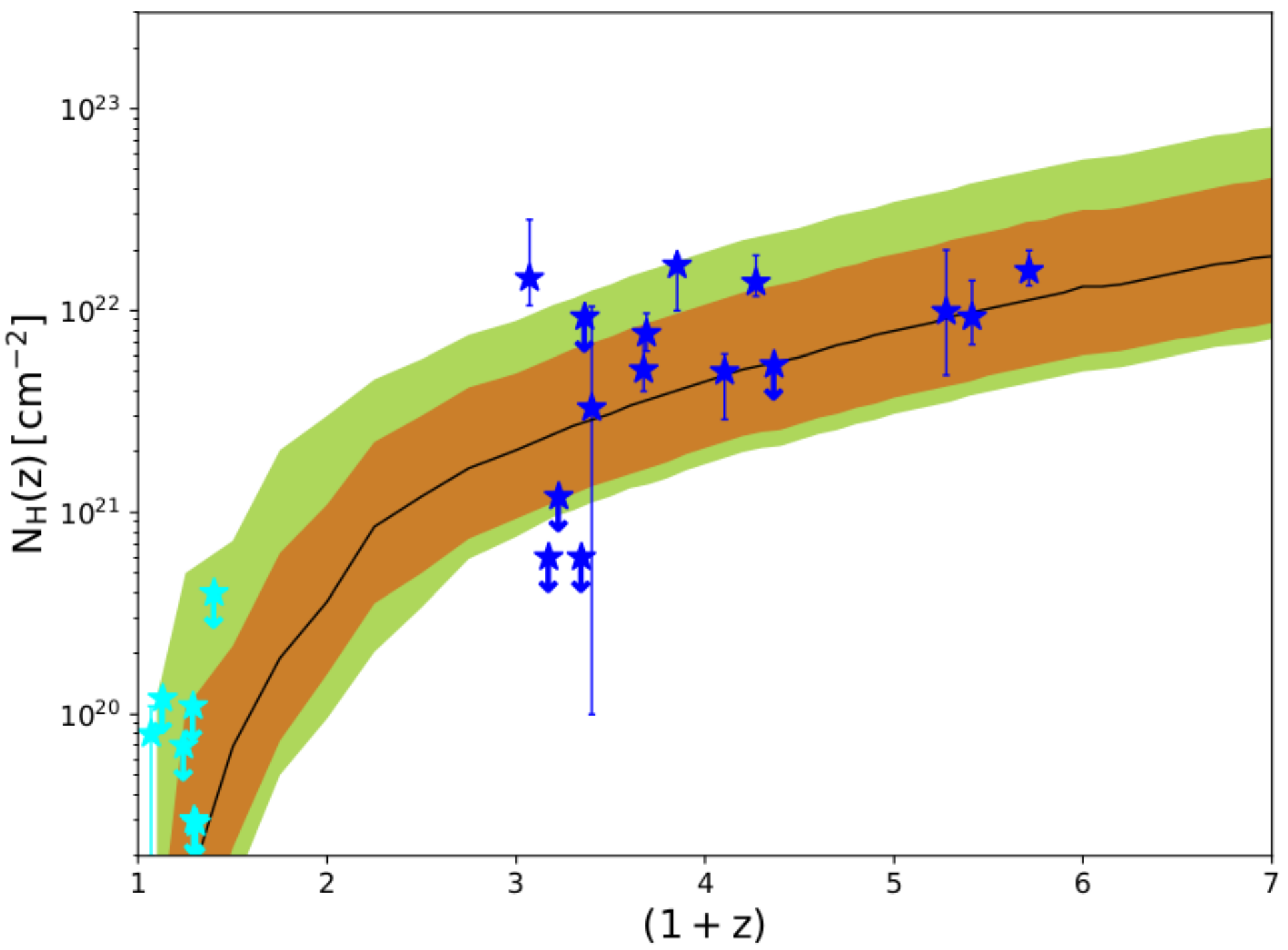}%
	\caption{Distribution of $N_H(z)$ with redshift for blazars from this work. \emph{Blue} objects belong to the Silver sample, while in \emph{cyan} low-$z$ data are shown. Error bars were computed within XSPEC at $90\%$ confidence level ($\Delta\chi^2=2.71$), as were the upper limits (denoted with arrows). All curves and coloured areas were obtained from \citet{Campana:missing}: the solid line, along with its corresponding 1- and 2-sigma envelopes in \emph{brown} and \emph{green}, respectively, is the median of the absorbed LOS distribution, representing the mean absorption contribution from both a diffuse WHIM and additional intervening over-densities. We take the 2-sigma lower envelope as the minimum contribution from a diffuse WHIM alone.}
	\label{fig:nhzrelation}
\end{figure*}

At the beginning of this paper we introduced the $N_H(z)-z$ relation, only apparently describing the increase of intrinsic absorption with redshift, since the IGM absorption component was neglected. Even considering the existence of the IGM contribution, a definitive direct detection is probably beyond the reach of current instruments \citep[e.g.][]{Nicastro16:WHIMcontrov,Nicastro17:WHIMupdated}, thus it is not possible to "subtract" its cumulative effect from the single source, along with the Galactic component, to produce a real $N_H(z_{source})-z$ relation. \citet{Campana:missing} indirectly included the IGM absorption component from a cosmological simulation (in which they pierced through a number of line of sights), matching it to the observed $N_H(z)-z$ relation. This was achieved by attributing, for each redshift bin, the IGM absorption to a host galaxy at a given redshift, erroneously on purpose. This produced the curves and coloured areas \citep[see Fig. 2 of][]{Campana:missing}, that we use in our paper. In particular, among their 100 simulated LOS, the median of the absorbed LOS distribution (solid line in Figure~\ref{fig:nhzrelation}, along with its corresponding 1- and 2-sigma envelopes in \emph{brown} and \emph{green}, respectively) is dominated by 2 or more intervening over-densities (with density contrast\footnote{The density contrast of each cell is defined as the ratio between the gas density in the cell and the mean cosmic gas density.} $\Delta>300$ and temperature $T>10^6\,$K) that can be associated to, e.g., circumgalactic gas within small galaxy groups. The true IGM, i.e. the diffuse "metal fog" that is thought to compose the WHIM, produces a minimum absorbing contribution, here represented in Fig.~\ref{fig:nhzrelation} by the lower 2-sigma curve of the median LOS. This simulated least absorbed LOS is free from any absorber with $\Delta>100$ and it is relative to hot $10^{5-7}\,$K regions far from being collapsed.

\begin{table}[tb]
	\tiny
	\renewcommand{\arraystretch}{1.2}
	\caption{Intrinsic column densities used in Figure~\ref{fig:nhzrelation}. For blazars in bold, $N_H(z)$ was obtained with the broadband $0.2-79\,$keV fit with additional \emph{Swift-XRT}+\emph{NuSTAR} data. Errors and upper limits were computed within \texttt{XSPEC} at $90\%$ confidence level.}
	\label{tab:nhzrelation}
	\centering
	\begin{tabular}{ccc}
		\toprule
		\multicolumn{1}{c}{Name} &
		\multicolumn{1}{c}{$z$} &
		\multicolumn{1}{c}{$N_H(z)/10^{22}$cm$^{-2}$} \\
		\midrule
		\textbf{7C 1428+4218} & 4.715 & $1.52^{+0.44}_{-0.46}$ \\
		QSO J0525-3343 & 4.413 & $0.93^{+0.48}_{-0.25}$ \\
		QSO B1026-084 & 4.276 & $0.99^{+1.01}_{-0.51}$ \\ 
		\textbf{QSO B0014+810} & 3.366 & $<0.54$ \\
		PKS 2126-158 & 3.268 & $1.38^{+0.50}_{-0.20}$ \\
		QSO B0537-286 & 3.104 & $0.50^{+0.11}_{-0.21}$ \\
		QSO B0438-43 & 2.852 & $1.68^{+0.17}_{-0.68}$ \\
		\textbf{RBS 315} & 2.69 & $0.77^{+0.20}_{-0.14}$\\
		QSO J2354-1513 & 2.675 & $0.51^{+0.30}_{-0.11}$ \\
		\textbf{PBC J1656.2-3303} & 2.4	& $0.33^{+0.72}_{-0.32}$ \\
		QSO J0555+3948 & 2.363 & $<0.93$ \\
		\textbf{PKS 2149-306} & 2.345 & $<0.06$ \\
		QSO B0237-2322 & 2.225 & $<0.12$ \\
		\textbf{4C 71.07} & 2.172 & $<0.06$ \\
		PKS 0528+134 & 2.07	& $1.45^{+1.38}_{-0.39}$ \\
		\midrule
		TXS 2331+073 & 0.401 & $<0.04$ \\
		4C +31.63 & 0.295 & $<0.003$ \\
		B2 1128+31 & 0.29 & $<0.011$ \\ 
		\textbf{PKS 2004-447} & 0.24 & $<0.007$ \\
		PMN J0623-6436 & 0.129 & $<0.012 $\\
		PKS 0521-365 & 0.055 & $<0.0004$ \\
		OJ 287 & 0.306 & $<0.003$ \\
		BL Lacertae & 0.069 & $0.008^{+0.003}_{-0.006}$ \\
		\bottomrule
	\end{tabular}
\end{table}

We explored the $N_H(z)-z$ relation with our results for low- and high-$z$ blazars,  reported in Table~\ref{tab:nhzrelation} and shown in Figure~\ref{fig:nhzrelation}. The column densities obtained from our analysis seem to follow the increasing trend with redshift, consistently with the IGM curves simulated by \citet{Campana:missing}. Only PKS 0528+134 ($z=2.07$) showed a moderately high column density above the 2-sigma upper boundary of the IGM mean contribution (the upper green region in Figure~\ref{fig:nhzrelation}). This outlier could be explained with a particularly absorbed LOS, starting with its high Galactic column density ($N_H^{Gal}=38.5\times10^{20}$ cm$^{-2}$), due to its low Galactic latitude and to the intervening outer edge of the molecular cloud Barnard 30 in the $\lambda$ Orion ring of clouds \citep{Liszt93:COGalactic,Hogerheijde95:COcloud0528}. Consequently, the tabulated value \citep{Willingale13:GalacticH2}, even if it includes the contribution from molecular hydrogen, could be underestimating the amount of absorbing matter within our Galaxy. As a matter of fact, in the PL+EX fit the fitted Galactic value, free to vary between $\pm15\%$ uncertainties, was a lower limit, hinting a preference for Galactic columns close to the upper boundary (see Table~\ref{tab:spectral_analysis}). Besides, this source would also be compatible with the 3-sigma superior limit of the mean envelope, thus we consider it consistent with our proposed scenario.

Two outliers, namely 4C 71.07 ($z=2.172$) and PKS 2149-306 ($z=2.345$), happened to be below the 2$\sigma$ lower simulated curve, that represent the minimum absorption contribution due to a diffuse WHIM. In our work we used higher Galactic values \citep{Willingale13:GalacticH2} with respect to the earlier literature \citep{Ferrero03:XMM2126and2149,Page05:XMMquasar,Foschini06:XMMquasar,Eitan:2013}. However, these sources were already known for their low excess absorption column densities obtained with \emph{XMM-Newton} data. In fact, even using the LAB Galactic value \citep{Kalberla:LAB} for the two outliers did not solve the issue, yielding excess column density upper limits of $<0.09$ and $<0.07\times10^{22}$\,cm$^{-2}$, respectively. These two outliers should not be taken as a confutation of the excess absorption scenario emerged through the years for all extragalactic sources, although they cannot be ignored. They could be used as a "worst case" to re-build the lower envelope. However, this should not imply a dramatical change in the simulated characteristics of the IGM, since lowering the metallicity by less than a factor 2 would be probably enough.

\subsubsection{The role of the instrument's limits}

\begin{figure*}[tb]
	\centering	
	\includegraphics[width=0.75\columnwidth]{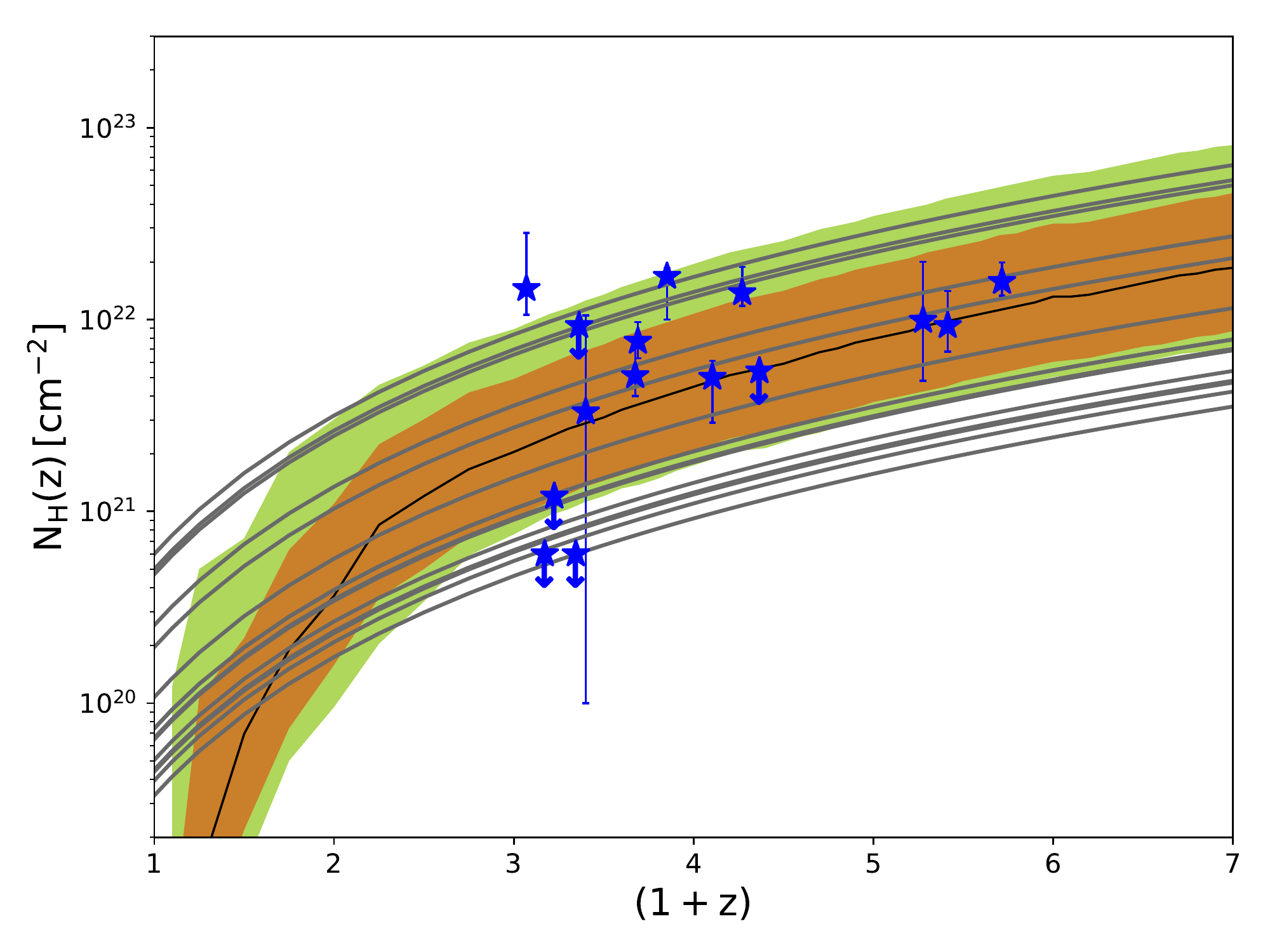}%
	\includegraphics[width=0.75\columnwidth]{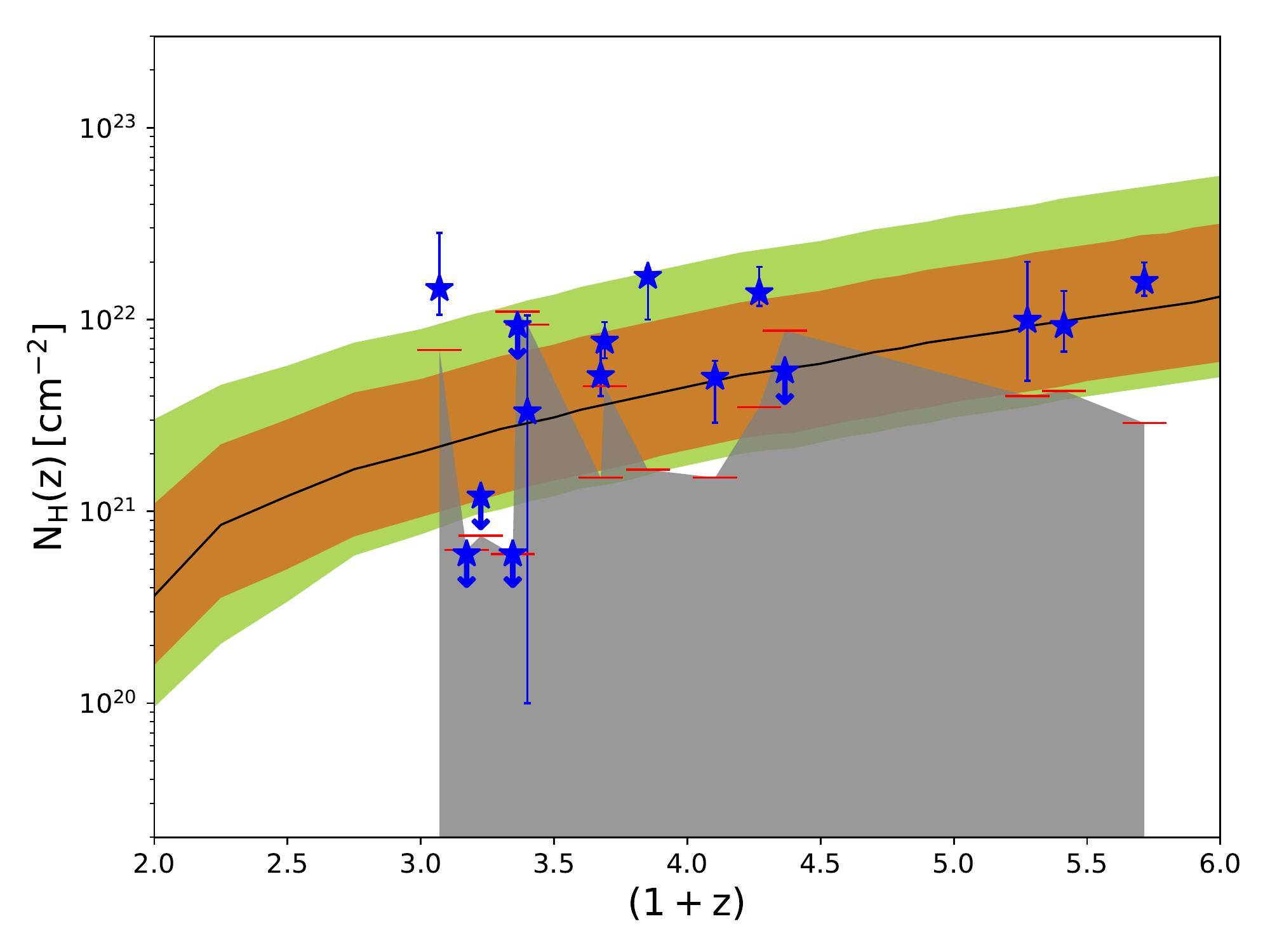}%
	\\
	\includegraphics[width=0.75\columnwidth]{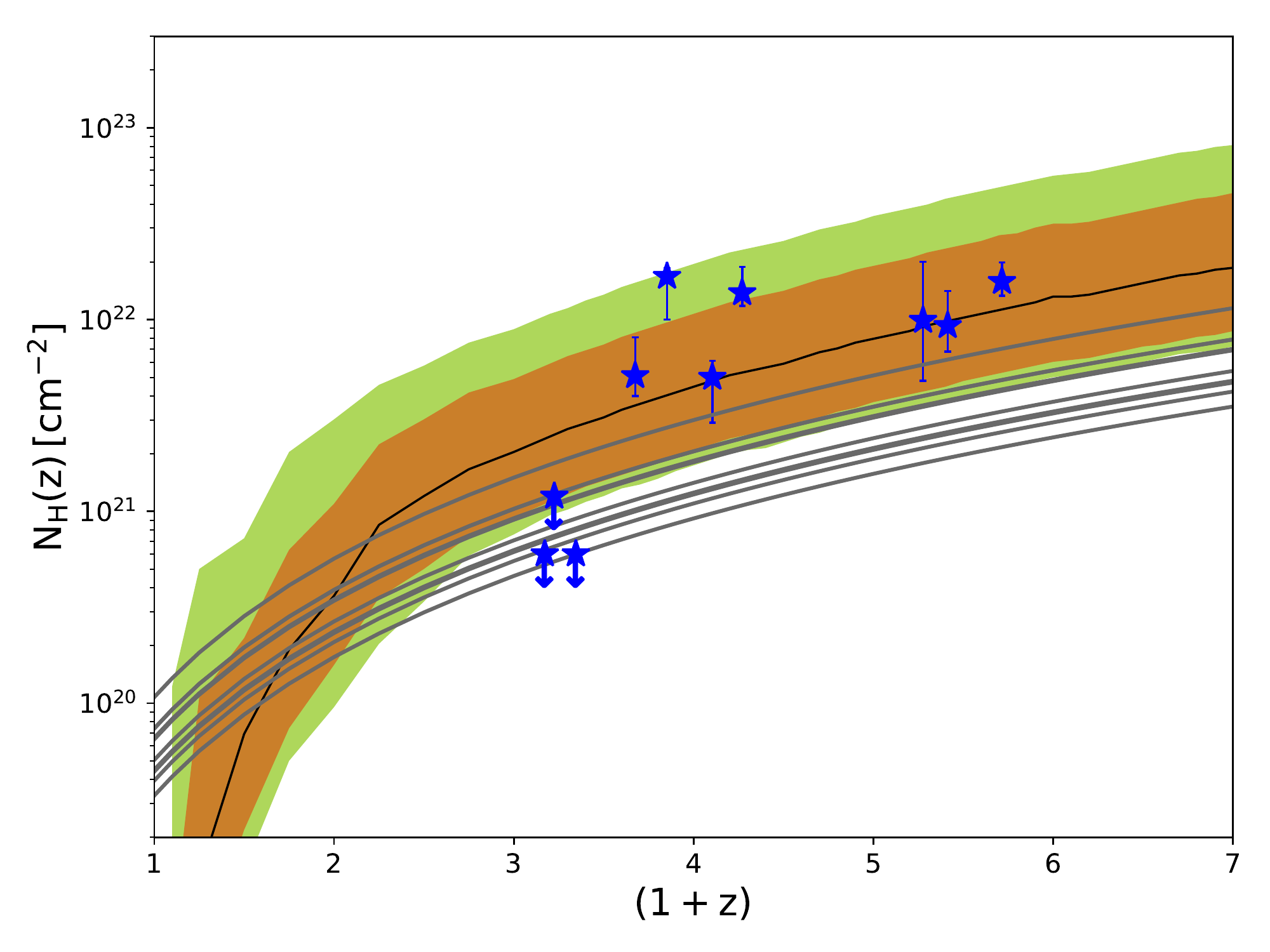}%
	\includegraphics[width=0.75\columnwidth]{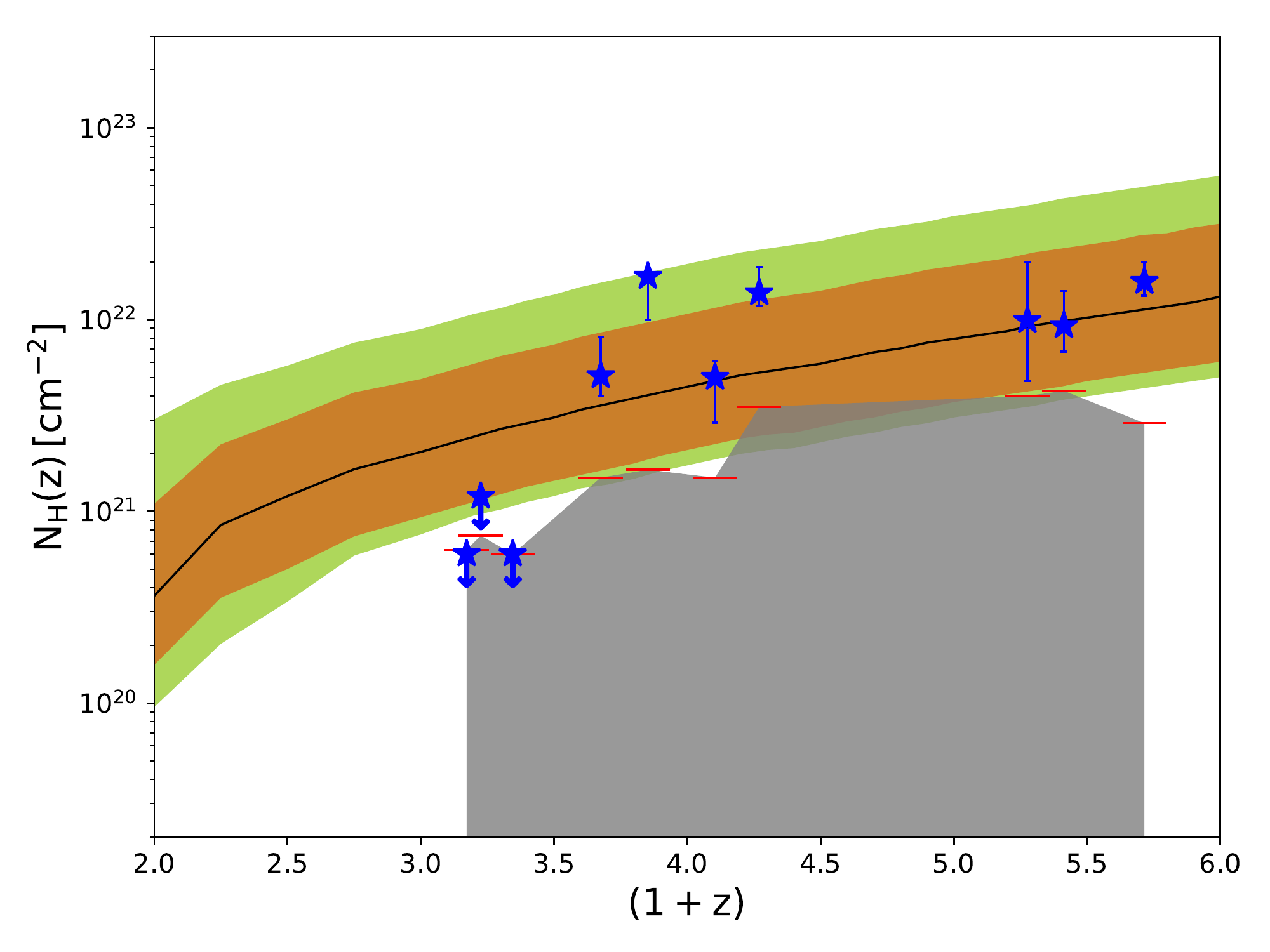}%
	\caption{$N_H(z)-z$ relation, see Figure~\ref{fig:nhzrelation} for the general description. In the \emph{left} panels, the solid \emph{grey} curves represent the extrapolation of the last-detection limit of each blazar, providing an overall sensitivity range of the instrument for excess absorption detections. On the \emph{right}, the same last-detection limits are shown for each blazar with \emph{red horizontal dashes}. The underlying \emph{grey} area is the "upper-limit" region for each blazar. Observing a blazar with a specific \emph{XMM-Newton} response, exposure time, absorbed by its $N_H^{Gal}$ and a putative excess absorption column density, one would be sensitive for detections of the latter only above its red dash, while below it one would fit an upper limit. \emph{Bottom} panels are analogous, with the exclusion of any blazar with a Galactic column density greater than $10^{21}$ cm$^{-2}$.}
	\label{fig:nhz_limits}
\end{figure*}

Typical fair objections can be arisen, e.g. it could be argued whether this observed increasing $N_H(z)-z$ relation is real. The validity of the increasing trend was already verified, also with statistical tests \citep[e.g.][]{Campana12:nhxcomplete,Starling:evoluzIGM,Eitan:2013,Arcodia16:GRBsNhz}. Moreover, high-$z$ column densities only increase as more realistic, lower metallicity values\footnote{The hydrogen equivalent column density is computed within \texttt{XSPEC} assuming solar abundancies.} are used, thus the trend would be enhanced.  

It could be also questioned the physical origin of the increasing trend. The lack of unabsorbed sources at high-redshift would be then only due to the incapability of measuring relatively low column densities towards distant sources. In principle, the minimum $N_H$ that can be detected is expected to increase with redshift, as more of the absorbed sub-keV energies are shifted below the observed band. There are indeed instrumental limits, but they influence regions in the $N_H(z)-z$ plot way below the observed impressive high-$z$ column densities \citep[e.g.][]{Starling:evoluzIGM}. This gap between the instrument's limits and the observed column densities is considered significant for validating the physical origin of the increasing $N_H(z)-z$ trend. Nonetheless, the presence of instrumental limits provides a fair argument against our conclusions and it should be verified also for our sources.

The instrumental incapability of detecting an (existing) high-$z$ excess column density can be enhanced, e.g., by a high Galactic absorption value and by a low photon statistic. In principle, the role of latter could be confidently excluded, since we analysed sources with more than $\sim10\,000$ photons. Then, our aim was to compute for each blazar what we called its \emph{last-detection limit}, namely the excess absorption column density value below which only upper limits can be fitted, due to instrumental limits. This purpose was fulfilled with the \texttt{fake} task within \texttt{XSPEC}, simulating for each blazar the spectrum that would have been extracted by \emph{XMM-Newton}, given its response and the observation(s) exposure time and its absorbing $N_H^{Gal}$ with $\pm15\%$ boundaries\footnote{This clarification is necessary, since with a different exposure time, or Galactic column density, the last-detection limit would drastically change.}. The input values for the simulations were obtained from our spectral fits (Table~\ref{tab:spectral_analysis}) with the PL+EX scenario. Each simulated spectrum was then fitted with a PL+EX model to compute the errors of the fitted excess column density. Different spectra were simulated for each source, using decreasing arbitrary excess column densities in input, down to the value that yielded an upper limit in the subsequent spectral fit. All three cameras were used to compute the final last-detection limit.

The left panels of Figure~\ref{fig:nhz_limits} show the $N_H(z)-z$ relation, along with the last-detection limit of each blazar, that was extrapolated with the scaling relation $(1+z)^{2.4}$ \citep[see][]{Campana14:scaling}. These curves provided an overall sensitivity range for the detection of excess column densities for our sources. In our sample, no selection criteria on Galactic column density values were included, leading to $N_H^{Gal}$ ranging from $1.22$ to $42.5\times10^{20}$ cm$^{-2}$. Excluding sources with a Galactic column density greater than $10^{21}$ cm$^{-2}$, the instrument reaches sensitivity for excess column detections well below the simulated lower IGM absorption contribution (see the bottom left panel). In the right panels of Figure~\ref{fig:nhz_limits} last-detection limits are shown for each source with a red horizontal dash, along with the underlying "upper-limit" area in grey. Again, the bottom panel was obtained excluding any blazar with a Galactic column density greater than $10^{21}$ cm$^{-2}$.

\begin{figure*}[tb]
	\centering
	\includegraphics[width=0.97\columnwidth]{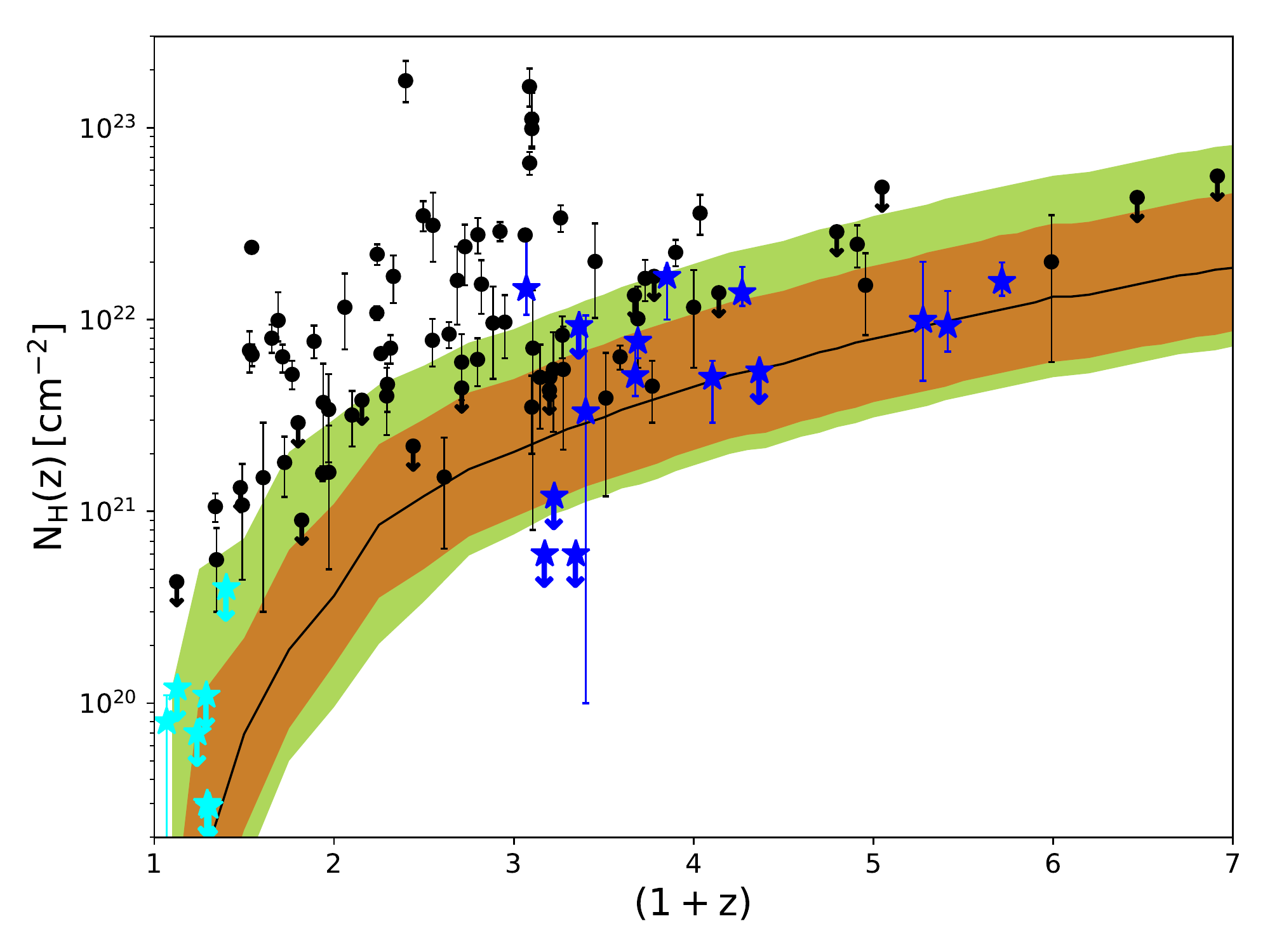}%
	\includegraphics[width=0.97\columnwidth]{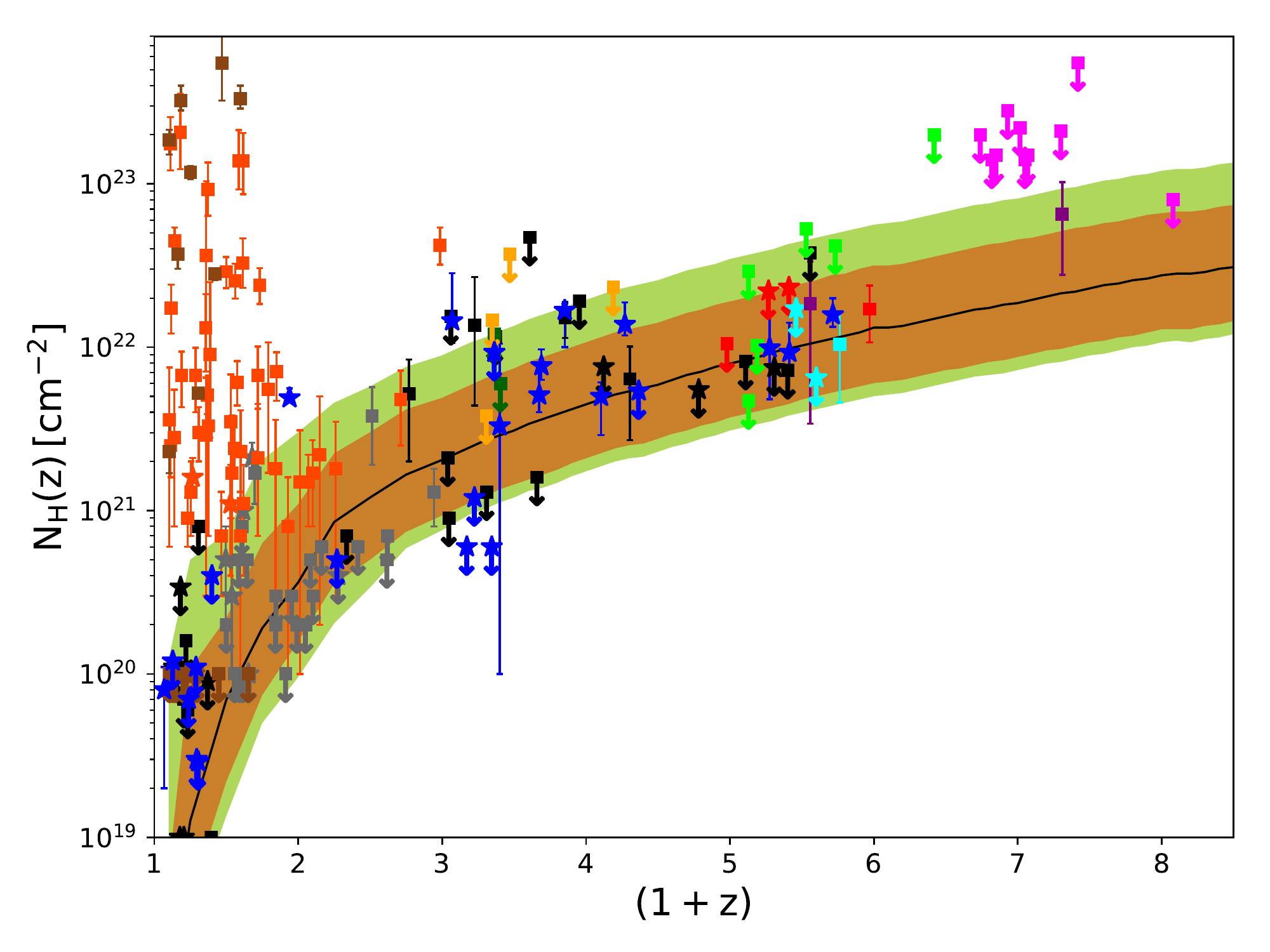}%
	\caption{See Figure~\ref{fig:nhzrelation} for a general description of the plot. On the \emph{left}, additional GRB data from \citet{Arcodia16:GRBsNhz} are plotted. On the \emph{right}, the distribution of $N_H(z)$ with redshift for blazars from this work (\emph{blue stars}) and for quasars from the literature (\emph{black} symbols are related to \citet{Page05:XMMquasar}; \emph{magenta} to \citet{Nanni17:highzRQQs}; \emph{grey} to \citet{Eitan:2013}; \emph{darkgreen} and \emph{orange} to \citet{Shemmer06:RQQs} and \citet{Shemmer08:RQQs}, respectively; \emph{red} to \citet{Saez11:RLQshighz}; \emph{cyan} to \citet{Yuan06:RLQshighz}; \emph{purple} to \citet{Grupe06:XMMhighzquasar}; \emph{lime} to \citet{Shemmer05:highzAGN}; \emph{darkred} to \citet{Campana:missing}; \emph{brown} to \citet{Ricci17:quasarXABS}; \emph{dark orange} to \citet{Corral11:someAGN}). \emph{Squares} stand for generic non-blazars AGN, while \emph{stars} for blazars. Error bars were computed within XSPEC at $90\%$ confidence level ($\Delta\chi^2=2.71$), as were the upper limits (denoted with arrows below the related symbol).}
	\label{fig:nhzrelation_literature}
\end{figure*}

Both left and right panels lead to the same conclusion. It is true that any instrument has its limits in detecting low column densities at high-$z$ and we are not incredibly sensitive to very low column densities \emph{per se}. Nonetheless, we are sensitive enough to conclude that our high-$z$ column densities are high for physical reasons, since our fitted values are significantly above the minimum values reachable by \emph{XMM-Newton} for each source (red dashes in Figure~\ref{fig:nhz_limits}). If the increasing trend was only produced by the instrument's limits, we would expect upper limits consistent with the upper edge of the grey area and not, as we observe at high-$z$, clear detections above it. This is more evident in bottom panels of Figure~\ref{fig:nhz_limits}, where only blazars with Galactic column densities below $10^{21}$ cm$^{-2}$ were considered (actually, it is below $6.15\times10^{20}$ cm$^{-2}$). Hence, selecting sources with a relatively low Galactic absorption component is extremely important to reach sufficient sensitivity to probe the diffuse IGM. Moreover, longer exposures with current instruments, e.g. \emph{XMM-Newton}, should be adopted to provide even lower last-detection limits.

\subsubsection{Comparison with previous works}
\label{sec:comparison_analysis}
Our results are generally in accordance with the literature involving the same sources and instruments \citep[e.g.][]{Reeves01:0537XMM,Ferrero03:XMM2126and2149,Worsley04:7Cflatt,Worsley04:0525warm,Brocksopp04:XMMquasars-0438,Page05:XMMquasar,Piconcelli05:RBS315,Yuan05:abs1026,Grupe06:XMMhighzquasar,Foschini06:XMMquasar,Tavecchio07:RBS315Suzaku-intrinsicbreak,Bottacini10:0537broadband,Eitan:2013,Tagliaferri15:NuSTAR-2149and4C,Paliya15:4Cbroadband,Dammando16:PKS2149study,Paliya16:broadband,Sbarrato16:NUSTARblazars}, of course taking into account the possible differences (e.g. the Galactic absorption model).

It is worth discussing that \citet{Paliya16:broadband} obtained a large disagreement between column densities measured from \emph{XMM-Newton} spectra and from broadband \emph{Swift-XRT}+\emph{NuSTAR} spectra, the latter several times larger (up to an order of magnitude). From this, they concluded that spectral curvature in high-$z$ blazars is not caused by excess absorption, but it is due to spectral breaks intrinsic to the blazar's emission, better investigated with a broadband analysis. Actually, our broadband fits, in which excess absorption was also constrained by \emph{XMM-Newton}, yielded column densities compatible with the narrowband fits (see Table~\ref{tab:spectral_analysis}). Hence, while a broadband spectrum does provide an extensive view on the curved spectral continuum, their claim is possibly driven by a misleading comparison between \emph{XMM-Newton}'s and \emph{Swift-XRT}'s performances. The former, with its larger effective area, allows us to assess the soft X-ray properties better than the latter can do. As a matter of fact, removing \emph{XMM-Newton} from our broadband analysis, \emph{Swift-XRT}+\emph{NuSTAR} data alone \emph{did} yield higher intrinsic column densities, e.g. $\sim5^{+3}_{-2}\times10^{22}\,\text{cm}^{-2}$ in QSO B0014+810 or $12^{+5}_{-4}\times10^{22}\,\text{cm}^{-2}$ in PBC J1656.2-3303. Moreover, fitting only \emph{Swift-XRT} data of QSO B0014+810 with a PL+EX model yielded a column density upper limit ($N_H(z)<3.81\times10^{22}\,\text{cm}^{-2}$) around half order of magnitude higher than the \emph{XMM-Newton} results ($N_H(z)<0.84\times10^{22}\,\text{cm}^{-2}$). We attribute the difference in the fitted column densities to \emph{Swift-XRT}'s lower photon counts (e.g. $\sim500-600$ for the two observations of QSO B0014+810) compared to the larger statistic provided by \emph{XMM-Newton}. The same conclusion is valid for the discrepancies obtained in 7C 1428+4218, RBS 315 and PBC J1656.2-3303.

The most complete window on the X-ray spectra would be provided by simultaneous \emph{XMM-Newton} and \emph{NuSTAR} observations. In the absence of this possibility, \emph{XMM-Newton} should be added anyway to \emph{Swift-XRT}+\emph{NuSTAR} data in a broadband analysis (see Section~\ref{sec:analysis}).

Then, we compared our results with other extragalactic sources from the literature. GRBs typically show a large scatter in $N_H(z)$, particularly at low $z$, due to their known prominent intrinsic absorption component, although a lower contribution, increasing with $z$ and enclosing all sources, is evident and was attributed to the diffuse IGM \citep[see][]{Behar:baryons,Campana10:nhzswift,Campana12:nhxcomplete,Starling:evoluzIGM,Campana:missing,Arcodia16:GRBsNhz}. In the left panel of Figure~\ref{fig:nhzrelation_literature}, we show intrinsic column densities from this work, along with GRB data from \citet{Arcodia16:GRBsNhz}. Both types of sources seem to agree with the simulated IGM absorption contributions, reported from \citet{Campana:missing}. In particular, GRBs are distributed upwards from the simulated areas, in agreement with having both intervening and intrinsic absorption contributions, while blazars clearly follow specifically the coloured areas representing the IGM contribution, in accordance with the idea of a missing absorption component within the host galaxy.

The $N_H(z)-z$ relation was previously studied also in quasars, although mostly focused on the greater amount of X-ray absorption detected in radio-loud\footnote{The distinction between radio-quiet and radio-loud AGN may be obsolete \citep[see the discussion in][]{Padovani16:jetted-nonjetted,Padovani17:jetted-nonjetted}.} (RLQs) rather than in radio-quiet quasars (RQQs), perhaps suggesting that it was due to the presence of the relativistic jet \citep[see discussions in, e.g.,][and references therein]{Elvis94:absquasar,Cappi97:Xspectraquasar,Fiore98:ROSATquasar,Reeves00:ASCAquasar,Page05:XMMquasar,Eitan:2013}. Nonetheless, at the time there was no clear distinction between blazars and other jetted AGN, in which the relativistic jet is pointing at wider angles ($\theta_{view}\gtrsim1/\Gamma$, where $\Gamma$ is the bulk Lorentz factor of the jet emitting region) with respect to the LOS. In the former the SED is dominated by the beamed non-thermal emission of the jet, while the latter shows, for increasingly wide angles, an X-ray spectrum always more similar to non-jetted AGN \citep[see Fig. 3 in][]{Sbarrato15:blazcandidates2,Dermer95:beaming}. Moreover, in these works RLQs (in the way they interpreted it, e.g. jetted AGN regardless of the jet direction) typically had better statistics with respect to RQQs and/or were observed up to larger distances. The reason is that most of their RLQs were later identified as blazars and then benefited of the relativistic beaming. On the contrary, RQQ-samples of these earlier works consisted mainly in lower-redshift sources \citep[e.g. 12 RQQs out of 286 at $z>2.2$ in][]{Fiore98:ROSATquasar,Eitan:2013}, for which negligible IGM excess absorption is expected, and/or in quasars with lower counts statistic \citep[e.g.][]{Page05:XMMquasar}, for which a column density detection cannot be clearly established. Hence, the lack, in the above-mentioned works, of clear detections of excess column densities in RQQs, with respect to the corresponding RLQ-samples, is perfectly understandable. 

Here we promoted a different point of view, attributing the observed hardening to absorption in excess of the Galactic value, occurring along the IGM. This would solve the paradox of the incomparably lower amount of intrinsic absorption detected in the optical/UV compared to X-ray analysis \citep[e.g. see the discussions in][]{Elvis94:absquasar,Cappi97:Xspectraquasar,Fabian01:0525warmabs,Fabian01:7Cwarmabsorber,Worsley04:0525warm,Worsley04:7Cflatt,Page05:XMMquasar}, that through the years led to preferring the intrinsic spectral breaks scenario. Nonetheless, our suggested scenario needed to be tested with quasars of the previous works, obtained by selecting only sources observed with \emph{XMM-Newton}\footnote{Only a few sources from \citet{Nanni17:highzRQQs} were analysed with \emph{Chandra} \citep{Weisskopf02:CHANDRA}.} and excluding sources below $z=0.1$ (the first redshift bin of the simulated IGM). References for intrinsic column density values are reported in the description of Figure~\ref{fig:nhzrelation_literature}. If the same object was studied in different works, we favoured the literature in which the analysis was performed with all EPIC cameras. We excluded quasars when clear evidence of lensing was found in the literature. We only reported from \citet{Corral11:someAGN} quasars with a detection in $N_H(z)$ and with a power-law as best-fit model \citep[see][for a complete comparison between all their results and the simulated IGM]{Campana:missing}. 

The right panel of Figure~\ref{fig:nhzrelation_literature} shows the $N_H(z)-z$ plot filled with our low- and high-$z$ blazars (blue stars) and with AGN from the literature, divided between blazars (stars) and non-blazars (squares). The latter are observed also above the simulated IGM curves, as for some generic AGN an intrinsic absorption component is expected \citep[e.g.][]{Ricci17:quasarXABS}, while the former are again consistent with having only the IGM absorption component. Few outliers, some of which were reprocessed, are discussed in Appendix~\ref{sec:app_literaturesources}. 

\subsection{The warm-hot IGM absorption contribution}
\label{sec:igmabs}

Within \texttt{XSPEC}, it is possible to directly model the IGM absorption component with \texttt{igmabs}\footnote{See \href{http://www.star.le.ac.uk/zrw/xabs/readme.html}{http://www.star.le.ac.uk/zrw/xabs/readme.html}.}. This model computes the X-ray absorption expected from a WHIM with a uniform medium (expressed in hydrogen density $n_0$, at solar metallicity), constant temperature $T$ and ionisation state $\xi$. Other parameters involved are the redshift of the source and the photon index of the photo-ionising spectrum, typically estimated with the measured cosmic X-ray background (CXRB). If all the main parameters of the WHIM, i.e. $T$, $\xi$ and $n_0$, are left free to vary some degeneracy is expected \citep[see][]{Starling:evoluzIGM}. Some constraints can be adopted, e.g. $n_0$ can be fixed to $1.7\times10^{-7}$ cm$^{-3}$ \citep[][and references therein]{Behar:baryons}, or the temperature can be constrained to be $10^6$ K \citep{Starling:evoluzIGM,Campana:missing}. We chose to tie the ionization parameter to $n_0$, leaving the latter and $T$ as the only free parameters of the \texttt{igmabs} model. The ionization parameter of the IGM is given by:\begin{equation*}\xi\approx\frac{4\pi F_{CXRB}}{n_e}
\end{equation*} where the electron density $n_e$ is $\sim1.2n_0$ and $F_{CXRB}=2.9\times10^{-7}$ erg cm$^{-2}$ s$^{-1}$ sr$^{-1}$ \citep{DeLuca04:CXRB,Starling:evoluzIGM,Campana:missing}. Hence, we constrained $\log\xi=1.48-\log(n_0)$ throughout all the \texttt{igmabs} fits.

According to Figure~\ref{fig:nhzrelation}, only a few sources showed an intrinsic column density compatible with the lower IGM absorption curve, proper of a diffuse WHIM, namely 4C 71.07, QSO B0237-2322, PKS 2149-306, PBC J1656.2-3303 and QSO B0014+810. Nonetheless, spectral fits performed with $0.2-10\,$keV \emph{XMM-Newton} spectra were incapable to constrain both $n_0$ and $T$. However, using also \emph{Swift-XRT}+\emph{NuSTAR}, together with \emph{XMM-Newton}, the fits started to be sensitive to those parameters. Luckily, among the 5 blazars compatible with the lower envelope, 4 had broadband data available (QSO B0237-2322 is the only source cut out). 

We then performed $0.2-79\,$keV spectral fits for individual sources, as described in Section~\ref{sec:analysis}, but fixing the Galactic absorption\footnote{We decided to freeze the Galactic column density given the many free parameters involved and the difficulties emerged in the narrow-band \emph{XMM-Newton} fits.} and modelling the excess absorption with a WHIM component. The continuum of the sources in the broadband LGP+\texttt{igmabs} fit was constrained with the results of the LGP+EX model. Hence, \emph{XMM-Newton} continua of QSO B0014+810, PKS 2149-306 and PBC J1656.2-3303 were constrained to a simple power-law, while for 4C 71.07 a fixed curvature of $b=0.05^{+0.03}_{-0.02}$ was assumed. In addition, \emph{Swift-XRT}+\emph{NuSTAR} continua were left free to vary with LGP parameters except for QSO B0014+810, in which a power-law continuum was used. This is a reasonable approximation, arisen to obviate experimental and computational limits of the model, that allowed us to better constrain the fitted parameters. A more rigorous fit with free continuum parameters would probably lead to upper limit measures for the WHIM characteristics. Results are shown in Table~\ref{tab:igmabs} and they are quite consistent with each other, due to the huge errors. The remaining parameters were fully compatibles, within the errors, with the values obtained in the LGP+EX scenario and were not reported.

\begin{table}[tb]
\scriptsize
	\setlength{\tabcolsep}{5.5pt}
	\renewcommand{\arraystretch}{1.2}
	\caption{Fit results with \texttt{igmabs}. Errors and upper/lower limits were computed within \texttt{XSPEC} at $90\%$ confidence level.}
	\label{tab:igmabs}
	\centering
	\begin{ThreePartTable}
		\begin{tabular}{cccccc}
			\toprule
			\multicolumn{1}{c}{Name} &
			\multicolumn{1}{c}{$z$} &
			\multicolumn{1}{c}{$n_0/10^{-7}$cm$^{-3}$} &
			\multicolumn{1}{c}{$\log(T/\text{K})$} &
			\multicolumn{1}{c}{$\log\xi$\tnote{a}} &
			\multicolumn{1}{c}{$\chi^2_{\nu}/\nu$} \\
			\midrule
			QSO B0014+810 & 3.366 & $4.07^{+1.37}_{-3.03}$ & $>6.84$\tnote{b} & $0.87\pm0.24$ & $1.01/842$ \\
			PBC J1656.2-3303 & 2.4	& $1.75^{+1.60}_{-1.09}$ & $5.73^{+1.22}_{-4.12}$ & $1.24\pm0.33$ & $1.10/687$ \\
			PKS 2149-306 & 2.345 & $1.16^{+1.75}_{-0.55}$ & $6.94^{+0.75}_{-4.77}$ & $1.41\pm0.43$ & $0.98/2737$ \\
			4C 71.07 & 2.172 & $0.30^{+0.60}_{-0.21}$ & $3.70^{+2.10}_{-3.20}$ & $2.00\pm0.45$ & $1.05/2118$ \\
			\midrule
			All & - & $1.01^{+0.53}_{-0.72}$ & $6.45^{+0.51}_{-2.12}$ & $1.47\pm0.27$ & $1.03/6392$ \\
		\bottomrule
		\end{tabular}
		\begin{tablenotes}
			\item[a] The parameter was tied to the hydrogen density with the relation $\log\xi=1.48-\log(n_0)$. Through this relation, asymmetric errors of $n_0$ were first averaged \citep[e.g. see][chap. 12]{Dagostini:asymmerrors} and then propagated.
			\item[b] $\log T$ was left free to vary between 0 and 8, the best value being in this case $7.81$.
		\end{tablenotes}
	\end{ThreePartTable}
\end{table}

\begin{figure}[tb]
	\centering	
	\includegraphics[width=0.6\columnwidth,angle=-90]{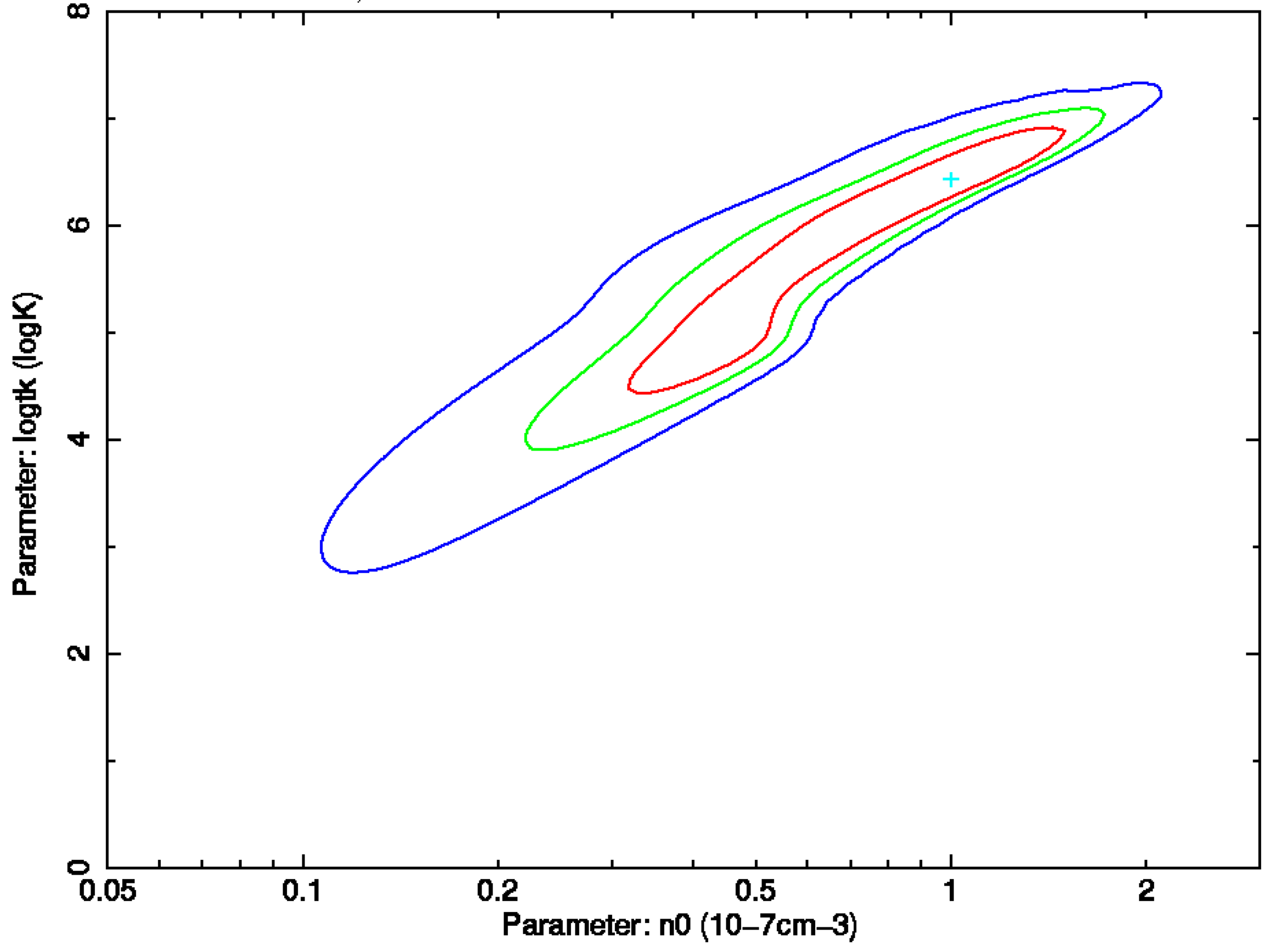}%
	\caption{Results of the joint \texttt{igmabs} fit with QSO B0014+810, PBC J1656.2-3303, PKS 2149-306, and 4C 71.07 fitted together. Confidence contours are shown for the hydrogen density $n_0$ (x-axis) and the temperature (y-axis) of the IGM. We display 68, 95 and 99 percent contours in \emph{red}, \emph{green} and \emph{blue}, respectively.}
	\label{fig:plocont_igmabs_joint}
\end{figure}

We then performed a joint fit with all four sources to obtain an overall measurement of the WHIM characteristics. The IGM absorption parameters (namely $n_0$, $T$ and $\xi$) were tied together among all the different observations. Results are displayed in Table~\ref{tab:igmabs} and the related contour plot is reported in Figure~\ref{fig:plocont_igmabs_joint}. The overall fitted values are consistent with the expected properties of the WHIM, i.e. an average hydrogen density $\approx10^{-7}$ cm$^{-3}$ and a temperature $\approx10^6\,$K \citep[e.g.][and references therein]{Cen06:WHIMsimul,Bregman07:IGMreview,Starling:evoluzIGM,Campana:missing}. 

Furthermore, the hydrogen density obtained within \texttt{XSPEC} spectral fits is expressed using solar abundances and metallicity. Hence, if the estimate $n_0=1.7\times10^{-7}$ cm$^{-3}$ \citep[][and references therein]{Behar:baryons} is to be trusted, we can infer the metallicity of the WHIM comparing it with our fitted value of $n_0$. The inferred metallicity is:\begin{equation*}Z=0.59_{-0.42}^{+0.31}Z_{\astrosun}\end{equation*}

This should be only considered as an important consistency check. What is more, we provided suitable candidates for deeper exposures with current instruments. Among the four sources used for this analysis, 4C 71.07 and PKS 2149-306 are the best candidates, given the higher Galactic column densities of the other two sources. The deleterious effect of such high Galactic values was shown in Figure~\ref{fig:nhz_limits}. A long simultaneous \emph{XMM-Newton}+\emph{NuSTAR} should provide more stringent limits.

Moreover, our work stands as a valid supporting alternative to methods involving direct detections of (extremely weak) absorption signals from the WHIM towards distant sources \citep[e.g.][and references therein]{Nicastro13:WHIMblaz1553,Ren14:WHIMsupercl}, in which, however, definitive detections can not be easily obtained with current instruments, yet \citep[see][and references therein]{Nicastro16:WHIMcontrov,Nicastro17:WHIMupdated}. 

\section{Conclusions}
\label{sec:conclusions}

The role played by the IGM in X-ray absorption, obviously increasing with redshift and likely dominating above $z\sim2$, was first inferred empirically \citep[see][and references therein]{Behar:baryons,Campana12:nhxcomplete,Starling:evoluzIGM,Eitan:2013,Arcodia16:GRBsNhz} and then confirmed through dedicated cosmological simulations \citep{Campana:missing}. We tested it studying a sample of high-redshift blazars. Since blazars are characterised by a kpc-scale relativistic jet pointing towards us, the host X-ray absorption component along the LOS have been likely swept. Hence, detecting the signature of X-ray absorption in excess to the Galactic value in the X-ray spectra of distant blazars provided strong insights in favour of the IGM absorption scenario. 

Our sample of blazars consisted in 15 sources selected above $z=2$ and observed by \emph{XMM-Newton} with at least $\sim10\,000$ photons detected (by all the three EPIC cameras combined). Moreover, 6 of these blazars boasted additional \emph{NuSTAR} (and simultaneous \emph{Swift-XRT}) observations, thus providing a large broadband spectrum that allowed a more detailed analysis. In all sources an additional curvature term was required by data, in excess to a Galactic absorption component. It was first characterised in terms of \emph{either} an intrinsic extra-absorber (the easiest way to assess the presence of excess absorption) or an intrinsic spectral break. Both alternatives separately improved the fits, although often yielding statistically undistinguishable results for the single source. Then, for the first time we included both terms and this description was assessed to be the best-fit model. In particular, we obtained that excess absorption was fitted in all sources, while the continuum curvature terms were consistent with a power-law in 11 sources out of 15.

Hence, thanks to an overall sample analysis, with the additional help of a low-redshift sample used for comparisons, we were able to conclude that excess absorption is preferred to explain the observed soft X-ray spectral hardening. The intrinsic excess column densities obtained were compatible with the $N_H(z)-z$ relation and the simulated IGM absorption contributions \citep{Campana:missing}, along with the other extragalactic sources. Only a couple of outliers lied below the simulated envelopes and should perhaps be considered for a slight re-adaptation of the IGM characteristics.

In addition, we performed spectral fits directly modelling a WHIM contribution, finding agreement with its expected characteristics \citep[e.g.][]{Bregman07:IGMreview}. A joint fit with 4 sources (consistent with the IGM lowest absorption contribution) yielded a WHIM with average density $n_0=1.01^{+0.53}_{-0.72}\times10^{-7}$\,cm$^{-3}$ (at solar metallicity) and temperature $\log(T/\text{K})=6.45^{+0.51}_{-2.12}$. In deriving these parameters some of the continua in the spectral models were constrained to power-laws, so that a more flexible spectral analysis would probably yield upper limit measures for the WHIM characteristics. Then, the fitted hydrogen density value corresponds to an ionisation parameter of $\log\xi=1.47\pm0.27$, if a constant CXRB flux is used \citep[from][]{DeLuca04:CXRB}, and to an IGM metallicity of $Z=0.59^{+0.31}_{-0.42}Z_{\astrosun}$, if an hydrogen density of $1.7\times10^{-7}$\,cm$^{-3}$ is assumed \citep[from][and references therein]{Behar:baryons}. This is an important consistency check for our scenario.

Furthermore, by attributing the X-ray spectral hardening in high-$z$ blazars uniquely to excess absorption along the IGM in 11 sources, we were necessarily suggesting that intrinsic spectral breaks, predicted by emission models, were "missed" within the observed band. We thoroughly checked for each source that the observed parameters, e.g. photon indexes, were consistent with such an explanation. We proved that, in principle, our proposed scenario is valid and does not contradict blazars' emission models, short of a condition on the product $\gamma_{cool}\Gamma$ (or $\gamma_{min}\Gamma$).

Future prospects are aimed to obtain deeper exposures with current instruments of the best candidates, i.e. sources with a low Galactic column and compatible with the IGM absorbing envelope (e.g. the two outliers 4C 71.07 and PKS 2149-306). Simultaneous \emph{XMM-Newton}+\emph{NuSTAR} observations are suggested for a thorough and reliable spectral analysis. Looking beyond, our work can be used as a stepping stone for more meticulous studies involving \emph{Athena} \citep{Nandra13:Athena}.

\begin{acknowledgements}
	 We thank Fabrizio Tavecchio for useful discussions and Tullia Sbarrato for her precious help in building the samples. We thank the anonymous referees for helpful comments. This work made use of data from the \emph{NuSTAR} mission, a project led by the California Institute of Technology, managed by the Jet Propulsion Laboratory and funded by the National Aeronautics and Space Administration. This work made use of data supplied by the UK \emph{Swift} Science Data Centre at the University of Leicester. Additionally, we acknowledge the use of the matplotlib package \citep{Hunter07:matplotlib}.
\end{acknowledgements}

%
%
\bibliographystyle{aa} 
\bibliography{bibliography} 

\onecolumn
	\input{Table}
	\input{Tab_lowz}
\twocolumn

\begin{appendix}
\section{Individual sources processing}
\label{sec:appA}
\subsection{7C 1428+4218 ($z=4.715$)}
\emph{XMM-Newton} observed 7C 1428+4218 three times: on 2002 December 9 (ObsID 0111260101) for a total exposure time of $18.9\,$ks; on 2003 January 17 (ObsID 0111260701) for a total exposure time of $14.6\,$ks; on 2005 June 5 (ObsID 0212480701) for a total exposure time of $19.7\,$ks. All three observations were performed with thin filter and Full Frame mode for all the EPIC cameras. 

The first observation was discarded after being processed due to low photon counts ($\approx3000$), compared to the others. The second observation, hereafter Obs2003, was regularly processed, i.e. we selected the default limit rate choice ($<0.35\,$c$\,$s$^{-1}$ for MOS1 and MOS2, $<0.4$\,c$\,$s$^{-1}$ for pn) for FPB filtering. The source spectrum was extracted from a circular region with a radius of $38''$ for all the three EPIC cameras (paying particular attention in avoiding the faint X-ray source 2XMM J143020.9+420529, located $\sim62''$ away). The background was derived from a circular source-free region with the same radius ($54''$ only for MOS2) near the selected source (this is valid throughout this section, unless otherwise stated). Using the \texttt{epatplot} task, we observed no pile-up for this observation. Also, according to \texttt{WebPIMMS}, the expected pile-up fraction was $0.6\%$ for EPIC-MOS cameras, $0.7\%$ for EPIC-pn (values under $5\%$ are typically fine). At the end of the processing for Obs2003, the result was a series of spectra with a count rate of 0.57 c$\,$s$^{-1}$ in 11.5 ks, 0.15 c$\,$s$^{-1}$ in 14.2 ks and 0.16 c$\,$s$^{-1}$ in 14.2 ks for EPIC-pn, MOS1 and MOS2 respectively. They provide a total number of $\sim11000$ photons.

The same procedure was adopted for the third observation (hereafter Obs2005), with only a few slightly different choices. First of all, the EPIC-pn event list was filtered from FPB below $<0.5\,$c$\,$s$^{-1}$. The source and background regions were selected similarly to Obs2003 ($38''$ circle for MOS2 and pn, $36''$ for MOS1). No pile-up is apparent for this observation as well. The result is a series of spectra with a count rate of 0.52 c$\,$s$^{-1}$ in 11.7 ks, 0.14 c$\,$s$^{-1}$ in 17.4 ks and 0.14 c$\,$s$^{-1}$ in 17.3 ks for EPIC-pn, MOS1 and MOS2 respectively. They provide a total number of $\sim11000$ photons, as much as in Obs2003.

7C 1428+4218 has been observed by \emph{NuSTAR} on 2014 July 14 (ObsID 60001103002) for a total exposure of $49.2\,$ks. The source spectrum has been extracted for FPMA (FPMB) from a circular region with a radius of $\sim32''$ ($\sim31''$). At the end of the processing, FPMA and FPMB spectra provided a count rate of 0.023 c$\,$s$^{-1}$ in 49.2 ks and 0.022 c$\,$s$^{-1}$ in 48.9 ks, respectively and a total number of $\sim2200$ photons.

\emph{Swift-XRT} observed the source on 2014 July 13 (ObsID 00080752002) for a total exposure of $7.5\,$ks. The resulting spectrum showed a count rate of 0.033 c$\,$s$^{-1}$ in 7.4 ks, providing $\sim250$ photons.

In Table~\ref{tab:processing_overall} we report a summary of exposure times and photon counts for each observation of every source.

\subsection{QSO J0525-3343 ($z=4.413$)}
\emph{XMM-Newton} observed QSO J0525-3343 eight times: first, on 2001 February 11 and on 2001 September 15; then, a series of six more was made between 2003 February 14 and 2003 August 8. All observations were performed with thin filter and Full Frame imaging mode for all the EPIC cameras.

Following \citet{Worsley04:0525warm}, we discarded the first observation because badly affected by background flaring. The second (ObsID 0050150301, Rev. 324; hereafter Obs324) was processed and analysed. Then, three of the remaining observations (Rev. 593, 598 and 603) were discarded, before processing, since their expected photon counts \citep[see][Table 1]{Worsley04:0525warm} were deemed to be non influential on the total amount of photons expected from the combined analysis. In addition, another observation was discarded after being processed for the same reason (Rev. 671). Three observations remained: Obs324, for a total exposure time of 28.4 ks; Obs583 (ID 0149500601, Rev. 583) and Obs588 (ID 0149500701, Rev. 588), both for a total exposure time of 12.2 ks.

For each observation, the event file was filtered from FPB: for Obs324 below $<0.45\,$c$\,$s$^{-1}$ for MOS1 and MOS2, $<0.5$\,c$\,$s$^{-1}$ for pn; Obs583 and 588 were filtered with the default limit rate choice. For all the three EPIC cameras of Obs324, the source spectrum was extracted from a circular region with a radius of $42''$. Regarding Obs583, the radius of the source and background region was $42''$ for the MOS cameras, $38''$ for the pn. In Obs588, circular regions with radius $34''$ (MOS1) and $36''$ (MOS2, pn) were chosen. No pile-up effect was observed for any observation of QSO J0525-3343.

At the end of the processing for Obs324, the result was three spectra with a total number of $\sim12000$ photons. Obs583 and 588 consist in a total number of $\sim6500$ photons each.

\subsection{QSO B1026-084 ($z=4.276$)}

\emph{XMM-Newton} observed QSO B1026-084 twice: on 2002 May 15 (ObsID 0093160701) for a total exposure time of $7.9\,$ks and on 2003 June 13 (ObsID 0153290101) for a total exposure time of $43.4\,$ks. Both observations were performed with thin filter and Full Frame mode for all the EPIC cameras.

Only the second (longer) observation was processed. The event file was filtered from FPB selecting the default limit rate choice. The source spectrum was extracted from a circular region with a radius of $36''$ for the EPIC-MOS cameras. Circular regions with a radius of $34''$ were taken for the EPIC-pn camera, paying particular attention in avoiding the couple of unidentified faint X-ray sources located few arc-seconds away. No pile-up effect was apparent for this blazar.

The result was a series of spectra with a count rate of 0.31 c$\,$s$^{-1}$ in 15.6 ks, 0.09 c$\,$s$^{-1}$ in 21.5 ks and 0.09 c$\,$s$^{-1}$ in 22.1 ks for pn, MOS1 and MOS2 respectively. They provide a total number of $\sim9000$ photons. It is slightly below the threshold of our selection criterion, but it was included in the analysis.

\subsection{QSO B0014+810 ($z=3.366$)}

\emph{XMM-Newton} observed QSO B0014+810 on 2001 August 23 (ObsID 0112620201) for a total exposure time of $42.9\,$ks. The observation was performed with medium filter and Full Frame mode for all the EPIC cameras.

The event file was filtered from FPB selecting the default limit rate choice. The source spectrum was extracted from a circular region with a radius of $42''$ for the MOS cameras, $35''$ for the EPIC-pn. No pile-up was observed for this blazar. Also, the tool \texttt{WebPIMMS} yielded an expected pile-up fraction of $1.3\%$ for EPIC-MOS cameras, $1.1\%$ for EPIC-pn. The result was a series of spectra with a count rate of 0.96 c$\,$s$^{-1}$ in 13.4 ks, 0.31 c$\,$s$^{-1}$ in 19.8 ks and 0.31 c$\,$s$^{-1}$ in 20.4 ks for pn, MOS1 and MOS2 respectively. They provide a total number of $\sim25000$ photons.

QSO B0014+081 has been observed three times by \emph{NuSTAR}: on 2014 December 21 (ObsID 60001098002, hereafter Obs2014) for a total exposure of $31.0\,$ks; on 2015 January 23 (ObsID 60001098004, hereafter Obs2015) for a total exposure of $36.4\,$ks; the most recent and shortest observation (ObsID 90201019002, 2016 April 12) was not processed. 

In Obs2014 (Obs2015), the source spectrum has been extracted for both FPMA and FPMB from a circular region with a radius of $\sim42''$ ($\sim41''$). In Obs2014 (Obs2015), FPMA and FPMB spectra have a count rate of 0.08 c$\,$s$^{-1}$ in 31.0 ks (0.07 in 36.1) and 0.07 c$\,$s$^{-1}$ in 31.0 ks (0.07 in 36.3), respectively, providing a total number of $\sim4700$ (5000) photons.

\emph{Swift-XRT} performed two observations of blazar QSO B0014+810, simultaneously with \emph{NuSTAR}, on 2014 December 21 (ObsID 00080003001), and on 2015 January 23 (ObsID 00080003002). The resulting spectrum for Obs2014 (Obs2015) showed a count rate of 0.09 c$\,$s$^{-1}$ in 6.5 ks (0.07 in 6.6), providing $\sim590$ (460) photons.

\subsection{PKS 2126-158 ($z=3.268$)}

\emph{XMM-Newton} observed PKS 2126-158 on 2001 May 1 (ObsID 0103060101) for a total exposure time of $23.4\,$ks. The observation was performed with medium filter and Full Frame mode for the EPIC-MOS cameras, with the same filter but in Extended Full Frame mode for the EPIC-pn.

The event file of the EPIC-pn has been filtered from FPB selecting the default limit rate choice, while for the EPIC-MOS cameras the event file has been filtered below $<0.18\,$c$\,$s$^{-1}$. The source spectrum was first extracted from a circular region with a radius of $40''$ for the MOS cameras, $36''$ for the EPIC-pn. However, a possible pile-up contamination was found using the SAS task \texttt{epatplot}, as the expected pattern distributions seemed to be discrepant from the observed ones. A more conservative annular region was then opted for all the EPIC cameras. We excised a $5''$ core from the MOS1 circular source region, obtaining a better fit with \texttt{epatplot}. The same internal radius was then adopted for the EPIC-MOS2 annular region. We then excised the core from the EPIC-pn source region up to $12''$ before finding an adequate result with \texttt{epatplot}. A greater excision was expected because EPIC-pn operated in Extended Full Frame mode, for which the image collection time is longer than in the normal Full Frame mode. Thus, pile-up becomes non negligible at a lower count rate. No pile-up effects were observed for this blazar after the selection of annular regions. Also the tool \texttt{WebPIMMS} was used for consistency, confirming the result: the expected pile-up fraction is $2.5\%$ for EPIC-MOS cameras, $3.9\%$ for EPIC-pn.

At the end of the processing, the result was a series of spectra with a count rate of 0.78 c$\,$s$^{-1}$ in 13.1 ks, 0.55 c$\,$s$^{-1}$ in 19.9 ks and 0.56 c$\,$s$^{-1}$ in 19.9 ks for pn, MOS1 and MOS2 respectively. They provide a total number of $\sim32000$ photons.

\subsection{QSO B0537-286 ($z=3.104$)}

\emph{XMM-Newton} observed QSO B0537-286 first on 2000 March 19 (ObsID 0114090101, hereafter Obs00) for a total exposure time of $53.0\,$ks. The observation was separated in two consecutive exposures for each EPIC camera. The first exposure was performed with medium filter and Full Frame mode for MOS1 and pn, in Large Window mode for MOS2. During the second exposure, MOS1 operated in Large Window mode, whilst MOS2 and pn in Full Frame mode. 

For each exposure, the event file has been filtered from FPB selecting the default limit rate choice. For each EPIC camera, the two consecutive exposures were then merged through the XMM SAS task \texttt{merge}.

\emph{XMM-Newton} observed QSO B0537-286 a second time on 2005 March 20 (ObsID 0206350101, hereafter Obs05) for a total exposure time of $81.9\,$ks. This observation was performed with thin filter and all the EPIC cameras operated in Full Frame mode. The event file of the EPIC-pn was filtered from FPB below $<0.35\,$c$\,$s$^{-1}$, while the EPIC-MOS cameras were filtered below $<0.15\,$c$\,$s$^{-1}$. 

The source spectrum of Obs00 was extracted from a circular region with a radius of $38''$ for the MOS cameras, $33''$ for the EPIC-pn. Note that the two EPIC-MOS cameras operated alternatively in Large Window and Full frame mode in the two consecutive exposures. Thus, the outer regions of the merged central chip have lower background. We then selected the background region from an external chip. The same radii were adopted for Obs05, in which QSO B0537-286 was detected at a moderately large off-axis angle. We note that no pile-up contamination was apparent in both observations.

At the end of the processing, the result for Obs00 (Obs05) was a series of spectra with a count rate of 0.81 (0.76) c$\,$s$^{-1}$ in 32.2 (13.5) ks, 0.25 (0.23) c$\,$s$^{-1}$ in 38.1 (23.4) ks and 0.25 (0.23) c$\,$s$^{-1}$ in 38.4 (19.1) ks for pn, MOS1 and MOS2 respectively. They provide a total number of $\sim45000$ (20000) photons.

\subsection{QSO B0438-43 ($z=2.852$)}

\emph{XMM.Newton} observed QSO B0438-43 on 2002 April 6 (ObsID 0104860201) for a total exposure time of $12.9\,$ks. The observation was performed with thin filter and Full Frame mode for all the EPIC cameras.

The event file was filtered from FPB selecting the default limit rate choices. The source spectrum was extracted from a circular region with a radius of $46''$ for the MOS cameras, $35''$ for the EPIC-pn. The result was a series of spectra with a count rate of 0.89 c$\,$s$^{-1}$ in 8.8 ks, 0.27 c$\,$s$^{-1}$ in 12.2 ks and 0.28 c$\,$s$^{-1}$ in 12.2 ks for pn, MOS1 and MOS2 respectively. They provide a total number of $\sim14500$ photons.

\subsection{RBS 315 ($z=2.69$)}

\emph{XMM-Newton} observed RBS 315 three times: the first on 2003 July 25 (ObsID 0150180101, hereafter Obs2003) for a total exposure time of $22.2\,$ks; then twice (two days apart) on 2013, on January 13 (ObsID 0690900101, hereafter Obs2013a) and on January 15 (ObsID 0690900201, hereafter Obs2013b), for a total exposure time of $108.0\,$ks and $96.7\,$ks,respectively. All the observations were performed with thin filter and Full Frame mode for all the EPIC cameras.

The event files of Obs2003 have been filtered from FPB selecting the default limit rate choice. Obs2013a was quite affected by FPB, particularly at the beginning and at the end of the exposure. In the EPIC-MOS event files the background rate was lower than the default selection, hence we filtered below $0.15$ and $0.2\,$c$\,$s$^{-1}$ for MOS1 and 2, respectively. Moreover, we excluded any $t<4.74516E08\,$s (the first $\sim17\,$ks of the exposure) from the Good Time Intervals (GTI) for both MOS cameras. The EPIC-pn event list was filtered with the default limit rate choice and in addition any $t\ge4.74588E08\,$s was excluded. Obs2013b was comparably affected by FPB and the event lists were filtered with a similar modus operandi: all the EPIC event files were filtered with the same rate thresholds of Obs2013a; any $t>4.7475E08\,$s was excluded from all the EPIC event files, whilst EPIC-pn was additionally shortened excising any $t<4.74718E08\,$s.

In all the observations, the source spectrum was first extracted from circular regions. However, some pile-up contamination was found using the SAS task \texttt{epatplot}, as the expected pattern distributions seemed to be discrepant from the observed ones. A more conservative annular region was then opted for all the EPIC cameras. In Obs2003, we excised a $10''$ core from all the source regions, obtaining a better fit with \texttt{epatplot}. The selected external radius is $44''$ for MOS cameras, and $37''$ for EPIC-pn. Same for Obs2013a [Obs2013b], selecting annuli with $R_i,R_e=(10'',40'')$ [$(13'',44'')$] for MOS1/2 and annuli with $R_i,R_e=(10'',36'')$ [$(13'',38'')$] for pn. In all observations, the background was extracted from a circular (source-free) region with radius equal to $R_e$, for each EPIC camera. No pile-up was apparent for this blazar after the selction of annular regions. Also the tool \texttt{WebPIMMS} was used for consistency, confirming the result: the expected pile-up fraction in Obs2013a (the one with the highest count rate) is $1.9\%$ for all EPIC cameras. For Obs2003 and Obs2013b, with a lower count rate, the pile-up fraction is expected to be even smaller.

At the end of the processing, Obs2003 consists in a series of spectra with a count rate of 1.24 c$\,$s$^{-1}$ in 18.2 ks, 0.42 c$\,$s$^{-1}$ in 21.6 ks and 0.41 c$\,$s$^{-1}$ in 21.7 ks for pn, MOS1 and MOS2 respectively. They provide a total number of $\sim40500$ photons. In Obs2013a (Obs2013b), count rates of 1.32 in 40.4 ks (0.93 in 28.4 ks), 0.43 in 54.7 ks (0.31 in 66.8 ks) and 0.43 in 56.6 ks (0.32 in 67.6 ks) for pn, MOS1 and MOS2 respectively, provide a total number of $\sim101000$ ($\sim69000$) photons.

RBS 315 was observed twice by \emph{NuSTAR}: on 2014 December 24 (ObsID 60001101002, hereafter Obs2014) for a total exposure of $37.4\,$ks and on 2015 January 18 (ObsID 60001101004, hereafter Obs2015) for a total exposure of $31.9\,$ks. In both observations the source spectrum was extracted for both FPMA and FPMB from a circular region with a radius of $\sim70''$. At the end of the processing for Obs2014 (Obs2015), FPMA and FPMB spectra showed a count rate of 0.38 c$\,$s$^{-1}$ in 31.5 ks (0.26 in 37.4) and 0.35 c$\,$s$^{-1}$ in 31.7 ks (0.24 in 37.4), respectively. They provide a total number of $\sim23000$ (19000) photons.

\emph{Swift-XRT} performed two observations of blazar RBS 315, simultaneously with \emph{NuSTAR}, on 2014 December 24 (ObsID 00080243001), and on 2015 January 18 (ObsID 00080243002). The resulting spectrum for Obs2014 (Obs2015) showed a count rate of 0.21 c$\,$s$^{-1}$ in 4.9 ks (0.15 in 5.1), providing $\sim1000$ (800) photons.

\subsection{QSO J2354-1513 ($z=2.675$)}

\emph{XMM-Newton} observed QSO J2354-1513 on 2004 December 5 (ObsID 0203240201) for a total exposure time of $86.9\,$ks. The observation was performed with thin filter and Full Frame mode for all the EPIC cameras.

The event file was filtered from FPB below $<0.4\,$c$\,$s$^{-1}$ for MOS1 and MOS2, $<0.7$\,c$\,$s$^{-1}$ for pn. Moreover, a time-threshold was added to the MOS2 event file and we excluded any exposure time after $2.18641E08\,$s. The source spectrum was extracted from a circular region with a radius of $38''$ for the MOS cameras, $30''$ for the EPIC-pn. At the end of the processing, the result was a series of spectra with a count rate of 0.32 c$\,$s$^{-1}$ in 33.2 ks, 0.09 c$\,$s$^{-1}$ in 49.7 ks and 0.09 c$\,$s$^{-1}$ in 31.5 ks for pn, MOS1 and MOS2 respectively. They provide a total number of $\sim18000$ photons.

\subsection{PBC 1656.2-3303 ($z=2.4$)}

\emph{XMM-Newton} observed PBC 1656.2-3303 on 2009 September 11 (ObsID 0601741401) for a total exposure time of $22.6\,$ks. The observation was performed with medium filter and Full Frame mode for all the EPIC cameras.

The event file was filtered from FPB below the default limit-rate for MOS1 and MOS2, $<0.55$\,c$\,$s$^{-1}$ for pn. The source spectrum was extracted from a circular region with a radius of $50''$ for the MOS cameras, $43''$ for the EPIC-pn. No pile-up contamination was noted for this blazar. At the end of the processing, the result was a series of spectra with a count rate of 0.98 c$\,$s$^{-1}$ in 17.9 ks and 0.34 c$\,$s$^{-1}$ in 22.0 ks for EPIC-pn and EPIC-MOS cameras, respectively. They provide a total number of $\sim32500$ photons.

PBC 1656.2-3303 was observed by \emph{NuSTAR} on 2015 September 27 (ObsID 60160657002) for a total exposure of $21.1\,$ks. The source spectrum was extracted from a circular region with a radius of $\sim63''$ ($\sim60''$) for FPMA (FPMB). At the end of the processing, FPMA and FPMB spectra showed a count rate of 0.14 c$\,$s$^{-1}$ in 20.5 ks and 0.13 c$\,$s$^{-1}$ in 20.8 ks, respectively. They provide a total number of $\sim5600$ photons.

\emph{Swift-XRT} observed PBC 1656.2-3303, simultaneously with \emph{NuSTAR}, on 2015 September 27 (ObsID 00081202001). The processed spectrum showed a count rate of 0.09 c$\,$s$^{-1}$ in 6.9 ks, providing $\sim620$ photons. 

\subsection{QSO J0555+3948 ($z=2.363$)}

\emph{XMM-Newton} observed QSO J0555+3948 on 2005 April 1 (ObsID 0300630101) for a total exposure time of $31.2\,$ks. The observation was performed with medium filter and Full Frame mode for all the EPIC cameras.

The event file was filtered from FPB below the default limit rate for MOS1 and MOS2, $<0.5$\,c$\,$s$^{-1}$ for pn. The source spectrum was extracted from a circular region with a radius of $45''$ for the MOS cameras, $40''$ for the EPIC-pn. No pile-up was observed in this processing analysis for QSO J0555+3948. The result was a series of spectra with a count rate of 0.40 c$\,$s$^{-1}$ in 12.4 ks, 0.13 c$\,$s$^{-1}$ in 19.3 ks and 0.14 c$\,$s$^{-1}$ in 19.0 ks for pn, MOS1 and MOS2 respectively. They provide a total number of $\sim10100$ photons.

\subsection{PKS 2149-306 ($z=2.345$)}

\emph{XMM-Newton} observed PKS 2149-306 on 2001 May 1 (ObsID 0103060401) for a total exposure time of $24.9\,$ks. The observation was performed with medium filter for all the EPIC cameras and the EPIC-MOS operated in Full Frame mode, while EPIC-pn in Large Window mode. The event file was filtered from FPB selecting the default limit rate choice. The source spectrum was extracted from a circular region with a radius of $46''$ for MOS1, $42''$ for MOS2 and $38''$ for the EPIC-pn (paying particular attention in avoiding the near faint X-ray source, likely 2XMM J215159.2-302735 located $\gtrsim50''$ away). No pile-up contamination was apparent in this analysis and \texttt{WebPIMMS} yielded an expected pile-up fraction of $2.9\%$ for MOS cameras, $1.8\%$ for pn. At the end of the processing, \emph{XMM-Newton} spectra showed a count rate of 2.06 c$\,$s$^{-1}$ in 19.6 ks, 0.60 c$\,$s$^{-1}$ in 23.9 ks and 0.59 c$\,$s$^{-1}$ in 23.9 ks for pn, MOS1 and MOS2 respectively. They provide a total number of $\sim69000$ photons.

PKS 2149-306 was observed twice by \emph{NuSTAR}: on 2013 December 17 (ObsID 60001099002, hereafter Obs2013) for a total exposure of $38.5\,$ks and on 2014 April 18 (ObsID 60001099004, hereafter Obs2014) for a total exposure of $44.2\,$ks. In both observations, the source spectrum was extracted for FPMA (FPMB) from a circular region with a radius of $\sim83''$ ($\sim76''$). FPMA and FPMB spectra yielded a count rate of 0.78 c$\,$s$^{-1}$ in 38.4 ks (0.67 in 44.0) and 0.78 c$\,$s$^{-1}$ in 38.3 ks (0.58 in 43.9), respectively, providing a total number of $\sim60000$ (54900) photons. 

\emph{Swift-XRT} performed two observations of blazar PKS 2149-306, simultaneously with \emph{NuSTAR}, on 2013 December 16 (ObsID 00031404013), and on 2014 April 18 (ObsID 00031404015). The processed Obs2013 (Obs2014) resulted in a count rate of 0.36 c$\,$s$^{-1}$ in 7.1 ks (0.33 in 6.4), providing $\sim2600$ (2100) photons.

\subsection{QSO B0237-2322 ($z=2.225$)}

\emph{XMM-Newton} observed QSO B0237-2322 on 2006 January 20 (ObsID 0300630301) for a total exposure time of $26.9\,$ks. The observation was performed with thin filter and Full Frame mode for all the EPIC cameras.

The event file was filtered from FPB below $<0.2\,$c$\,$s$^{-1}$ for the two EPIC-MOS cameras and $<0.45\,$c$\,$s$^{-1}$ for EPIC-pn. In the latter case, we also opted for a time selection excluding the exposure up to $t=2.54094E08\,$s (corresponding roughly to the first $10\,$ks of the observation). The source spectrum was extracted from a circular region with a radius of $40''$ for MOS1, $36''$ for MOS2 and $30''$ for the EPIC-pn (paying particular attention in avoiding the unidentified faint X-ray source located $\approx60''$ away). No pile-up contaminated this XMM-Newton observation, according to our analysis with \texttt{epatplot}. The result was a series of spectra with a count rate of 0.89 c$\,$s$^{-1}$ in 9.7 ks, 0.24 c$\,$s$^{-1}$ in 19.1 ks and 0.24 c$\,$s$^{-1}$ in 17.6 ks for pn, MOS1 and MOS2 respectively, with a total number of $\sim17400$ photons.

\subsection{4C 71.07 ($z=2.172$)}

\emph{XMM-Newton} observed 4C 71.07 on 2001 April 13 (ObsID 0112620101) for a total exposure time of $36.7\,$ks. The observation was performed with medium filter and Large Window mode for the EPIC-MOS cameras, with the same filter but in Full Frame mode for the EPIC-pn.

The event file of the EPIC-MOS cameras was filtered from FPB below $<0.4\,$c$\,$s$^{-1}$, below $<0.8\,$c$\,$s$^{-1}$ for EPIC-pn. In the latter case, we also excluded any exposure time prior to $t=1.03494E08\,$s (roughly the first 7 ks of the observation). The source spectrum was first extracted from a circular region with a radius of $55''$ for the MOS cameras, $41''$ for the EPIC-pn. Since the MOS cameras operated in Large Window mode, the background region was selected on an outer chip. However, some pile-up contamination was found using the SAS task \texttt{epatplot}. A more conservative annular region was then opted for all the EPIC cameras excising a $12''$ core. At the end of the processing, count rates of 3.65 c$\,$s$^{-1}$ in 22.3 ks, 1.22 c$\,$s$^{-1}$ in 28.0 ks and 1.18 c$\,$s$^{-1}$ in 28.0 ks were obtained for pn, MOS1 and MOS2 respectively. They provide a total number of $\sim149000$ photons.

4C 71.07 was observed twice by \emph{NuSTAR}: on 2013 December 15 (ObsID 60002045002, hereafter Obs2013) for a total exposure of $30.0\,$ks and on 2014 January 18 (ObsID 60002045004, hereafter Obs2014) for a total exposure of $36.4\,$ks. In Obs2013 (Obs2014), the source spectrum has been extracted for both FPMs from a circular region with a radius of $\sim65''$ ($\sim69''$). In Obs2013 (Obs2014), FPMA and FPMB spectra resulted in a count rate of 0.37 c$\,$s$^{-1}$ in 29.6 ks (0.68 in 36.2) and 0.40 c$\,$s$^{-1}$ in 29.6 ks (0.64 in 36.2), respectively, with a total number of $\sim23000$ (48000) photons.

Swift-XRT performed two observations of blazar 4C 71.07, simultaneously with NuSTAR, on 2013 December 16 (ObsID 00080399001), and on 2014 January 18 (ObsID 00080399002). The processed spectrum for Obs2013 (Obs2014) showed a count rate of 0.29 c$\,$s$^{-1}$ in 5.0 ks (0.33 in 4.7), providing $\sim1450$ (1550) photons.

\subsection{PKS 0528+134 ($z=2.07$)}

\emph{XMM-Newton} observed PKS 0528+134 four times within six days on 2009. The first $30.2\,$ks observation (ObsID 0600121401) was performed on September 8, followed by a $29.2\,$ks observation (ObsID 0600121501, hereafter Obs501) on September 10, a $27.6\,$ks observation (ObsID 0600121601, hereafter Obs601) on September 11 and a $39.3\,$ks observation (ObsID 0600121701, hereafter Obs701) on September 14. All observations were performed with thin filter and Full Frame imaging mode for all the EPIC cameras, except the EPIC-pn exposure of Obs501 performed in Small Window mode.

Obs401 was heavily affected by background flaring. The EPIC-pn event file was shortened up to a $\sim6\,$ks residual duration and the processed observation, that has the highest count rate among the EPIC cameras, would have provided just $\approx1500$ photons. This observation was then discarded. In Obs501 the EPIC-pn exposure (performed in Small Window mode) was conservatively discarded after processing due to heavy background contamination resulting in a low number of photons and in an ambiguous pile-up \texttt{epatplot} check. For the remaining observations (Obs601 and 701) all EPIC exposures were processed and analysed. For each observation, the event file of the MOS cameras was filtered from FPB below the default limit choice, while the EPIC-pn event file was filtered below $0.6\,$c$\,$s$^{-1}$ (0.55) for Obs601 (Obs701). In addition to the rate threshold, the MOS cameras in Obs701 were filtered excluding all the exposure time above $t=3.693E08\,$s. For Obs501, the source spectrum was extracted from a circular region with a radius of $42''$ for MOS1 and $40''$ for MOS2, while for Obs601 a $45''$- and $30''$-radius was opted for the source region of the MOS cameras and pn, respectively. In Obs701, $38''$, $36''$ and $38''$ for MOS1, MOS2 and pn, respectively.

No pile-up contamination was observed in any observation of this blazar. At the end of the processing a total number of $\sim3900$, $\sim9900$ and $\sim4000$ photons was provided by Obs501, Obs601 and 701, respectively.

\begin{table}[tb]
	\scriptsize
	\setlength{\tabcolsep}{3.8pt}
	\caption{Summary for individual sources processing. We report the exposure time (in ks) and the number of photons for each observation of every source, referring to different cameras when needed. Each observation ID is labelled with "XMM", "NU", or "XRT", for \emph{XMM-Newton}, \emph{NuSTAR} and \emph{Swift-XRT} observations, respectively.}
	\label{tab:processing_overall}
	\centering
	\begin{tabular}{cccc}
		\toprule
		\multicolumn{1}{c}{Source}& 
		\multicolumn{1}{c}{Obs. ID}& 
		\multicolumn{1}{c}{Exposure time (ks)/camera}& 
		\multicolumn{1}{c}{Tot. counts} \\
		\midrule
		7C 1428+4218& XMM 0111260701&	11.5/pn  14.2/MOS & 11000 \\
		& XMM 0212480701&	11.7/pn  17.4/MOS1  17.3/MOS2 & 11000 \\
		& NU 60001103002&	49.2/FPMA  48.9/FPMB & 2200 \\
		& XRT 00080752002&	7.4 & 250 \\
		QSO J0525-3343	& XMM 0050150301&	16.1/pn  24.5/MOS1 24.3/MOS2 & 12000 \\
		& XMM 0149500601&	8.4/pn  11.7/MOS1  11.6/MOS2 & 6500 \\
		& XMM 0149500701&	8.0/pn  11.8/MOS1  11.9/MOS2 & 6500 \\
		QSO B1026-084	& XMM 0153290101&	15.6/pn  21.5/MOS1  22.1/MOS2 & 9000 \\
		QSO B0014+810	& XMM 0112620201&	13.4/pn  19.8/MOS1  20.4/MOS2 & 25000 \\
		& NU 60001098002&	31.0/FPMA  31.0/FPMB & 4700 \\
		& NU 60001098004&	36.1/FPMA  36.3/FPMB & 5000 \\
		& XRT 00080003001&	6.5 & 600 \\
		& XRT 00080003002&	6.6 & 500 \\
		PKS 2126-158	& XMM 0103060101&	13.1/pn  19.9/MOS & 32000 \\						
		QSO B0537-286	& XMM 0114090101&	32.2/pn  38.1/MOS1 38.4/MOS2 & 45000 \\
		& XMM 0206350101&	13.5/pn  23.4/MOS1 19.1/MOS2 & 20000 \\
		QSO B0438-43	& XMM 0104860201&	8.8/pn  12.2/MOS & 14500 \\
		RBS 315			& XMM 0150180101&	18.2/pn  21.6/MOS1  21.7/MOS2 & 40500 \\
		& XMM 0690900101&	40.4/pn  54.7/MOS1  56.6/MOS2 & 101000 \\
		& XMM 0690900201&	28.4/pn  66.8/MOS1  67.6/MOS2 & 69000 \\
		& NU 60001101002&	31.5/FPMA  31.7/FPMB & 23000 \\
		& NU 60001101004&	37.4/FPMA  37.4/FPMB & 19000 \\
		& XRT 00080243001&	4.9 & 1000 \\
		& XRT 00080243002&	5.1 & 800 \\
		QSO J2354-1513	& XMM 0203240201&	33.2/pn  49.7/MOS1 31.5/MOS2 & 18000 \\
		PBC 1656.2-3303	& XMM 0601741401&	17.9/pn  22.0/MOS & 32500 \\
		& NU 60160657002&	20.5/FPMA  20.8/FPMB & 5600 \\
		& XRT 00081202001&	6.9 & 600 \\
		QSO J0555+3948	& XMM 0300630101&	12.4/pn  19.3/MOS1 19.0/MOS2 & 10000 \\
		PKS 2149-306	& XMM 0103060401&	19.6/pn  23.9/MOS & 69000 \\
		& NU 60001099002&	38.4/FPMA  38.3/FPMB & 60000 \\
		& NU 60001099004&	44.0/FPMA  43.9/FPMB & 55000 \\
		& XRT 00031404013&	7.1 & 2600 \\
		& XRT 00031404015&	6.4 & 2100 \\
		QSO B0237-2322	& XMM 0300630301&	9.7/pn  19.1/MOS1 17.6/MOS2 & 17000 \\
		4C 71.07		& XMM 0112620101&	22.3/pn  28.0/MOS & 149000 \\
		& NU 60002045002&	29.6/FPMA  29.6/FPMB & 23000 \\
		& NU 60002045004&	36.2/FPMA  36.2/FPMB & 48000 \\
		& XRT 00080399001&	5.0 & 1500 \\
		& XRT 00080399002&	4.7 & 1600 \\
		PKS 0528+134	& XMM 0600121501&	21.3/MOS1 21.5/MOS2 & 3900 \\
		& XMM 0600121601&	20.0/pn  26.5/MOS1 26.6/MOS2 & 9900 \\
		& XMM 0600121701&	7.3/pn  12.7/MOS1 12.6/MOS2 & 4000 \\
		\bottomrule
	\end{tabular}
\end{table}

\section{Comparison with previous works: outliers of the $N_H(z)-z$ relation}
\label{sec:app_literaturesources}

The $N_H(z)-z$ relation is confirmed by data of the literature (see Section~\ref{sec:nhzrelation}). Only one object showed an incompatibly low column density, i.e. 4C 06.41 ($z=1.27$), for which \citet{Eitan:2013} reported from EPIC-pn data an intrinsic column density upper limit of $N_H(z)<0.5\times10^{20}$ cm$^{-2}$. We reprocessed the \emph{XMM-Newton} observation (ObsID 0151390101) for all EPIC cameras, following the standard procedures explained in Section~\ref{sec:observations}. Fitting the X-ray spectrum with a PL model yielded a reduced chi-square of $1.17$ (with 394 dof). Adding an excess absorption component did not improve the fit ($\chi^2_{\nu}=1.18/393$) and an even lower column density was obtained ($N_H(z)<0.2\times10^{20}$ cm$^{-2}$). Nonetheless, the data-model ratio showed some residual curvature in the data, thus we switched to a LGP continuum. The fit was significantly improved to a reduced chi-square of $\chi^2_{\nu}=0.92/392$ (F-test $p$-value of $\sim10^{-23}$), with an excess column density upper limit of $N_H(z)<0.5\times10^{21}$ cm$^{-2}$ and a non-zero curvature term. This column density is perfectly consistent with our proposed scenario. 

Furthermore, different Galactic absorption models were used in the literature. However, in 4C 71.07 (PKS 2149-306) the upper limit changed from $<0.09\times10^{22}$ cm$^{-2}$ (<0.07) to $<0.06\times10^{22}$ cm$^{-2}$ (0.06) when the Galactic model was switched from LAB to Willingale's \citep[see also][]{Arcodia16:GRBsNhz}. Hence, even in the most critical case, we can confidently take 4C 71.07 and PKS 2149-306 as the lowest absorbed extragalactic objects.

In addition, a few objects apparently yielded a column density detection above the 2-sigma upper boundary of the mean IGM absorption contribution. As a matter of fact, non-blazar quasars are allowed to be observed above the simulated IGM curves, since some intrinsic absorption could be present and it is typically observed \citep[e.g.][]{Ricci17:quasarXABS}. In blazars, the lack of an intrinsic component invokes full consistency with the simulated mean IGM curve and its 1-sigma and 2-sigma boundaries, although it is also true that, by definition, a few objects are allowed to lie outside of the coloured areas. One outlier, PKS 0528+134, is ours and was already discussed. Three other blazars from the literature showed a relatively high detection of $N_H(z)$, namely QSO B0235+1624 ($z=0.94$), PKS 0838+133 ($z=0.68$) and QSO B0607+710 ($z=0.27$). 

In QSO B0235+1624, an excess column density of $N_H(z)=(0.60\pm0.03)\times10^{22}$ cm$^{-2}$ was obtained with EPIC-pn data by \citet{Eitan:2013} using a Galactic column density of $7.7\times10^{20}$ cm$^{-2}$ \citep{Kalberla:LAB}. Nonetheless, switching to the Willingale's Galactic model yields a greater value, i.e. $10.9\times10^{20}$ cm$^{-2}$. This is a moderately high increase of a factor $\sim42\%$, hence it could contribute to the excessively high intrinsic column density fitted by \citet{Eitan:2013}. We reprocessed the four archival \emph{XMM-Newton} observations of the source (i.e. ObsIDs 0110990101, 0206740101, 0206740501 and 0206740701). A PL model yielded a poor fit, with reduced chi-square of $6.73$ (387 dof), $2.39$ (332 dof), $1.55$ (248 dof) and $1.51$ (197 dof), for the four ObsIDs as ordered above, respectively. Adding excess absorption, the fit was significantly improved for all the four observations, although in Obs. 0110990101 a curved (concave) continuum was required. We then fitted all the observations simultaneously with a LGP+EX tying the excess absorption, while the source parameters were left free to vary in order to model the different states of blazar QSO B0235+1624 \citep[see][]{Raiteri06:0235analysis}. The fit yielded a good result ($\chi^2_{\nu}=1.01/1165$), with an absorbing excess column density of $N_H(z)=0.49^{+0.07}_{-0.03}\times10^{22}$ cm$^{-2}$. This blazar was confirmed to be slightly above the upper 2-sigma mean envelope, although to a lesser extent and still compatibly with a 3-sigma envelope. Its line of sight could be considered particularly absorbed \citep[e.g. a DLA with $\log N_{HI}=21.79\pm0.09$,][]{Junkkarinen04:absAO0235,Kanekar14:absDLAs}. Moreover, its Galactic column density is relatively high and the fitted value was a lower limit ($>10.5\times10^{20}$ cm$^{-2}$), perhaps implying that some more Galactic matter is required to better fit the data. 

In PKS 0838+133 and QSO B0607+710, the detection is really close to the 2-sigma superior limit of the mean envelope. Thus, this source would probably become consistent with our envelopes just including the Galactic molecular hydrogen in the X-ray spectral analysis.

\section{Intrinsic spectral breaks}
\label{sec:appB}

Several parameters are needed to shape intrinsic spectral breaks, i.e. the bulk Lorentz factor of the relativistic jet, the minimum energy of the electrons distribution or, in general, its shape and its cooling efficiency, the peak frequency of the photons distribution and the redshift of the source. The large number of parameters introduces some degeneracy among them, and if on the one hand it is possible to model any break in the X-ray band, on the other hand these breaks can easily occur out of the observing band. In this scenario the hardening would be attributed to excess absorption along the IGM. 

The physics of the jet emission is complex \citep[see, e.g.,][and references therein]{Sikora94:blazarsmodels,Sikora97:blazarsmodels,Sikora09:blazarsmodels,Tavecchio08:BLRspectrum,Ghisellini09:CanonicalBlazars-7Csed,Ghisellini15:blazarsMODELS}. We here briefly report the concepts essential for our discussion. The electrons injection function in the emitting region (supposed spherical with radius $R$) is a smoothly joining broken power-law:\begin{equation*}\label{eq:Qgamma}Q(\gamma)=Q_0\frac{\left(\gamma/\gamma_b\right)^{-s_1}}{1+\left(\gamma/\gamma_b\right)^{-s_1+s_2}}\,\,\,\,\,\,\,\,\,\,\,\,\,\,\,[\text{s}^{-1}\,\text{cm}^{-3}]
\end{equation*}and it is assumed constant in time within $\sim R/c$ \citep[e.g.][]{Ghisellini09:CanonicalBlazars-7Csed}. This is the time when the electrons energy distribution $N(\gamma)$ is computed:\begin{equation*}\label{eq:Ngamma}N(\gamma)=\frac{\int_{\gamma}^{\gamma_{max}}\left[Q(\gamma)+P(\gamma)\right]d\gamma}{\dot{\gamma}}\,\,\,\,\,\,\,\,\,\,\,\,\,\,\,[\text{cm}^{-3}]
\end{equation*} This holds above a particular energy, named $\gamma_{cool}$, while below it is proportional to the underlying $Q(\gamma)$. $P(\gamma)$ is the term representing the electron-positron pairs produced in photon-photon collisions, here neglected for simplicity.

For our purposes, it is sufficient to say that energy breaks in the emission spectrum are linked to the shape of the electrons energy distribution, i.e. to breaks between different slopes (refer to, e.g., \citet{Ghisellini15:blazarsMODELS} for a detailed explanation). However, this is valid provided that the photons spectrum can be approximated by a peaked distribution. In the BLR (torus) case, it is considered to be peaked at $\nu_{ext}$, equal to $2.46\times10^{15}\,$Hz ($7.7\times10^{13}\,$Hz) and it is typically approximated by a black-body \citep{Tavecchio08:BLRspectrum}.

\begin{figure}[tb]
	\centering	
	\includegraphics[width=0.9\columnwidth]{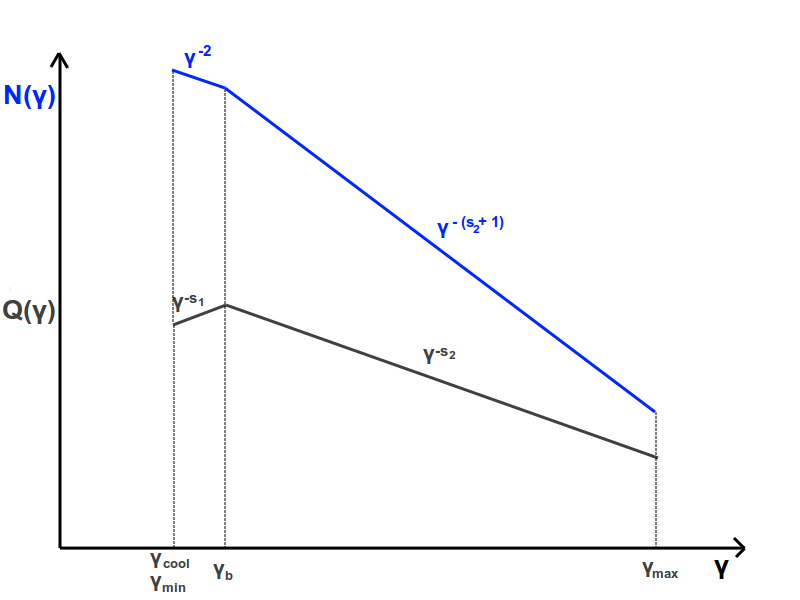}%
	\\
	\includegraphics[width=0.9\columnwidth]{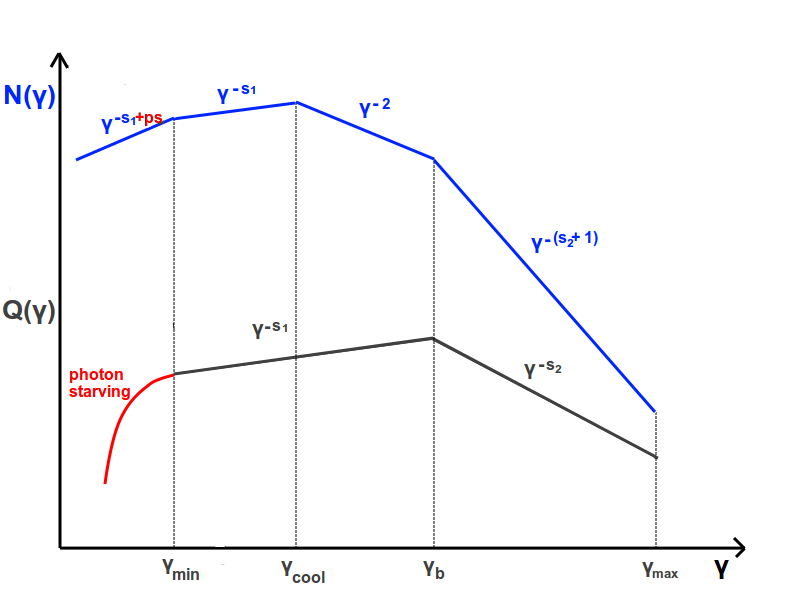}%
	\caption{Slopes of $N(\gamma)$ related to the underlying $Q(\gamma)$. The scenario in the \emph{top} panel is suggested for all blazars in which the observed photon index is significantly greater than $\sim1.5$. It represents a $N(\gamma)$ in which electrons are injected down to a $\gamma_b$ close to the minimum energy, so that we would be observing the emission related to the $N(\gamma)\propto\gamma^{-(s_2+1)}$ region.	The \emph{bottom} panel is suggested to explain all blazars in which the observed photon index is significantly smaller than $\sim1.5$, being related to $N(\gamma)\propto\gamma^{-s_1}$, plus a possible additional hardening due to photon starving. Then, depending on $\gamma_{min}$ and $\gamma_{cool}$ (multiplied by $\Gamma$), spectral breaks can be detected or not within the observing band.}
	\label{fig:electrondistrib}
\end{figure}

Breaks in the electrons distribution can be due to incomplete cooling and/or to photon starving. A break due to incomplete cooling arises if electrons efficiently cool down to a $\gamma_{cool}$ (with $N(\gamma)\propto\gamma^{-2}$), while below this energy electrons reflect the underlying (harder) injection function (see Figure~\ref{fig:electrondistrib}). This break reverberates in the emission spectrum\footnote{In the frame comoving with the emitting blob, all seed photons are observed as coming frontally (if the blob is located inside the BLR or the torus) within a $\sim1/\Gamma$ cone. The radiation is then seen boosted by the relativistic Doppler factor $\delta$, that ranges between $2\Gamma$, for axis-photons at $\theta=0$, and $\Gamma$, for edge-photons at $\theta\sim1/\Gamma$. For several reasons, equally acceptable, a factor $\Gamma$, 1.5$\Gamma$ or 2$\Gamma$ has been used in the literature \citep[see, e.g.,][]{Tavecchio07:RBS315Suzaku-intrinsicbreak,Tavecchio08:BLRspectrum,Ghisellini09:CanonicalBlazars-7Csed,Ghisellini15:blazarsMODELS}. We adopted a factor $\Gamma\nu_{ext}$ because we aimed to provide an upper limit to the product $\gamma_{cool}\Gamma$ in most cases. The scattered photons are then seen Doppler boosted by $\delta$ by the external observer (us). Throughout this section we make the general assumption that $\delta\sim\Gamma$, thus that the viewing angle of the observer is $\sim1/\Gamma$.}:\begin{equation}\label{eq:Ebcool}
E_{b,cool}=\frac{\gamma_{cool}^2\Gamma^2\nu_{ext}}{1+z}
\end{equation}In addition, there could be an energy break due to photon starving, as electrons with the minimum energy will scatter the photon distribution mostly at an energy:\begin{equation}\label{eq:Eb_gammamin}
E_{b,min}=\frac{\gamma_{min}^2\Gamma^2\nu_{ext}}{1+z}
\end{equation}related to the peak of the black-body that approximates the photons energy distribution. Below this energy, since $\gamma_{min}$ is the minimum electron energy and below $\nu_{ext}$ there is a lack of photons, the spectrum becomes harder.

The absence of a break within the observing band can be explained if the expected breaks occur below the observing band, i.e. if $E_b\lesssim0.3\,$keV. We take this values to be conservative, even if a break at $0.4-0.6\,$keV would be likely missed within the absorption features. Alternatively, a break can be shifted above the observing band, i.e. we take as a limit $E_b\gtrsim8-10\,$keV for \emph{XMM-Newton}. Where a break is expected to be shifted depends on a match between observed photon indexes and slopes predicted by emission models.

The BKN+EX scenario was explored for each blazar, since photon indexes are necessary to test our hypothesis, although it confirmed what emerged from the LGP+EX scenario. Only in 4 sources out of 15 (namely QSO B0537-286, RBS 315, QSO J0555+3948 and 4C 71.07) both absorption and intrinsic curvature terms coexisted. From the lack or presence of energy breaks in the spectrum (in addition to the excess absorption component) we obtained limits on $\gamma_{cool}^2\Gamma^2$. When available, bulk Lorentz factors where obtained from the $\beta_{app}$ values of the literature. Formally $\beta_{app}\lesssim\Gamma$, although we assumed $\theta_{v}\sim1/\Gamma$ and $\beta_{app}\sim\Gamma$, unless otherwise stated. 

We proved that emission models hold, short of a condition on the product $\gamma_{cool}^2\Gamma^2$, varying from source to source, even from different observations of the same object. In general, since in the torus case the typical frequency is a factor $\sim10^{-2}$ with respect to the BLR case, the torus likely represents a safer explanation if breaks are shifted below the observing band, since it requires loose conditions on $\gamma_{cool}^2\Gamma^2$. Nonetheless, beware that $\gamma_{cool}$ would be probably significantly different between the two scenarios, since the jet conditions change. Future VLBA/VLBI studies will be able to provide more accurate bulk Lorentz factors and this will possibly rule out one of the emission models for some specific sources, or our explanation.

Furthermore, this scenario is consistent with the absence of intrinsic breaks in our low-$z$ FSRQs. For most high-$z$ blazars, we proved that intrinsic breaks can easily occur below $\sim0.3\,$keV. For a high-$z$ source at $z\sim2$ ($z\sim5$) this is equivalent in shifting the break below $\sim0.9\,$keV ($\sim1.8\,$keV) in a local FSRQ, where they could be in principle observed. Nonetheless, all low-$z$ blazars show concave spectra, in accord with the presence of a SSC component (or of the upturn from synchrotron to IC emission, see Section~\ref{sec:lowzanalysis}) at low energies, observed up to $1-3\,$keV. Hence, any spectral break around $0.3-1.8\,$keV would be covered. This is in agreement with observing high-energy $\Gamma$ compatibles with, or softer than, 1.5 in our low-$z$ blazars (see Table~\ref{tab:lowz_XMM}), indicating that we are indeed observing the emission above the covered break.

\subsection{7C 1428+4218 ($z=4.715$)}
Blazar 7C 1428+4218 was consistent with a PL+EX model, thus the absence of a break in the broadband $0.2-79\,$keV has to be explained.

Fitted photon indexes are:
\begin{equation*}\Gamma_{obs}=\begin{cases}
		\Gamma_{XMM03}=1.72\pm0.03 \\
		\Gamma_{XMM05}=1.53\pm0.03 \\
		\Gamma_{NU14}=1.54\pm0.07
	\end{cases}\end{equation*}

XMM05 and NU14 show consistency with a complete cooling and a $N(\gamma)\propto\gamma^{-2}$ from $\gamma_{cool}$ up to $\gamma_{b}$. This corresponds to $F(\nu)\propto\nu^{-0.5}$, hence to photon indexes $\sim1.5$. XMM03 is apparently in disagreement, but it can be explained with an expedient. In particular, a varying electrons injection among different observations may have played the required role. In XMM03 a $\Gamma>1.5$ requires the electrons to be injected down to a $\gamma_b\approx\gamma_{cool}\sim\gamma_{min}$. In this case the observed index would be related to the slope above the break in the electrons distribution. Figure~\ref{fig:electrondistrib} (top panel) show this case with simplicity. Being $s_2$ the slope of $Q(\gamma)$ above $\gamma_b$, the slope of $N(\gamma)$ is $n_2=s_2+1$. This slope should be $>2$ to be steeper than the typical cooling slope, but also $<3$ to produce the high-energy peak of the IC hump (in $\nu F_{\nu}$). This states that the above-mentioned explanation holds as long as photon indexes between $1.5-2$ are observed. In XMM03, this requirement is fulfilled.

The fitted photon indexes are compatible with an energy break occurring below the observing band. Hence, from $E_b\lesssim0.3\,$keV we inferred that:
\begin{equation*}
\gamma_{cool}\Gamma<\begin{cases}
13 & \text{BLR} \\
73 & \text{torus}
\end{cases}
\end{equation*}

\citet{Veres10:VLBI7C} studied jet properties of 7C 1428+4218 from the brightness temperature measured with VLBI. The authors actually inferred $\delta$ and provided $\Gamma$ assuming a viewing angle of $\sim3\deg$, reported from a SED fit performed by \citet{Celotti07:Bulkcompton}. We decided to neglect this fitted viewing angle and to approximate $\delta\sim\Gamma$, varying in the range $8.6-12.0$. According to our inferred limits, the BLR emission model holds provided that $\gamma_{cool}\sim\gamma_{min}\sim1$, but the torus is probably a safer bet, since it allows a considerable margin in the estimate of the product $\gamma_{cool}\Gamma$. Hence, unless otherwise stated, throughout this section we will focus on the compatibility of our proposed scenario with the BLR emission model.

\subsection{QSO J0525-3343 ($z=4.413$)}

The reference model for QSO J0525-3343 is PL+EX. No spectral break is then required within the observed \emph{XMM-Newton} band. 

Fitted photon indexes are:
\begin{equation*}\Gamma_{obs}=\begin{cases}
\Gamma_{Obs324}=1.59\pm0.03 \\
\Gamma_{Obs583}=1.62\pm0.03 \\
\Gamma_{Obs588}=1.60\pm0.03
\end{cases}\end{equation*}

They are all compatibles, within the errors, and slightly higher than the expected photon index in case of a complete cooling. $N(\gamma)$ is then required to be similar to the scenario outlined for blazar 7C 1428+4218 (see Figure~\ref{fig:electrondistrib}), namely the electrons have to be injected down to a $\gamma_b$ close to $\gamma_{cool}$ and to the minimum of the distribution. We inferred:

\begin{equation*}
\gamma_{cool}\Gamma<\begin{cases}
13 & \text{BLR} \\
71 & \text{torus}
\end{cases}
\end{equation*}

Future VLBI studies involving QSO J0525-3343 will help to validate or exclude the BLR emission model.

\subsection{QSO B1026-084 ($z=4.276$)}

The \emph{XMM-Newton} X-ray spectrum of blazar QSO B1026-084 is consistent with a PL+EX.

The fitted photon index is $\Gamma=1.46\pm0.04$, hence it is compatible, within the errors, with the slope expected for a complete cooling of the emitting electrons, i.e. $\sim1.5$. Consequently, the energy break due to incomplete cooling occurred below the observing band if:

\begin{equation*}
\gamma_{cool}\Gamma<\begin{cases}
13 & \text{BLR} \\
70 & \text{torus}
\end{cases}
\end{equation*}
No bulk Lorentz factor is available in the literature.

\subsection{QSO B0014+810 ($z=3.366$)}

The broadband $0.3-79\,$keV X-ray spectrum of QSO B0014+810 is, similarly to 7C 1428+4218, consistent with a PL+EX model. The fitted photon indexes are:
\begin{equation*}\Gamma_{obs}=\begin{cases}
		\Gamma_{XMM01}=1.50^{+0.02}_{-0.03} \\
		\Gamma_{Nu14}=1.72^{+0.04}_{-0.02} \\
		\Gamma_{Nu15}=1.61\pm0.04
	\end{cases}\end{equation*}
	
In \emph{XMM-Newton} the photon index is compatible with a complete cooling of the electrons population, whilst the two \emph{Swift-XRT}+\emph{NuSTAR} observations show instead steeper photon indexes. They can be easily explained with a scenario analogous to 7C 1428+4218, in which the electrons have to be injected down to a $\gamma_b$ close to the minimum of the electrons energy distribution (see Figure~\ref{fig:electrondistrib}). All observations suggest that energy breaks occurred below the observing band, thus we can provide upper limits on the product $\gamma_{cool}\Gamma$ from equation~\ref{eq:Ebcool}:

\begin{equation*}
\gamma_{cool}\Gamma<\begin{cases}
11 & \text{BLR} \\
64 & \text{torus}
\end{cases}
\end{equation*}

Apparent velocities were obtained for several jet components of blazar QSO B0014+810 by \citet{Britzen08:VLBI_B0014}. Approximating $\Gamma\sim\beta_{app}$ and conservatively taking the jet component with the higher apparent velocity, namely $\beta_{app}\sim12.0\pm1.9$, the BLR model seems to be barely validated, even with $\gamma_{cool}\sim1$. The torus case is likely to be a safer bet for QSO B0014+810.

\subsection{PKS 2126-158 ($z=3.268$)}

The reference model for the \emph{XMM-Newton} spectrum of blazar PKS 2126-158 is a PL+EX scenario.

The fitted photon index is $\Gamma=1.45\pm0.02$. Although it is not precisely compatible with the $\sim1.5$ index expected for a complete cooling of the electron, it is reasonably close to it. Then, blazars' emission models hold, provided that energy breaks occurred below the observing band. Thus, the product $\gamma_{cool}\Gamma$ should be smaller than 11 and 63, for the BLR and torus case, respectively.

No bulk Lorentz factor was available in the literature.

\subsection{QSO B0537-286 ($z=3.104$)}

The two \emph{XMM-Newton} observations showed different continua, but were fitted with a common excess column density. The spectrum in Obs00 is concave, while Obs05 is consistent with a power-law continuum. Nonetheless, both observations show a rather hard photon index of $\Gamma\sim1.18$ ($\Gamma_{high}$ for Obs00). A similar case was reported in \citet{Ajello16:PMNJ0641} and also the herewith proposed explanation is similar. The hard photon index can be explained with a moderate value of $\gamma_{cool}$, that would shift the energy break above the observed \emph{XMM-Newton} band. This break can be due to inefficient cooling alone, in which case the observed slope in $F(\nu)$ should be related to the underlying electrons distribution, that goes as $s_1$. Alternatively, the break can be due to a combined effect of inefficient cooling plus photon-starving, in which case an even harder photon index is expected. See Figure~\ref{fig:electrondistrib} for more details on the behaviour of $Q(\gamma)$ and $N(\gamma)$ in a similar case.

The concave continuum in Obs00 can be explained invoking an underlying component in the SED ($\nu F_{\nu}$), seen thanks to the very hard X-ray spectrum. The presence of this underlying component in one of the two \emph{XMM-Newton} observations could be due to a different state of the source, e.g. see Figure 4 of \citet{Bottacini10:0537broadband}. In any case, both observations are consistent with an energy break occurring above the observing band, thus we can provide lower limits on the product $\gamma_{cool}\Gamma$ imposing the break to be above $8-10\,$keV:

\begin{equation*}
\gamma_{cool}\Gamma>\begin{cases}
57-64 & \text{BLR} \\
320-358 & \text{torus}
\end{cases}
\end{equation*}

Hence, for QSO B0537-286 the BLR case represents the safer bet, provided $\gamma_{cool}\sim\text{a few}$.


\subsection{QSO B0438-43 ($z=2.852$)}

The \emph{XMM-Newton} spectrum of blazar QSO B0438-43 is consistent with a PL+EX.

The fitted photon index is $\Gamma=1.71\pm0.03$. It is steeper than the $\sim1.5$ index expected for a complete cooling of the electrons, hence the explanation is analogous to blazar 7C 1428+4218 (see Figure~\ref{fig:electrondistrib}). Energy breaks may occur below the observing band, provided that the product $\gamma_{cool}\Gamma$ is smaller than 11 and 60, for the BLR and torus case, respectively.

No bulk Lorentz factor was available in the literature.

\subsection{RBS 315 ($z=2.69$)}

The reference model for both the $0.2-10\,$keV and the $0.2-79\,$keV spectrum of blazar RBS 315 is provided by the LGP+EX scenario. Hence, we performed also the $0.2-79\,$keV fit with the BKN+EX model, in order to provide photon indexes and to distinguish the energy breaks involved. In Table~\ref{tab:RBS315_BKNEX} results are shown for the three \emph{XMM-Newton} archive observations and the two \emph{Swift-XRT}+\emph{NuSTAR} spectra, respectively. The Galactic value was left free to vary between the $\pm15\%$ boundaries of the tabulated value \citep{Willingale13:GalacticH2}. The fitted lower limit ($>17.6\times10^{20}\,\text{cm}^{-2}$) has also the upper $+15\%$ bound at $18.7\times10^{20}\,\text{cm}^{-2}$. An excess column density of $0.56^{+0.27}_{-0.22}\times10^{22}\,\text{cm}^{-2}$ was fitted, compatible, within the errors, with LGP+EX results.

\begin{table}[ht]
	\small
	\renewcommand{\arraystretch}{1.2}
	\caption{Broadband $0.2-79\,$keV spectral fit for blazar RBS 315, using the BKN+EX model.}
	\label{tab:RBS315_BKNEX}
	\centering
		\begin{tabular}{ccccc}
			\toprule
			\multicolumn{1}{c}{Obs.}& 
			\multicolumn{1}{c}{$\Gamma_{low}$}& 
			\multicolumn{1}{c}{$\Gamma$}& 
			\multicolumn{1}{c}{$E_b$}&
			\multicolumn{1}{c}{$\chi^2_{\nu}/\nu$}\\
			&
			&
			& 
			(keV)& 
			\\
			\midrule
			$\text{XMM}2003$ & $1.05^{+0.09}_{-0.10}$ & $1.25\pm0.02$ & $1.38^{+0.35}_{-0.24}$  & $1.03/2714$\\
			$\text{XMM}2013a$& $1.21^{+0.09}_{-0.08}$ & $1.46^{+0.02}_{-0.01}$ & $1.32^{+0.18}_{-0.12}$  & \\
			$\text{XMM}2013b$& $1.20^{+0.15}_{-0.09}$ & $1.43^{+0.05}_{-0.02}$ & $1.28^{+1.00}_{-0.14}$  & \\
			$\text{Nu}2014$	& $1.31^{+0.10}_{-0.15}$ & $1.59^{+0.04}_{-0.06}$ & $6.76^{+1.09}_{-2.64}$ & \\
			$\text{Nu}2015$	& $1.12^{+0.15}_{-0.13}$ & $1.69^{+0.06}_{-0.04}$ & $4.67^{+1.17}_{-0.66}$ & \\
			\bottomrule
		\end{tabular}
\end{table}

\emph{XMM-Newton} Obs2003 shows a hard spectrum down to a spectral break at $\sim1.4\,$keV, below which it becomes even harder. An excess column density was already considered in the BKN+EX fit, but it was not enough. The observed hardening has to be modelled with an additional spectral break, produced by the minimum Lorentz factor of the electrons energy distribution (see Equation~\ref{eq:Eb_gammamin}). Moreover, the energy break for an incomplete cooling has to occur above the observed \emph{XMM-Newton} band, because the high-energy photon index is hard ($\Gamma\sim1.25$). The explanation is similar to blazar QSO B0537-286, with the addition, within the observing band, of a break due to photon starving that hardens even more ($\Gamma\sim1.05$) the soft X-ray spectrum below $E_{b,min}$. In Figure~\ref{fig:electrondistrib} (bottom panel) we display $N(\gamma)$ with the underlying injection function required to produce the observed X-ray spectrum in Obs2003. The conditions of the BLR emission model are:
\begin{equation*}
\begin{cases}
\gamma_{min}\Gamma=23^{+3}_{-2} & \text{    for   } E_{b,min}=1.38^{+0.35}_{-0.24}\,\text{keV} \\
\gamma_{cool}\Gamma>54-61& \text{    for   } E_{b,cool}>8-10\,\text{keV}
\end{cases}
\end{equation*}that can be adapted to the torus case simply changing the external frequency at which the seed photons distribution peaks.

\emph{XMM-Newton} Obs2013a and b can be treated similarly. They both show an energy break at $\sim1.3\,$keV, above which the photon index is slightly harder than $1.5$, but can be easily considered consistent with a complete cooling scenario. Referring to Figure~\ref{fig:electrondistrib}, this $\sim1.5$ slope would be related to the $N(\gamma)\propto\gamma^{-2}$ region, while below the break the hardening can be adequately accounted for with the slope of the underlying injection function. Hence, $\gamma_{min}$ can be considered $\sim1$. A constraint on $\gamma_{cool}\Gamma$ can be obtained for Obs2013a:
\begin{equation*}
\gamma_{cool}\Gamma=22^{+2}_{-1}\,\,\,\,\,\,\,\,\text{for}\,\,\,\,\,\,\,\,E_{b,cool}=1.32^{+0.18}_{-0.12}\,\text{keV}
\end{equation*}and for Obs2013b: \begin{equation*}
\gamma_{cool}\Gamma=22^{+8}_{-1}\,\,\,\,\,\,\,\,\text{for}\,\,\,\,\,\,\,\,E_{b,cool}=1.28^{+1.00}_{-0.14}\,\text{keV}
\end{equation*}

\emph{Swift-XRT}+\emph{NuSTAR} data show a break around 5 and 6 keV, in Obs2014 and 2015 respectively. Below the break, they both show a hard photon index, while above they are softer than $\sim1.5$ (see Table~\ref{tab:RBS315_BKNEX}). We already highlighted for blazar 7C 1428+4218 that a soft high-energy photon index can be explained with a particular injection function. Comparing to Figure~\ref{fig:electrondistrib}, this would be obtained with $\gamma_{cool}\sim\gamma_b\sim\text{a few}$:\begin{equation*}
\gamma_{cool}\Gamma=47^{+4}_{-10}\,\,\,\,\,\,\,\,\text{for}\,\,\,\,\,\,\,\,E_{b,cool}=6.76^{+1.09}_{-2.64}\,\text{keV}
\end{equation*}for Obs2014, while for Obs2015: \begin{equation*}
\gamma_{cool}\Gamma=41^{+5}_{-3}\,\,\,\,\,\,\,\,\text{for}\,\,\,\,\,\,\,\,E_{b,cool}=4.67^{+1.17}_{-0.66}\,\text{keV}
\end{equation*}

No bulk Lorentz factor was found in the literature for RBS 315.
\subsection{QSO J2354-1513 ($z=2.675$)}

The \emph{XMM-Newton} spectrum of blazar QSO J2354-1513 was adequately fitted with a PL+EX.

The fitted photon index is $\Gamma=1.62^{+0.02}_{-0.03}$. It is slightly steeper than the $1.5$ index expected for a complete cooling of the electrons, hence the explanation is analogous to blazar 7C 1428+4218 (see Figure~\ref{fig:electrondistrib}). Being the \emph{XMM-Newton} spectrum consistent with a power-law continuum, the product $\gamma_{cool}\Gamma$ is required to be smaller than 11 and 59, for the BLR and torus case, respectively. No bulk Lorentz factor was available in the literature.

\subsection{PBC J1656.2-3303 ($z=2.4$)}

The \emph{XMM-Newton} observation of blazar PBC J1656.2-3303 showed a simple power-law continuum, with a soft X-ray hardening adequately explained by an excess column density. The observed photon index is extremely hard, being $\Gamma=1.21\pm0.02$. This spectrum can be explained similarly to QSO B0537-286. A moderate value of $\gamma_{cool}$ would shift the energy break due to incomplete cooling above the observed \emph{XMM-Newton} band. Thus:\begin{equation*}
\gamma_{cool}\Gamma>\begin{cases}
52-58 & \text{BLR} \\
292-326 & \text{torus}
\end{cases}
\end{equation*}No bulk Lorentz factor was obtained in the literature for blazar PBC J1656.2-3303.

In the broadband $0.2-79\,$keV fit \emph{Swift-XRT}+\emph{NuSTAR} data showed some additional curvature in the LGP+EX model, along with a column density in excess of the Galactic value. We then fitted the broadband spectrum with a BKN+EX model, although \emph{XMM-Newton} data were constrained to a power-law continuum. We obtained:\begin{equation*}
\begin{cases}
\Gamma_{low}=0.63^{+0.32}_{-0.35}\\
E_b=2.26^{+0.70}_{-0.37}\\
\Gamma_{high}=1.61\pm 0.05
\end{cases}
\end{equation*}The high-energy photon index is softer than 1.5, while the low-energy slope is extremely hard, to the point of requiring both an incomplete cooling and photon starving. Hence, the fitted energy break should be produced by $\gamma_{min}\sim\gamma_{cool}\sim\gamma_b\sim\text{a few}$ (re-adapt Figure~\ref{fig:electrondistrib} to picture this scenario):
\begin{equation*}
\gamma_{min}\Gamma\sim\gamma_{cool}\Gamma=28^{+4}_{-2}
\end{equation*}

\subsection{QSO J0555+3948 ($z=2.363$)}

QSO J0555+3948 is one of the 4 sources in which a curvature term was fitted in the LGP+EX scenario, in addition to an excess column density. The BKN+EX fit was performed to obtain a value of the energy break and the related photon indexes, obtaining:\begin{equation*}
\begin{cases}
\Gamma_{low}=1.66^{+0.08}_{-0.04}\\
E_b=3.42^{+0.95}_{-0.90}\\
\Gamma_{high}=1.35^{+0.13}_{-0.24}
\end{cases}
\end{equation*}

The \emph{XMM-Newton} spectrum is concave, although the explanation seems to be totally different with respect to Obs00 of blazar QSO B0537-286, in which the break was from a very hard to a hard spectrum. In QSO J0555+3948 the low-energy index is soft, while the high-energy slope is slightly harder than 1.5. This concavity can be easily explained with a SSC component, provided the bulk Lorentz factor is low, otherwise a high-$z$ FSRQ would typically show a naked IC component. The source shows $\Gamma=1.6\pm0.1$ \citep{Homan15:MOJAVE}, thus the SSC scenario is a favourable explanation. 

Still, the high-energy photon index has to be accounted for. In principle, it should be considered as a harder-than-1.5 case, hence it would require the energy break due to $\gamma_{cool}$ to be shifted above the observed \emph{XMM-Newton} band. Nonetheless, we mentioned that the bulk Lorentz factor is low, then this would imply a $\gamma_{cool}\approx30$ (following the usual upper/lower limits computation). However, note that an intervening SSC component would likely produce a smooth transition and the fitted high-energy photon index is compatible, within the errors, with a $\Gamma=1.48$ slope. Thus, we do not deem necessary for this source to invoke a shift of the energy break due to incomplete cooling. The SSC component is likely covering any intrinsic spectral break.

\subsection{PKS 2149-306 ($z=2.345$)}

The \emph{XMM-Newton} X-ray spectrum of blazar PKS 2149-306 is consistent with a PL with marginal excess absorption.

The fitted photon index is $\Gamma=1.46\pm0.01$, hence it can be considered consistent with a complete cooling of the emitting electrons. Consequently, the BLR scenario holds if the energy break due to incomplete cooling occurred below the observing band, i.e. if:\begin{equation*}
\gamma_{cool}\Gamma<\begin{cases}
10 & \text{BLR} \\
56 & \text{torus}
\end{cases}
\end{equation*}

In the broadband fit \emph{Swift-XRT}+\emph{NuSTAR} data required, instead, some additional curvature in the LGP+EX model. We then tested the BKN+EX model, with \emph{XMM-Newton} data constrained to a power-law continuum, obtaining:\begin{equation*}
\begin{cases}
\Gamma_{low}=0.93^{+0.09}_{-0.12}\\
E_b=2.48^{+0.65}_{-0.28}\\
\Gamma_{high}=1.37\pm 0.01
\end{cases}
\end{equation*} for Obs2013. In the same fit, Obs2014 was fitted by:\begin{equation*}
\begin{cases}
\Gamma_{low}=1.00^{+0.07}_{-0.08}\\
E_b=3.17^{+0.56}_{-0.55}\\
\Gamma_{high}=1.46\pm 0.01
\end{cases}
\end{equation*}

In Obs2013 the high-energy photon index is harder than 1.5, while the lower-energy slope is extremely hard. Hence this observation can be explained with the observed break being produced by the $\gamma_{min}$ of the electrons energy distribution, while the hard photon index above the break can be accounted for if the break due to incomplete cooling is shifted above the observing band. Thus, from this observation we inferred the following conditions:\begin{equation*}
\begin{cases}
\gamma_{min}\Gamma=29^{+4}_{-2} & \text{    for    } E_{b,min}=3.17^{+0.56}_{-0.55}\,\text{keV} \\
\gamma_{cool}\Gamma>142 & \text{    for   } E_{b,cool}\gtrsim 60\,\text{keV}
\end{cases}
\end{equation*}

The second condition is quite severe for the BLR model and the torus case would provide an ever higher limit\footnote{As for the 8\,keV limit for \emph{XMM-Newton} spectra, we arbitrarily (and empirically) adopted a $\sim60$\,keV limit for \emph{NuSTAR}.}. No bulk Lorentz factor is available in the literature for PKS 2149-306, hence the disproof for the proposed scenario is up to future VLBI observations. However, note that in this fit the break was uniquely fitted by \emph{Swift-XRT} (being below 3\,keV) and we already stressed that it probably yield unreliable results compared to \emph{XMM-Newton} (see the discussion in Section~\ref{sec:comparison_analysis}). Hence, whenever a break below 3\,keV is fitted with \emph{Swift-XRT}+\emph{NuSTAR} data, results are to be taken with caution.

In Obs2014 the situation is similar, except that the high-energy photon index is consistent with a complete cooling scenario, thus the only condition that has to be respected is on the observed energy break. Given the extremely hard low-energy photon index, this break should be due to both an incomplete cooling and photon starving, thus $\gamma_{min}\Gamma\sim\gamma_{cool}\Gamma=33\pm3$.

\subsection{QSO B0237-2322 ($z=2.225$)}

The \emph{XMM-Newton} spectrum of blazar QSO B0237-2322 is consistent with a PL with a marginal excess absorption component.

The fitted photon index is $\Gamma=1.73\pm0.03$. It is steeper than the $\sim1.5$ index expected for a complete cooling of the electrons, hence the explanation is similar, e.g., to blazar 7C 1428+4218 (see Figure~\ref{fig:electrondistrib}). Hence, the product $\gamma_{cool}\Gamma$ is expected to be smaller than 10 and 55, for the BLR and torus case, respectively. No bulk Lorentz factor was available in the literature.

\subsection{4C 71.07 ($z=2.172$)}

In blazar 4C 71.07, both the $0.2-10\,$keV and the $0.2-79\,$keV spectrum were better fitted with a LGP+EX model, with marginal evidence of an excess column density. Hence, we performed also a broadband fit with the BKN+EX model, in order to provide photon indexes and energy break values. In Table~\ref{tab:4C_BKNEX} we show the results for the \emph{XMM-Newton} observation (first row) and the two \emph{Swift-XRT}+\emph{NuSTAR} spectra (second and third row). The Galactic value was left free to vary between the $\pm15\%$ boundaries of the tabulated value \citep{Willingale13:GalacticH2}. The fitted lower limit ($>2.85\times10^{20}\,\text{cm}^{-2}$) has also the upper $+15\%$ bound at $3.63\times10^{20}\,\text{cm}^{-2}$. A marginal excess column was fitted ($<0.07\times10^{22}\,\text{cm}^{-2}$), compatibly with the LGP+EX case.

\begin{table}[ht]
	\small
	\renewcommand{\arraystretch}{1.3}
	\caption{Broadband $0.2-79\,$keV spectral fit for blazar 4C 71.07, using the BKN+EX model.}
	\label{tab:4C_BKNEX}
	\centering
		\begin{tabular}{ccccc}
			\toprule
			\multicolumn{1}{c}{Obs.}& 
			\multicolumn{1}{c}{$\Gamma_{low}$}& 
			\multicolumn{1}{c}{$\Gamma$}& 
			\multicolumn{1}{c}{$E_b$}&
			\multicolumn{1}{c}{$\chi^2_{\nu}/\nu$}\\
			&
			&
			& 
			(keV)& 
			\\
			\midrule
			$\text{XMM}$	& $1.30\pm0.03$ & $1.38^{+0.02}_{-0.01}$ & $1.62^{+0.51}_{-0.23}$  & $1.04/2115$\\
			$\text{Nu}2013$	& $1.12^{+0.11}_{-0.06}$ & $1.68\pm0.02$ & $2.14\pm0.39$ & \\
			$\text{Nu}2014$	& $1.22^{+0.07}_{-0.08}$ & $1.65^{+0.01}_{-0.02}$ & $3.69^{+0.60}_{-1.56}$ & \\
			\bottomrule
		\end{tabular}
\end{table}

The \emph{XMM-Newton} observation shows a hard spectrum with a mild break (the two photon indexes are slightly non-compatibles) at $\sim1.6\,$keV, that can be then attributed to the minimum Lorentz factor of the electrons energy distribution (equation~\ref{eq:Eb_gammamin}, provided that the following condition is satisfied:\begin{equation*}
\gamma_{min}\Gamma=23^{+4}_{-2}\text{\,\,\,\,\,\,\,\,\,\,\,for\,\,\,\,\,\,\,\,\,\,} E_{b,min}=1.62^{+0.51}_{-0.23}\,\text{keV}
\end{equation*}

A series of bulk Lorentz factors was obtained for different emitting regions of 4C 71.07 \citep{Homan15:MOJAVE}, ranging from $13\pm1$ to $20\pm1$. This is consistent with the averaged value of $\Gamma=17.0\pm2.2$ reported by \citet{Jorstad17:VLBAblaz}. Hence, the above condition is fulfilled with a minimum energy of the electrons distribution of $\sim1-2$.

Moreover, given the harder-than-1.5 spectrum observed, the energy break for an incomplete cooling has to be shifted above $\sim8-10\,$keV, thus providing the condition:\begin{equation*}
\gamma_{cool}\Gamma>50-56 \text{\,\,\,\,\,\,\,\,\,\,for\,\,\,\,\,\,\,\,\,\,} E_{b,cool}>8-10\,\text{keV}
\end{equation*}

With the constraints on $\Gamma$ provided by \citet{Homan15:MOJAVE}, a $\gamma_{cool}\sim3-4$ is required by \emph{XMM-Newton} data.

The two \emph{Swift-XRT}+\emph{NuSTAR} observations were fitted with an energy break around 2 and 4\,keV, in Obs2013 and 2014 respectively. Below the break, a hard photon index is observed, while above it is steeper than $\sim1.5$. This results can be adequately explained with $\gamma_{cool}\sim\gamma_b$ constrained by the observed breaks:\begin{equation*}\begin{cases}
\gamma_{cool}\Gamma=26\pm 2 & \text{for Obs2013} \\
\gamma_{cool}\Gamma=34^{+3}_{-7} & \text{for Obs2014}\end{cases}
\end{equation*}

Furthermore, given the bulk Lorentz factors reported in the literature, the BLR emission model can be considered consistent provided $\gamma_{cool}\sim1-2$. In case the addition of photon starving is required by the hard low-energy photon indexes, the same condition would be extended also to $\gamma_{min}$.

\subsection{PKS 0528+134 ($z=2.07$)}

The reference model for PKS 0528+134 is a simple PL+EX. No spectral break is then observed within the \emph{XMM-Newton} band. The fitted photon indexes are:
\begin{equation*}\Gamma_{obs}=\begin{cases}
\Gamma_{Obs501}=1.56\pm0.08 \\
\Gamma_{Obs601}=1.55\pm0.04 \\
\Gamma_{Obs701}=1.56\pm0.05
\end{cases}\end{equation*}

They are all compatibles, within the errors, and consistent with a complete cooling. Hence, we can provide a condition on $\gamma_{min}\sim\gamma_{cool}$, that are required to be shifted below the \emph{XMM-Newton} band:\begin{equation*}
\gamma_{cool}\Gamma<\begin{cases}
9.6 & \text{BLR} \\
54 & \text{torus}
\end{cases}
\end{equation*}

\citet{Homan15:MOJAVE} reported two measures of $\Gamma$, i.e. $9.4\pm0.5$ and $17.3\pm0.5$, and \citet{Jorstad17:VLBAblaz} recently reported a compatible averaged $\Gamma=12.6\pm3.1$. Hence, the torus case has to be considered a safer bet for PKS 0528+134.
\end{appendix}

\end{document}

%% file: Table.tex
\begin{landscape}
\tiny
\renewcommand{\arraystretch}{1.3}
\begin{ThreePartTable}
		\begin{TableNotes}\footnotesize
		\item[a] The fit is insensitive to the Galactic value between the $\pm15\%$ boundaries.
		\item[b] The Galactic value was left free to vary between the $\pm15\%$ boundaries of the tabulated value \citep{Willingale13:GalacticH2}. The fitted upper/lower limit has also the lower/upper $15\%$ bound.
		\item[c] The excess column density can be better constrained to $0.48^{+0.11}_{-0.21}\times10^{22}\,\text{cm}^{-2}$ fixing $b_{obs05}=0$ in the LGP+EX fit.
		\item[d] An error calculation could not be computed due to the high reduced chi-square value.
		\item[e] This LGP+EX fit was performed fixing $b_{XMM}=0$, since the reference model for \emph{XMM-Newton} data is PL+EX.
		\item[f] These errors were computed at 1-sigma confidence level within \texttt{XSPEC}. 
		\item[g] Fit performed with RGS data added to the analysis, to improve the accuracy of the excess column. The F-test reported, if any, is computed without these data. 
		\end{TableNotes}
\begin{longtable}{cccccccccccccc}
\caption{Spectral fit results. \emph{XMM-Newton} data were analysed in the $0.2-10\,$keV energy range for the EPIC-pn detector and $0.3-10\,$keV for EPIC-MOS detectors. For blazars with also \emph{NuSTAR} observation(s) additional rows were added for the broadband $0.2-79\,$keV fits. The first two columns show the source and the specific observation, if more than one (see Appendix~\ref{sec:appA} for abbreviations), which is highlighted in bold if referred to \emph{Swift-XRT}+\emph{NuSTAR} data. The third column shows the models used (PL=simple power-law with Galactic absorption; BKN=broken power-law with Galactic absorption; LGP=log-parabola with Galactic absorption; EX=cold excess absorption fixed at the source's redshift). The reference model for each blazar is highlighted in italics and labelled with an asterisk, comments in the text. In the fourth column, the Galactic value was either fixed \citep{Willingale13:GalacticH2} or free to vary between a $\pm15\%$ of the fixed value (when errors are reported). When fitted, the excess column density (in unity of $10^{22}\,\text{cm}^{-2}$) is reported in the fifth column. Columns 6, 7 and 8 represent, in the BKN model, the low- and high-energy photon index and the energy break, respectively. The seventh column shows also the photon index in the simple PL scenario. The ninth column, in the LGP model, represents the slope at the pivot energy, fixed at 1\,keV in \emph{XMM-Newton} spectra and at 5\,keV for \emph{NuSTAR} spectra. The tenth column show the curvature term in the LGP scenario. In columns 11 and 12,	$C_1$ and $C_2$ are the two floating constants representing the cross-normalization parameter among the different cameras. They are referred to MOS1 and MOS2, respectively, in \emph{XMM-Newton} fits and to FPMB and \emph{Swift-XRT}, respectively, in the broadband fit. The last column shows the F-test p-value computed with respect to a specific model, reported aside. Errors and upper limits were computed at $90\%$ confidence level ($\Delta\chi^2=2.706$) within \texttt{XSPEC}, unless otherwise stated.}
\label{tab:spectral_analysis} \\
\toprule
\multicolumn{1}{c}{Source}& 
\multicolumn{1}{c}{Obs. ID}& 
\multicolumn{1}{c}{Model}& 
\multicolumn{1}{c}{$N_H^{Gal}$}& 
\multicolumn{1}{c}{$N_H(z)$}&
\multicolumn{1}{c}{$\Gamma_{low}$}& 
\multicolumn{1}{c}{$\Gamma$}& 
\multicolumn{1}{c}{$E_b$}&
\multicolumn{1}{c}{$a$} &
\multicolumn{1}{c}{$b$}& 
\multicolumn{1}{c}{$C_1$}& 
\multicolumn{1}{c}{$C_2$}&
\multicolumn{1}{c}{$\chi^2_{\nu}/\nu$}&
\multicolumn{1}{c}{F-test/model}\\
&
&
&
\multicolumn{1}{c}{$(10^{20}\,\text{cm}^{-2})$}&
\multicolumn{1}{c}{$(10^{22}\,\text{cm}^{-2})$}&
&
& 
(keV)&
(@1-5\,keV)&
&
& 
&
& 
\\
\midrule
\endfirsthead
\multicolumn{3}{l}{\footnotesize Continued} \\
\toprule
\multicolumn{1}{c}{Source}&
\multicolumn{1}{c}{Obs. ID}&  
\multicolumn{1}{c}{Model}& 
\multicolumn{1}{c}{$N_H^{Gal}$}& 
\multicolumn{1}{c}{$N_H(z)$}&
\multicolumn{1}{c}{$\Gamma_{low}$}& 
\multicolumn{1}{c}{$\Gamma$}& 
\multicolumn{1}{c}{$E_b$}&
\multicolumn{1}{c}{$a$} &
\multicolumn{1}{c}{$b$}& 
\multicolumn{1}{c}{$C_1$}& 
\multicolumn{1}{c}{$C_2$}&
\multicolumn{1}{c}{$\chi^2_{\nu}/\nu$}&
\multicolumn{1}{c}{F-test/model}\\
&
&
&
\multicolumn{1}{c}{$(10^{20}\,\text{cm}^{-2})$}&
\multicolumn{1}{c}{$(10^{22}\,\text{cm}^{-2})$}&
&
& 
(keV)&
(@1-5\,keV)&
&
& 
&
& 
\\
\midrule
\endhead
\midrule
\endfoot
\bottomrule
\insertTableNotes
\endlastfoot
7C 1428+4218& Obs2003 & PL	& $1.22$ & $\dots$ & $\dots$ & $1.59\pm0.02$ & $\dots$ & $\dots$ & $\dots$ & $1.03$ & $1.10$ & $1.21/503$&\\
			& Obs2005 & &  & $\dots$ & $\dots$ & $1.42\pm0.02$ & $\dots$ & $\dots$ & $\dots$ & $1.05$ & $1.01$ & &\\
			& Obs2003 & BKN	& $1.22$ & $\dots$ & $0.65^{+0.34}_{-0.79}$ & $1.70\pm0.03$ & $0.59\pm0.10$ 		 & $\dots$ & $\dots$ & $1.01$ & $1.07$ & $0.94/499$ & $10^{-29}$/PL\\
			& Obs2005 & &  & $\dots$ & $0.71^{+0.38}_{-0.40}$ & $1.49\pm0.03$ & $0.59^{+0.22}_{-0.07}$ & $\dots$ & $\dots$ & $1.03$ & $0.99$ & &\\
			& Obs2003 & LGP	& $1.22$ & $\dots$ & $\dots$ & $\dots$ & $\dots$ & $1.52\pm0.03$ & $0.28\pm0.06$ & $1.01$ & $1.07$ & $1.05/501$ &$10^{-17}$/PL\\
			& Obs2005 & &  		 & $\dots$ & $\dots$ & $\dots$ & $\dots$ & $1.36\pm0.03$ & $0.17\pm0.06$ & $1.03$ & $0.99$ & &\\
			& Obs2003 & \emph{PL+EX*}	& $1.22_{1.04}^{1.40}$\tnote{a} & $1.90^{+0.28}_{-0.36}$ & $\dots$ & $1.72\pm0.03$ & $\dots$ & $\dots$ & $\dots$ & $1.00$ & $1.07$ & $0.94/501$& \\
			& Obs2005 & &  					   & 						& $\dots$ & $1.53\pm0.03$ & $\dots$ & $\dots$ & $\dots$ & $1.03$ & $0.99$ & &\\ 
			& Obs2003 & LGP+EX	& $1.22_{1.04}^{1.40}$\tnote{a} & $2.28^{+0.29}_{-0.38}$ & $\dots$ & $\dots$ & $\dots$ & $1.76\pm0.06$ & $-0.05^{+0.08}_{-0.05}$ & $1.00$ & $1.07$ & $0.93/499$ & 0.14/PL+EX\\
			& Obs2005 & &  		 			   & 						  & $\dots$ & $\dots$ & $\dots$ & $1.59\pm0.06$ & $-0.1^{+0.07}_{-0.05}$ & $1.03$ & $0.99$ & &\\
\cmidrule{2-14}
			& Obs2005 &PL	&  	$1.22$	 & $\dots$ & $\dots$ & $1.53\pm0.07$ & $\dots$ & $\dots$ & $\dots$ & $1.05$ & $1.01$ &  $1.14/385$&\\
			& \textbf{Obs2014} &		&  		& $\dots$ & $\dots$ & $1.42\pm0.02$ & $\dots$ & $\dots$ & $\dots$ & $1.03$ & $0.80$ & &\\
			& Obs2005 &BKN	&  	$1.22$	 & $\dots$ & $0.70^{+0.38}_{-0.40}$ & $1.49\pm0.03$ & $0.59^{+0.22}_{-0.07}$ & $\dots$ & $\dots$ & $1.03$ & $0.99$ & $1.01/381$& $10^{-9}$/PL\\
			& \textbf{Obs2014} & &  		 & $\dots$ & $1.38\pm0.10$ & $2.02^{+1.36}_{-0.43}$ & $11.84^{+8.13}_{-7.19}$ & $\dots$ & $\dots$ & $1.03$ & $1.01$ & &\\
			& Obs2005 &LGP	&  	$1.22$	 & $\dots$ & $\dots$ & $\dots$ & $\dots$ & $1.36\pm0.03$ & $0.17\pm0.06$ & $1.03$ & $0.99$ & $1.05/383$ &$10^{-8}$/PL\\
			& \textbf{Obs2014} & &  		 & $\dots$ & $\dots$ & $\dots$ & $\dots$ & $1.46\pm0.08$ & $0.25\pm0.13$ & $1.03$ & $1.08$ & &\\
			& Obs2005 &\emph{PL+EX*}	& $1.22_{1.04}^{1.40}$\tnote{a} & $1.58^{+0.41}_{-0.25}$ & $\dots$ & $1.51\pm0.03$ & $\dots$ & $\dots$ & $\dots$ & $1.03$ & $0.99$ & $1.03/383$& \\
			& \textbf{Obs2014} & &  					   & 						& $\dots$ & $1.54\pm0.07$ & $\dots$ & $\dots$ & $\dots$ & $1.03$ & $0.82$ & &\\
			& Obs2005 &LGP+EX	& $1.22_{1.04}^{1.40}$\tnote{a} & $1.69^{+0.78}_{-0.71}$ & $\dots$ & $\dots$ & $\dots$ & $1.53\pm0.08$ & $-0.03\pm0.11$ & $1.03$ & $0.99$ & $1.01/381$& 0.01/PL+EX\\
			& \textbf{Obs2014} & &  		 			   & 						  & $\dots$ & $\dots$ & $\dots$ & $1.48\pm0.08$ & $0.22\pm0.13$ & $1.03$ & $1.06$ & &\\
\midrule
QSO J0525-3343	& Obs324 &PL	& $2.45$ & $\dots$ & $\dots$ & $1.51\pm0.02$ & $\dots$ & $\dots$ & $\dots$ & $1.05$ & $1.04$ & $1.09/675$ &\\
				& Obs583 &		&  		 & $\dots$ & $\dots$ & $1.54\pm0.03$ & $\dots$ & $\dots$ & $\dots$ & $1.05$ & $1.11$ & &\\
				& Obs588 &		&  		 & $\dots$ & $\dots$ & $1.52\pm0.03$ & $\dots$ & $\dots$ & $\dots$ & $1.07$ & $1.05$ & &\\
				& Obs324 &BKN	& $2.45$ & $\dots$ & $1.33^{+0.08}_{-0.11}$ & $1.60\pm0.05$ & $1.07^{+0.33}_{-0.22}$ & $\dots$ & $\dots$ & $1.03$ & $1.02$ & $1.00/669$ & $10^{-12}$/PL\\
				& Obs583 &		& 		 & $\dots$ & $0.64^{+0.71}_{-2.21}$ & $1.59\pm0.04$ & $0.52^{+0.67}_{-0.12}$ & $\dots$ & $\dots$ & $1.03$ & $1.10$ & &\\
				& Obs588 &		& 		 & $\dots$ & $1.02^{+0.27}_{-0.51}$ & $1.59\pm0.04$ & $0.66^{+0.20}_{-0.14}$ & $\dots$ & $\dots$ & $1.05$ & $1.03$ & &\\
				& Obs324 &LGP	& $2.45$ & $\dots$ & $\dots$ & $\dots$ & $\dots$ & $1.56\pm0.03$ & $0.18\pm0.06$ & $1.03$ & $1.02$ & $1.02/672$ & $10^{-12}$/PL\\
				& Obs583 &		&  		 & $\dots$ & $\dots$ & $\dots$ & $\dots$ & $1.59\pm0.04$ & $0.19\pm0.08$ & $1.03$ & $1.09$ & &\\
				& Obs588 &		&  		 & $\dots$ & $\dots$ & $\dots$ & $\dots$ & $1.57\pm0.04$ & $0.18\pm0.09$ & $1.05$ & $1.03$ & &\\
				& Obs324 &\emph{PL+EX*}	& $2.45_{2.08}^{2.82}$\tnote{a} & $0.93^{+0.48}_{-0.25}$ & $\dots$ & $1.59\pm0.03$ & $\dots$ & $\dots$ & $\dots$ & $1.03$ & $1.02$ & $1.00/673$ & \\
				& Obs583 &		&  					   & 							 	 & $\dots$ & $1.62\pm0.03$ & $\dots$ & $\dots$ & $\dots$ & $1.03$ & $1.10$ & &\\
				& Obs588 &		&  					   & 							 	 & $\dots$ & $1.60\pm0.03$ & $\dots$ & $\dots$ & $\dots$ & $1.05$ & $1.03$ & &\\				
				& Obs324 &LGP+EX	& $2.45_{2.08}^{2.82}$\tnote{a} & $0.63^{+0.45}_{-0.36}$ & $\dots$ & $\dots$ & $\dots$ & $1.54^{+0.05}_{-0.07}$ & $0.07^{+0.09}_{-0.08}$ & $1.03$ & $1.02$ & 	$1.01/670$ &\\
				& Obs583 &		&   			   & 						 		 & $\dots$ & $\dots$ & $\dots$ & $1.58^{+0.06}_{-0.07}$ & $0.07^{+0.11}_{-0.10}$ & $1.03$ & $1.09$ & &\\
				& Obs588 &		&	 			   & 						 		 & $\dots$ & $\dots$ & $\dots$ & $1.56^{+0.06}_{-0.07}$ & $0.05^{+0.12}_{-0.10}$ & $1.05$ & $1.03$ & &\\
\midrule
QSO B1026-084	& &PL	& $5.28$ & $\dots$ & $\dots$ & $1.39\pm0.02$ & $\dots$ & $\dots$ & $\dots$ & $1.00$ & $1.01$ & $1.18/257$ &\\
				& &BKN	& $5.28$ & $\dots$ & $1.16\pm0.09$ & $1.49\pm0.04$ & $1.17^{+0.23}_{-0.15}$ & $\dots$ & $\dots$ & $0.99$ & $0.99$ & $1.08/255$ & $10^{-6}$/PL\\
				& &LGP	& $5.28$ & $\dots$ & $\dots$ & $\dots$ & $\dots$ & $1.30\pm0.04$ & $0.19^{+0.08}_{-0.07}$ & $0.99$ & $0.99$ & $1.11/256$ & $10^{-6}$/PL\\
				& &\emph{PL+EX*}	& $5.28_{4.49}^{6.07}$\tnote{a} & $0.99^{+0.99}_{-0.53}$ & $\dots$ & $1.46\pm0.04$ & $\dots$ & $\dots$ & $\dots$ & $0.99$ & $0.99$ & $1.11/255$ & \\
				& &LGP+EX	& $5.28_{4.49}^{6.07}$\tnote{a} & $0.47^{+0.98}_{-0.42}$ & $\dots$ & $\dots$ & $\dots$ & $1.40^{+0.05}_{-0.09}$ & $0.09\pm0.13$ & $0.99$ & $0.99$ & $1.11/254$ &\\
\midrule
QSO B0014+810	& &PL	& $22.2$ & $\dots$ & $\dots$ & $1.50\pm0.02$ & $\dots$ & $\dots$ & $\dots$ & $1.01$ & $1.04$ & $1.03/376$ &\\
				& &	BKN	& $22.2$ & $\dots$ & insensitive & $1.49\pm0.02$ & $0.42^{+0.11}_{-0.03}$ & $\dots$ & $\dots$ & $1.01$ & $1.04$ & $1.01/372$& 0.04/PL\\
				& &	LGP	& $22.2$ & $\dots$ & $\dots$ & $\dots$ & $\dots$ & $1.51\pm0.03$ & $-0.03\pm0.05$ & $1.01$ & $1.04$ & $1.03/373$& 0.27/PL\\
				& &	\emph{PL+EX*}	& $<22.1$\tnote{b} & $<0.84$ & $\dots$ & $1.48\pm0.03$ & $\dots$ & $\dots$ & $\dots$ & $1.01$ & $1.04$ & $1.03/374$&\\
\cmidrule{2-14}
				& &PL	&  	$22.2$	 & $\dots$ & $\dots$ & $1.50\pm0.02$ & $\dots$ & $\dots$ & $\dots$ & $1.01$ & $1.04$ &  $1.01/844$ &\\
				&\textbf{Obs2014} &		&  		 	 & $\dots$ & $\dots$ & $1.72\pm0.05$ & $\dots$ & $\dots$ & $\dots$ & $0.98$ & $0.76$ & &\\
				&\textbf{Obs2015} &		&  		 	 & $\dots$ & $\dots$ & $1.61\pm0.04$ & $\dots$ & $\dots$ & $\dots$ & $1.00$ & $0.90$ & &\\
				& &BKN	&  	$22.2$	 & $\dots$ & insensitive & $1.49\pm0.02$ & $0.44^{+0.09}_{-0.03}$  & $\dots$ & $\dots$ & $1.01$ & $1.04$ & $0.97/838$& $10^{-7}$/PL\\
				&\textbf{Obs2014} &		&  		 	 & $\dots$ & $1.30\pm0.26$ & $1.77\pm0.05$ & $2.42^{+3.08}_{-0.61}$ & $\dots$ & $\dots$ & $0.98$ & $0.90$ & &\\
				&\textbf{Obs2015} &		&  		 	 & $\dots$ & $0.88^{+0.41}_{-0.49}$ & $1.64\pm0.05$ & $1.68^{+2.09}_{-0.38}$ & $\dots$ & $\dots$ & $1.00$ & $1.03$ & &\\
				& &LGP	&  	$22.2$	 & $\dots$ & $\dots$ & $\dots$ & $\dots$ & $1.51\pm0.03$ & $-0.02\pm0.05$ & $1.01$ & $1.04$ & $0.98/841$& $10^{-7}$/PL\\
				&\textbf{Obs2014} &		&  			 & $\dots$ & $\dots$ & $\dots$ & $\dots$ & $1.66\pm0.05$ & $0.21\pm0.09$ & $0.98$ & $0.93$ & &\\
				&\textbf{Obs2015} &		&  			 & $\dots$ & $\dots$ & $\dots$ & $\dots$ & $1.54\pm0.06$ & $0.20\pm0.09$ & $1.00$ & $1.11$ & &\\
				& &\emph{PL+EX*}	& $22.4^{+0.7}_{-1.9}$ & $<0.54$ & $\dots$ & $1.50^{+0.02}_{-0.03}$ & $\dots$ & $\dots$ & $\dots$ & $1.01$ & $1.03$ & $1.02/842$&\\
				&\textbf{Obs2014} &		&  					   & 		 & $\dots$ & $1.72^{+0.04}_{-0.02}$ & $\dots$ & $\dots$ & $\dots$ & $0.98$ & $0.76$ & &\\
				&\textbf{Obs2015} &		&  					   & 		 & $\dots$ & $1.61\pm0.04$ & $\dots$ & $\dots$ & $\dots$ & $1.00$ & $0.90$ & &\\
				& &LGP+EX	& $<21.9$\tnote{b} & $<0.62$ & $\dots$ & $\dots$ & $\dots$ & $1.47\pm0.12$ & $0.02\pm0.11$ & $1.01$ & $1.04$ & $0.98/839$& $10^{-7}$/PL+EX\\
				&\textbf{Obs2014} &		&  		 		   & 		 & $\dots$ & $\dots$ & $\dots$ & $1.65\pm0.06$ & $0.22\pm0.09$ & $0.98$ & $0.94$ & &\\
				&\textbf{Obs2015} &		&  		 		   & 		 & $\dots$ & $\dots$ & $\dots$ & $1.53\pm0.06$ & $0.21\pm0.09$ & $1.00$ & $1.11$ & &\\
\midrule
PKS 2126-158	& & PL	& $6.15$ & $\dots$ & $\dots$ & $1.32\pm0.01$ & $\dots$ & $\dots$ & $\dots$ & $1.09$ & $1.12$ & $1.54/394$&\\
				& &BKN	& $6.15$ & $\dots$ & $0.93^{+0.08}_{-0.14}$ & $1.44\pm0.03$ & $1.07^{+0.12}_{-0.15}$ & $\dots$ & $\dots$ & $1.06$ & $1.09$ & $1.02/392$&$10^{-36}$/PL\\
				& &LGP	& $6.15$ & $\dots$ & $\dots$ & $\dots$ & $\dots$ & $1.18\pm0.02$ & $0.30\pm0.04$ & $1.06$ & $1.09$ & $1.11/393$&$10^{-30}$/PL\\
				& &\emph{PL+EX*}& $>5.26$\tnote{b} & $1.38^{+0.50}_{-0.20}$ & $\dots$ & $1.45\pm0.02$ & $\dots$ & $\dots$ & $\dots$ & $1.06$ & $1.09$ & $0.99/392$&\\
				& &LGP+EX& $6.15_{5.23}^{7.07}$\tnote{a} & $1.24^{+0.53}_{-0.35}$ & $\dots$ & $\dots$ & $\dots$ & $1.42^{+0.06}_{-0.07}$ & $0.03\pm0.07$ & $1.06$ & $1.09$ & $0.99/391$& 0.44/PL+EX\\
\midrule
QSO B0537-286	& Obs00 &PL	& $2.42$ & $\dots$ & $\dots$ & $1.30\pm0.01$ & $\dots$ & $\dots$ & $\dots$ & $1.03$ & $1.04$ & $1.11/776$ &\\
				& Obs05 &	&  		 & $\dots$ & $\dots$ & $1.18\pm0.02$ & $\dots$ & $\dots$ & $\dots$ & $1.05$ & $0.99$ & &\\
				& Obs00 &BKN& $2.42$ & $\dots$ & $1.35\pm0.02$ & $1.18\pm0.04$ & $2.39^{+0.40}_{-0.30}$ 		 & $\dots$ & $\dots$ & $1.04$ & $1.05$ & $1.01/770$&$10^{-14}$/PL\\
				& Obs05 &	&  		 & $\dots$ & $0.57^{+0.32}_{-0.48}$ & $1.21\pm0.02$ & $0.58^{+0.16}_{-0.07}$ & $\dots$ & $\dots$ & $1.04$ & $0.98$ & &\\
				& Obs00 &LGP	& $2.42$ & $\dots$ & $\dots$ & $\dots$ & $\dots$ & $1.34\pm0.02$ & $-0.10\pm0.03$ & $1.04$ & $1.05$ & $1.05/772$&$10^{-9}$/PL\\
				& Obs05 &	&  		 & $\dots$ & $\dots$ & $\dots$ & $\dots$ & $1.13\pm0.03$ & $0.11^{+0.05}_{-0.04}$ & $1.04$ & $0.98$ & &\\
				& Obs00 &PL+EX	& $<2.53$\tnote{b} & $<0.18$ & $\dots$ & $1.30\pm0.01$ & $\dots$ & $\dots$ & $\dots$ & $1.03$ & $1.04$ & $1.11/774$&\\
				& Obs05 &		&  				   & 		 & $\dots$ & $1.18\pm0.02$ & $\dots$ & $\dots$ & $\dots$ & $1.05$ & $0.99$ & &\\
				& Obs00 &\emph{LGP+EX*}	& $2.42_{2.06}^{2.78}$\tnote{a} & $0.47^{+0.15}_{-0.24}$\tnote{c} & $\dots$ & $\dots$ & $\dots$ & $1.43\pm0.04$ & $-0.19^{+0.05}_{-0.04}$ & $1.04$ & $1.05$ & $1.02/770$ & $10^{-13}$/PL+EX\\
				& Obs05 &		&  		 			   			& 						 & $\dots$ & $\dots$ & $\dots$ & $1.22\pm0.04$ & $0.01\pm0.06$ & $1.04$ & $0.98$ & &\\
\midrule
QSO B0438-43	& &PL	& $1.41$ & $\dots$ & $\dots$ & $1.45$\tnote{d} & $\dots$ & $\dots$ & $\dots$ & $1.14$ & $1.15$ & $2.48/286$&\\
				& &BKN	& $1.41$ & $\dots$ & $-0.69^{+0.56}_{-0.67}$ & $1.62\pm0.02$ & $0.57^{+0.05}_{-0.03}$ & $\dots$ & $\dots$ & $1.07$ & $1.08$ & $1.06/284$&$10^{-53}$/PL\\
				& &LGP	& $1.41$ & $\dots$ & $\dots$ & $\dots$ & $\dots$ & $1.29\pm0.03$ & $0.53\pm0.06$ & $1.07$ & $1.08$ & $1.45/285$&$10^{-35}$/PL\\
				& &\emph{PL+EX*}& $1.41_{1.20}^{1.62}$\tnote{a} & $1.60^{+0.17}_{-0.68}$ & $\dots$ & $1.71\pm0.03$ & $\dots$ & $\dots$ & $\dots$ & $1.06$ & $1.07$ & $0.97/284$&\\
				& &LGP+EX& $1.41_{1.20}^{1.62}$\tnote{a} & $1.78^{+0.27}_{-0.14}$ & $\dots$ & $\dots$ & $\dots$ & $1.80^{+0.05}_{-0.09}$ & $-0.11\pm0.09$ & $1.06$ & $1.07$ & $0.96/283$&0.07/PL+EX\\
\midrule
RBS 315			& Obs2003 &PL	& $16.3$ & $\dots$ & $\dots$ & $1.12$\tnote{d} & $\dots$ & $\dots$ & $\dots$ & $1.09$ & $1.10$ & $2.25/1376$&\\
				& Obs2013a &		&  		 & $\dots$ & $\dots$ & $1.30$\tnote{d} & $\dots$ & $\dots$ & $\dots$ & $1.10$ & $1.11$ & &\\
				& Obs2013b &		&  		 & $\dots$ & $\dots$ & $1.29$\tnote{d} & $\dots$ & $\dots$ & $\dots$ & $1.10$ & $1.14$ & &\\
				& Obs2003 &BKN	& $16.3$ & $\dots$ & $0.67^{+0.09}_{-0.08}$ & $1.23\pm0.02$ & $1.19^{+0.12}_{-0.08}$ & $\dots$ & $\dots$ & $1.07$ & $1.08$ & $1.10/1370$&$10^{-210}$/PL\\
				& Obs2013a &		& 		 & $\dots$ & $0.82\pm0.05$ & $1.43\pm0.01$ & $1.18^{+0.06}_{-0.05}$ & $\dots$ & $\dots$ & $1.08$ & $1.09$ & &\\
				& Obs2013b &		& 		 & $\dots$ & $0.82^{+0.06}_{-0.07}$ & $1.41\pm0.02$ & $1.17\pm0.07$ & $\dots$ & $\dots$ & $1.07$ & $1.11$ & &\\
				& Obs2003 &LGP	& $16.3$ & $\dots$ & $\dots$ & $\dots$ & $\dots$ & $0.91\pm0.03$ & $0.33\pm0.04$ & $1.06$ & $1.07$ & $1.21/1373$&$10^{-184}$/PL\\
				& Obs2013a&		&  		 & $\dots$ & $\dots$ & $\dots$ & $\dots$ & $1.08\pm0.02$ & $0.38\pm0.02$ & $1.07$ & $1.09$ & &\\
				& Obs2013b&		&  		 & $\dots$ & $\dots$ & $\dots$ & $\dots$ & $1.07\pm0.02$ & $0.36\pm0.03$ & $1.07$ & $1.11$ & &\\
				& Obs2003 &PL+EX	& $>18.4$\tnote{b} & $1.27^{+0.09}_{-0.08}$  & $\dots$ & $1.25\pm0.01$ & $\dots$ & $\dots$ & $\dots$ & $1.07$ & $1.08$ & $1.11/1374$&\\
				& Obs2013a&		&  				   &					 	 & $\dots$ & $1.45\pm0.01$ & $\dots$ & $\dots$ & $\dots$ & $1.08$ & $1.09$ & &\\
				& Obs2013b&		&  				   &					 	 & $\dots$ & $1.42\pm0.01$ & $\dots$ & $\dots$ & $\dots$ & $1.07$ & $1.11$ & &\\				
				& Obs2003 &\emph{LGP+EX*}	& $>17.3$\tnote{b} 	& $0.75^{+0.22}_{-0.14}$ & $\dots$ & $\dots$ & $\dots$ & $1.14\pm0.04$ & $0.11\pm0.04$ & $1.06$ & $1.07$ & $1.08/1371$&$10^{-10}$/PL+EX\\
				& Obs2013a&		&  		 			& 				 		 & $\dots$ & $\dots$ & $\dots$ & $1.32\pm0.03$ & $0.14\pm0.04$ & $1.07$ & $1.09$ & &\\
				& Obs2013b&		&  		 			& 				 		 & $\dots$ & $\dots$ & $\dots$ & $1.31\pm0.03$ & $0.13^{+0.02}_{-0.04}$ & $1.07$ & $1.11$ & &\\
\cmidrule{2-14}
				& Obs2003 &PL	&  	$16.3$	& $\dots$ & $\dots$ & $1.12\pm0.01$ & $\dots$ & $\dots$ & $\dots$ & $1.09$ & $1.10$ &  $1.70/2726$&\\
				& Obs2013a&&  		 	& $\dots$ & $\dots$ & $1.30\pm0.01$ & $\dots$ & $\dots$ & $\dots$ & $1.10$ & $0.12$ & &\\
				& Obs2013b&&  		 	& $\dots$ & $\dots$ & $1.29\pm0.01$ & $\dots$ & $\dots$ & $\dots$ & $1.10$ & $1.14$ & &\\
				&\textbf{Obs2014} &&  		 	& $\dots$ & $\dots$ & $1.49\pm0.02$ & $\dots$ & $\dots$ & $\dots$ & $0.99$ & $0.61$ & &\\
				&\textbf{Obs2015} &&  		 	& $\dots$ & $\dots$ & $1.61\pm0.02$ & $\dots$ & $\dots$ & $\dots$ & $1.01$ & $0.52$ & &\\
				& Obs2003 &BKN	& $16.3$ 	& $\dots$ & $0.67^{+0.09}_{-0.08}$ & $1.23\pm0.02$ & $1.19^{+0.12}_{-0.08}$ & $\dots$ & $\dots$ & $1.07$ & $1.08$ & $1.05/2716$&$10^{-278}$/PL\\
				& Obs2013a&& 		 	& $\dots$ & $0.82\pm0.05$ & $1.43\pm0.01$ & $1.18^{+0.06}_{-0.05}$ & $\dots$ & $\dots$ & $1.08$ & $1.09$ & &\\
				& Obs2013b&& 		 	& $\dots$ & $0.82^{+0.06}_{-0.07}$ & $1.41\pm0.02$ & $1.17\pm0.07$ & $\dots$ & $\dots$ & $1.07$ & $1.11$ & &\\
				&\textbf{Obs2014} &&  		 	& $\dots$ & $1.07\pm0.09$ & $1.53^{+0.03}_{-0.02}$ & $4.14^{+0.72}_{-0.47}$ & $\dots$ & $\dots$ & $0.99$ & $0.89$ & &\\
				&\textbf{Obs2015} &&  		 	& $\dots$ & $1.01\pm0.12$ & $1.68^{+0.04}_{-0.03}$ & $4.34^{+0.66}_{-0.45}$ & $\dots$ & $\dots$ & $1.01$ & $0.97$ & &\\
				& Obs2003 &LGP	& $16.3$ & $\dots$ & $\dots$ & $\dots$ & $\dots$ & $0.91\pm0.03$ & $0.33\pm0.04$ & $1.06$ & $1.07$ & $1.10/2721$&$10^{-255}$/PL\\
				& Obs2013a&&  		 & $\dots$ & $\dots$ & $\dots$ & $\dots$ & $1.08\pm0.02$ & $0.38\pm0.02$ & $1.07$ & $1.09$ & &\\
				& Obs2013b&&  		 & $\dots$ & $\dots$ & $\dots$ & $\dots$ & $1.07\pm0.02$ & $0.36\pm0.03$ & $1.07$ & $1.11$ & &\\
				&\textbf{Obs2014} &&  		 & $\dots$ & $\dots$ & $\dots$ & $\dots$ & $1.37\pm0.03$ & $0.25\pm0.05$ & $0.99$ & $0.86$ & &\\
				&\textbf{Obs2015} &&  		 & $\dots$ & $\dots$ & $\dots$ & $\dots$ & $1.45\pm0.03$ & $0.37\pm0.06$ & $1.01$ & $0.89$ & &\\
				& Obs2003 &PL+EX	& $>18.5$\tnote{b} & $1.32^{+0.09}_{-0.08}$  & $\dots$ & $1.25\pm0.01$ & $\dots$ & $\dots$ & $\dots$ & $1.07$ & $1.08$ & $1.09/2724$&\\
				& Obs2013a&&  				   &					 	 & $\dots$ & $1.45\pm0.01$ & $\dots$ & $\dots$ & $\dots$ & $1.08$ & $1.09$ & &\\
				& Obs2013b&&  				   &					 	 & $\dots$ & $1.43\pm0.01$ & $\dots$ & $\dots$ & $\dots$ & $1.07$ & $1.11$ & &\\	
				&\textbf{Obs2014} &&  					& 						 & $\dots$ & $1.51\pm0.02$ & $\dots$ & $\dots$ & $\dots$ & $1.03$ & $0.82$ & &\\
				&\textbf{Obs2015} &&  					& 						 & $\dots$ & $1.62\pm0.02$ & $\dots$ & $\dots$ & $\dots$ & $1.03$ & $0.82$ & &\\
				& Obs2003 &\emph{LGP+EX*}	& $>17.5$\tnote{b} 	& $0.77^{+0.20}_{-0.14}$ & $\dots$ & $\dots$ & $\dots$ & $1.15\pm0.04$ & $0.11\pm0.04$ & $1.06$ & $1.07$ & $1.03/2719$&$10^{-32}$/PL+EX\\
				& Obs2013a&&  			& 				 		 & $\dots$ & $\dots$ & $\dots$ & $1.33\pm0.03$ & $0.13\pm0.03$ & $1.07$ & $1.09$&&\\									
				& Obs2013b&&  		 			& 				 		 & $\dots$ & $\dots$ & $\dots$ & $1.31\pm0.03$ & $0.13\pm0.04$ & $1.07$ & $1.11$ & &\\
				&\textbf{Obs2014} &&  		 			& 						 & $\dots$ & $\dots$ & $\dots$ & $1.40\pm0.03$ & $0.20\pm0.05$ & $0.99$ & $0.85$ & &\\
				&\textbf{Obs2015} &&  		 			& 						 & $\dots$ & $\dots$ & $\dots$ & $1.47\pm0.04$ & $0.33\pm0.06$ & $1.01$ & $0.89$ & &\\
\midrule
QSO J2354-1513	& &PL	& $2.78$ & $\dots$ & $\dots$ & $1.51\pm0.02$ & $\dots$ & $\dots$ & $\dots$ & $1.05$ & $1.05$ & $1.29/326$&\\
				& &BKN	& $2.78$ & $\dots$ & $0.28^{+0.51}_{-1.47}$ & $1.57\pm0.02$ & $0.53\pm0.08$ & $\dots$ & $\dots$ & $1.04$ & $1.03$ & $1.04/324$&$10^{-16}$/PL\\
				& &LGP	& $2.78$ & $\dots$ & $\dots$ & $\dots$ & $\dots$ & $1.43\pm0.02$ & $0.23\pm0.05$ & $1.03$ & $1.02$ & $1.11/325$&$10^{-12}$/PL\\
				& &\emph{PL+EX*}	& $2.78_{1.81}^{3.20}$\tnote{a} & $0.51^{+0.30}_{-0.11}$ & $\dots$ & $1.62^{+0.02}_{-0.03}$ & $\dots$ & $\dots$ & $\dots$ & $1.03$ & $1.02$ & $1.02/324$&\\
				& &LGP+EX	& $2.78_{1.81}^{3.20}$\tnote{a} & $0.55^{+0.29}_{-0.20}$ & $\dots$ & $\dots$ & $\dots$ & $1.63^{+0.07}_{-0.06}$ & $-0.03\pm0.10$ & $1.03$ & $1.02$ & $1.03/323$&\\
\midrule
PBC J1656.2-3303& &PL	& $33.1$ & $\dots$ & $\dots$ & $1.13\pm0.01$ & $\dots$ & $\dots$ & $\dots$ & $1.05$ & $1.03$ & $1.24/420$&\\
				& &BKN	& $33.1$ & $\dots$ & $0.88^{+0.08}_{-0.11}$ & $1.19\pm0.02$ & $1.44\pm0.20$	 & $\dots$ & $\dots$ & $1.04$ & $1.02$ & $1.13/418$ &$10^{-9}$/PL\\
				& &LGP	& $33.1$ & $\dots$ & $\dots$ & $\dots$ & $\dots$ & $0.99\pm0.04$ & $0.18\pm0.05$ & $1.04$ & $1.02$ & $1.16/419$&$10^{-7}$/PL\\
				& &\emph{PL+EX*}	& $>33.7$\tnote{b} & $<1.14$ & $\dots$ & $1.20\pm0.02$ & $\dots$ & $\dots$ & $\dots$ & $1.04$ & $1.02$ & $1.14/418$&\\
				& &LGP+EX	& $>29.3$\tnote{b} & $<1.40$ & $\dots$ & $\dots$ & $\dots$ & $1.18^{+0.02}_{-0.06}$ & $0.04^{+0.10}_{-0.09}$ & $1.04$ & $1.02$ & $1.15/417$&\\
\cmidrule{2-14}
				& &PL	&  	$33.1$	 & $\dots$ & $\dots$ & $1.13\pm0.01$ & $\dots$ & $\dots$ & $\dots$ & $1.05$ & $1.03$ &  $1.25/690$ &\\
				&\textbf{Obs2015} &		&  		 	 & $\dots$ & $\dots$ & $1.53\pm0.04$ & $\dots$ & $\dots$ & $\dots$ & $1.04$ & $0.79$ & &\\
				& &BKN	&  	$33.1$	 & $\dots$ & $0.88^{+0.08}_{-0.11}$ & $1.19\pm0.02$ & $1.44^{+0.20}_{-0.21}$ & $\dots$ & $\dots$ & $1.04$ & $1.02$ & $1.08/686$ &$10^{-21}$/PL\\
				&\textbf{Obs2015} &		&  		 	 & $\dots$ & $0.49^{+0.30}_{-0.36}$ & $1.60\pm0.05$ & $2.24^{+0.50}_{-0.35}$ & $\dots$ & $\dots$ & $1.04$ & $1.07$ & &\\
				& &LGP	&  	$33.1$	 & $\dots$ & $\dots$ & $\dots$ & $\dots$ & $0.99\pm0.04$ & $0.18\pm0.05$ & $1.04$ & $1.02$ & $1.12/688$ &$10^{-17}$/PL\\
				&\textbf{Obs2015} &		&  		 	 & $\dots$ & $\dots$ & $\dots$ & $\dots$ & $1.40\pm0.05$ & $0.36\pm0.09$ & $1.04$ & $1.13$ & &\\
				& &PL+EX	& $>35.3$\tnote{b} & $0.53^{+0.58}_{-0.33}$ & $\dots$ & $1.21\pm0.02$ & $\dots$ & $\dots$ & $\dots$ & $1.04$ & $1.02$ & $1.15/688$&\\
				&\textbf{Obs2015} &		&  					   & 					& $\dots$ & $1.55\pm0.04$ & $\dots$ & $\dots$ & $\dots$ & $1.04$ & $0.84$ & &\\
				& &LGP+EX	& $>33.1$\tnote{b} & $<1.05$ & $\dots$ & $\dots$ & $\dots$ & $1.18\pm0.09$ & $0.02^{+0.05}_{-0.09}$ & $1.04$ & $1.02$ & $1.11/686$&\\
				&\textbf{Obs2015} &		&  		 		   & 		 & $\dots$ & $\dots$ & $\dots$ & $1.43\pm0.06$ & $0.31\pm0.09$ & $1.04$ & $1.12$ & &\\

				& &\emph{LGP+EX*}$^,$\tnote{e}	& $>34.5$\tnote{b} & $0.33^{+0.72}_{-0.32}$ & $\dots$ & $\dots$ & $\dots$ & $1.20\pm0.02$ & $0$ & $1.04$ & $1.02$ & $1.11/687$&$10^{-8}$/PL+EX\\
				&\textbf{Obs2015} &		&  		 		   &		 & $\dots$ & $\dots$ & $\dots$ & $1.43\pm0.06$ & $0.31\pm0.09$ & $1.04$ & $1.12$ & &\\				
\midrule
QSO J0555+3948	& &PL	& $42.5$ & $\dots$ & $\dots$ & $1.68\pm0.03$ & $\dots$ & $\dots$ & $\dots$ & $1.08$ & $1.09$ & $1.14/290$&\\
				& &BKN	& $42.5$ & $\dots$ & $1.96^{+0.09}_{-0.15}$ & $1.54^{+0.05}_{-0.14}$ & $1.67^{+1.18}_{-0.22}$ & $\dots$ & $\dots$ & $1.08$ & $1.09$ & $1.01/288$&$10^{-8}$/PL\\
				& &LGP	& $42.5$ & $\dots$ & $\dots$ & $\dots$ & $\dots$ & $1.94\pm0.07$ & $-0.39\pm0.09$ & $1.09$ & $1.10$ & $0.99/289$&$10^{-10}$/PL\\
				& &PL+EX	& $<37.0$\tnote{b} & $<0.16$ & $\dots$ & $1.58\pm0.03$ & $\dots$ & $\dots$ & $\dots$ & $1.08$ & $1.09$ & $1.03/288$&\\
				& &\emph{LGP+EX*}	& $<43.7$\tnote{b} & $<0.93$ & $\dots$ & $\dots$ & $\dots$ & $1.91^{+0.27}_{-0.13}$ & $-0.37^{+0.30}_{-0.22}$ & $1.09$ & $1.09$ & $0.99/287$&$10^{-4}$/PL+EX\\
\midrule
PKS 2149-306	& &PL	& $1.74$ & $\dots$ & $\dots$ & $1.44\pm0.01$ & $\dots$ & $\dots$ & $\dots$ & $1.02$ & $1.02$ & $1.01/444$&\\
				& &BKN	& $1.74$ & $\dots$ & $1.45\pm0.01$ & $0.97^{+0.41}_{-1.59}$ & $7.14^{+1.64}_{-2.08}$ & $\dots$ & $\dots$ & $1.03$ & $1.02$ & $1.00/442$&\\
				& &LGP	& $1.74$ & $\dots$ & $\dots$ & $\dots$ & $\dots$ & $1.45\pm0.01$ & $-0.02\pm0.02$ & $1.03$ & $1.02$ & $1.00/443$&\\
				& &\emph{PL+EX*}& $<1.89$\tnote{b} & $<0.04$ & $\dots$ & $1.44\pm0.01$ & $\dots$ & $\dots$ & $\dots$ & $1.03$ & $1.02$ & $1.01/444$&\\
\cmidrule{2-14}
				& &PL	&  	$1.74$	 & $\dots$ & $\dots$ & $1.44\pm0.01$ & $\dots$ & $\dots$ & $\dots$ & $1.02$ & $1.02$ &  $1.05/2741$&\\
				&\textbf{Obs2013} &&  		 	 & $\dots$ & $\dots$ & $1.35\pm0.01$ & $\dots$ & $\dots$ & $\dots$ & $1.06$ & $0.73$ & &\\
				&\textbf{Obs2014} &&  		 	 & $\dots$ & $\dots$ & $1.44\pm0.01$ & $\dots$ & $\dots$ & $\dots$ & $0.99$ & $0.68$ & &\\
				& &BKN	&  	$1.74$	 & $\dots$ & $1.45\pm0.01$ 	& $1.16^{+0.20}_{-1.42}$ & $6.24^{+2.41}_{-1.17}$ & $\dots$ & $\dots$ & $1.03$ & $1.02$ & $0.96/2735$&$10^{-49}$/PL\\
				&\textbf{Obs2013} &&  		 	 & $\dots$ & $0.94^{+0.09}_{-0.11}$	& $1.37\pm0.01$ & $2.49^{+0.66}_{-0.55}$ & $\dots$ & $\dots$ & $1.06$ & $0.93$ & &\\
				&\textbf{Obs2014} &&  		 	 & $\dots$ & $1.00^{+0.07}_{-0.08}$ & $1.46\pm0.01$ & $3.15^{+0.57}_{-0.54}$ & $\dots$ & $\dots$ & $0.99$ & $0.97$ & &\\
				& &LGP	&  	$1.74$	 & $\dots$ & $\dots$ & $\dots$ & $\dots$ & $1.45\pm0.01$ & $-0.02\pm0.02$ & $1.03$ & $1.02$ & $0.98/2738$&$10^{-39}$/PL\\
				&\textbf{Obs2013} &&  			 & $\dots$ & $\dots$ & $\dots$ & $\dots$ & $1.28\pm0.02$ & $0.13\pm0.02$ & $1.06$ & $0.91$ & &\\
				&\textbf{Obs2014} &&  			 & $\dots$ & $\dots$ & $\dots$ & $\dots$ & $1.35\pm0.02$ & $0.16\pm0.03$ & $0.99$ & $0.89$ & &\\
				& &PL+EX	& $>1.65$\tnote{b} & $<0.08$ & $\dots$ & $1.46\pm0.01$ & $\dots$ & $\dots$ & $\dots$ & $1.02$ & $1.02$ & $1.05/2739$&\\
				&\textbf{Obs2013} &&  				   & 		 & $\dots$ & $1.35\pm0.01$ & $\dots$ & $\dots$ & $\dots$ & $1.06$ & $0.74$ & &\\
				&\textbf{Obs2014} &&  				   & 		 & $\dots$ & $1.44\pm0.01$ 			& $\dots$ & $\dots$ & $\dots$ & $0.99$ & $0.69$ & &\\
				& &LGP+EX	& $>1.54$\tnote{b} & $<0.12$ & $\dots$ & $\dots$ & $\dots$ & $1.49\pm0.03$ & $-0.06\pm0.04$ & $1.03$ & $1.02$ & $0.98/2736$&\\
				&\textbf{Obs2013} &&  		 		   & 		 & $\dots$ & $\dots$ & $\dots$ & $1.28\pm0.02$ & $0.12\pm0.02$ & $1.06$ & $0.90$ & &\\
				&\textbf{Obs2014} &&  		 		   & 		 & $\dots$ & $\dots$ & $\dots$ & $1.36\pm0.02$ & $0.15\pm0.03$ & $0.99$ & $0.89$ & &\\
				& &\emph{LGP+EX*}$^,$\tnote{e}	& $1.74^{2.00}_{1.48}$\tnote{a} & $<0.06$ & $\dots$ & $\dots$ & $\dots$ & $1.44\pm0.01$ & $0$ 	& $1.02$  & $1.02$ & $0.98/2736$&$10^{-38}$/PL+EX\\
				&\textbf{Obs2013} &&  		 		   				& 		  & $\dots$ & $\dots$ & $\dots$ & $1.28\pm0.02$ & $0.13\pm0.02$ & $1.06$ & $0.91$ & &\\
				&\textbf{Obs2014} &&  		 		  				& 		  & $\dots$ & $\dots$ & $\dots$ & $1.35\pm0.02$ & $0.16\pm0.03$ & $0.99$ & $0.89$ & &\\
\midrule
QSO B0237-2322	& &PL	& $2.33$ & $\dots$ & $\dots$ & $1.74\pm0.02$ & $\dots$ & $\dots$ & $\dots$ & $1.01$ & $1.05$ & $0.95/304$&\\
				& &BKN	& $2.33$ & $\dots$ & $1.75\pm0.03$ & insensitive & $2.48^{+0.78}_{-0.52}$\tnote{f} & $\dots$ & $\dots$ & $1.01$ & $1.05$ & $0.95/302$&\\
				& &LGP	& $2.33$ & $\dots$ & $\dots$ & $\dots$ & $\dots$ & $1.74\pm0.02$ & $-0.02\pm0.05$ & $1.01$ & $1.05$ & $0.96/303$&\\
				& &\emph{PL+EX*}$^,$\tnote{g} & $<2.60$\tnote{b} & $<0.11$ & $\dots$ & $1.73\pm0.03$ & $\dots$ & $\dots$ & $\dots$ & $1.01$ & $1.05$ & $0.96/414$&\\
\midrule
4C 71.07		& &PL	& $3.16$ & $\dots$ & $\dots$ & $1.33\pm0.01$ & $\dots$ & $\dots$ & $\dots$ & $1.08$ & $1.08$ & $1.24/501$ &\\
				& &BKN	& $3.16$ & $\dots$ & $1.27^{+0.01}_{-0.02}$ & $1.38\pm0.01$ & $1.54^{+0.25}_{-0.20}$ & $\dots$ & $\dots$ & $1.07$ & $1.07$ & $1.09/499$ &$10^{-14}$/PL\\
				& &LGP	& $3.16$ & $\dots$ & $\dots$ & $\dots$ & $\dots$ & $1.29\pm0.01$ & $0.08\pm0.02$ & $1.07$ & $1.07$ & $1.09/500$ &$10^{-15}$/PL\\
				& &PL+EX	& $>3.30$\tnote{b} & $0.06^{+0.05}_{-0.03}$  & $\dots$ & $1.36\pm0.01$ & $\dots$ & $\dots$ & $\dots$ & $1.07$ & $1.08$ & $1.13/499$ &\\
				& &\emph{LGP+EX*}	& $3.16^{3.63}_{2.69}$\tnote{a} 	& $<0.07$ & $\dots$ & $\dots$ & $\dots$ & $1.30\pm0.03$ & $0.07\pm0.03$ & $1.07$ & $1.07$ & $1.09/498$ & $10^{-4}$/PL+EX\\
\cmidrule{2-14}
				& &PL	& $3.16$  & $\dots$ & $\dots$ & $1.33\pm0.01$ & $\dots$ & $\dots$ & $\dots$ & $1.08$ & $1.08$ &  $1.17/2123$ &\\
				&\textbf{Obs2013} &&  		 	& $\dots$ & $\dots$ & $1.64\pm0.02$ & $\dots$ & $\dots$ & $\dots$ & $1.04$ & $0.65$ & &\\
				&\textbf{Obs2014} &&  		 	& $\dots$ & $\dots$ & $1.62\pm0.01$ & $\dots$ & $\dots$ & $\dots$ & $1.02$ & $0.68$ & &\\
				& &BKN	& $3.16$ 	& $\dots$ & $1.27^{+0.01}_{-0.02}$ & $1.38\pm0.01$ & $1.55^{+0.25}_{-0.20}$ & $\dots$ & $\dots$ & $1.07$ & $1.07$ & $1.04/2117$&$10^{-54}$/PL\\
				&\textbf{Obs2013} &&  		 	& $\dots$ & $1.10^{+0.11}_{-0.13}$ & $1.68\pm0.02$ & $2.12^{+0.45}_{-0.37}$ & $\dots$ & $\dots$ & $1.04$ & $0.86$ & &\\
				&\textbf{Obs2014} &&  		 	& $\dots$ & $1.21^{+0.07}_{-0.20}$ & $1.65\pm0.02$ & $3.66^{+0.57}_{-1.66}$ & $\dots$ & $\dots$ & $1.02$ & $1.04$ & &\\
				& &LGP	& $3.16$ & $\dots$ & $\dots$ & $\dots$ & $\dots$ & $1.29\pm0.01$ & $0.08\pm0.02$ & $1.07$ & $1.07$ & $1.05/2120$&$10^{-49}$/PL\\
				&\textbf{Obs2013} &&  		 & $\dots$ & $\dots$ & $\dots$ & $\dots$ & $1.58\pm0.02$ & $0.17\pm0.04$ & $1.04$ & $0.85$ & &\\
				&\textbf{Obs2014} &&  		 & $\dots$ & $\dots$ & $\dots$ & $\dots$ & $1.53\pm0.02$ & $0.20\pm0.03$ & $1.02$ & $0.96$ & &\\
				& &PL+EX	& $>3.40$\tnote{b} & $0.10^{+0.03}_{-0.02}$  & $\dots$ & $1.37\pm0.01$ & $\dots$ & $\dots$ & $\dots$ & $1.07$ & $1.07$ & $1.12/2121$&\\
				&\textbf{Obs2013} &&  					& 						 & $\dots$ & $1.65\pm0.02$ & $\dots$ & $\dots$ & $\dots$ & $1.04$ & $0.67$ & &\\
				&\textbf{Obs2014} &&  					& 						 & $\dots$ & $1.63\pm0.02$ & $\dots$ & $\dots$ & $\dots$ & $1.02$ & $0.71$ & &\\
				& &\emph{LGP+EX*}$^,$\tnote{g}	& $>3.09$\tnote{b} 	& $<0.06$ & $\dots$ & $\dots$ & $\dots$ & $1.32\pm0.02$ & $0.05^{+0.03}_{-0.02}$ & $1.07$ & $1.08$ & $1.04/2973$&$10^{-30}$/PL+EX\\
				&\textbf{Obs2013} &&  		 			& 		  & $\dots$ & $\dots$ & $\dots$ & $1.58\pm0.03$ & $0.16\pm0.04$ & $1.04$ & $0.85$ & &\\
				&\textbf{Obs2014} &&  		 			& 		  & $\dots$ & $\dots$ & $\dots$ & $1.54\pm0.03$ & $0.19\pm0.03$ & $1.02$ & $0.96$ & &\\
\midrule
PKS 0528+134	& Obs501	& PL& $38.5$ & $\dots$ & $\dots$ & $1.41\pm0.05$ & $\dots$ & $\dots$ & $\dots$ & $\dots$ & $1.03$ & $1.18/620$ & \\
				& Obs601	&	&  		 & $\dots$ & $\dots$ & $1.37\pm0.03$ & $\dots$ & $\dots$ & $\dots$ & $1.02$ & $1.05$ & & \\
				& Obs701	&	&  		 & $\dots$ & $\dots$ & $1.39\pm0.04$ & $\dots$ & $\dots$ & $\dots$ & $0.94$ & $1.03$ & & \\
				& Obs501	&BKN& $38.5$ & $\dots$ & $0.82^{+0.36}_{-0.77}$ & $1.50^{+0.08}_{-0.07}$ & $1.27^{+0.50}_{-0.26}$ & $\dots$ & $\dots$ & $\dots$ & $1.04$ & $1.00/614$ & $10^{-21}$/PL\\
				& Obs601& 	& 	& $\dots$ & $0.70^{+0.19}_{-0.36}$ & $1.51^{+0.05}_{-0.04}$ & $1.29^{+0.16}_{-0.18}$ & $\dots$ & $\dots$ & $1.00$ & $1.04$ & \\
				& Obs701& 	& 	& $\dots$ & $0.75^{+0.31}_{-0.37}$ & $1.49^{+0.07}_{-0.06}$ & $1.24^{+0.31}_{-0.15}$ & $\dots$ & $\dots$ & $0.91$ & $1.00$ & \\
				& Obs501	&LGP	& $38.5$ & $\dots$ & $\dots$ & $\dots$ & $\dots$ & $1.13\pm0.15$ & $0.40\pm0.19$ & $\dots$ & $1.03$ & $1.01/617$ & $10^{-21}$/PL\\
				& Obs601& 	&  	& $\dots$ & $\dots$ & $\dots$ & $\dots$ & $1.01\pm0.08$ & $0.53\pm0.10$ & $1.00$ & $1.03$ & \\
				& Obs701& 	&  	& $\dots$ & $\dots$ & $\dots$ & $\dots$ & $1.14\pm0.12$ & $0.35\pm0.17$ & $0.92$ & $1.01$ & \\
				& Obs501	&\emph{PL+EX}*& $>35.3$\tnote{b} & $1.45^{+1.38}_{-0.39}$ & $1.56\pm0.08$ & $\dots$ & $\dots$ & $\dots$ & $\dots$ & $\dots$ & $1.04$ & $0.99/618$ & \\
				& Obs601& 	&  	  & 							&  $1.55\pm0.04$ & $\dots$ & $\dots$ & $\dots$ & $\dots$ & $1.00$ & $1.04$ & \\
				& Obs701& 	&  	  & 							&  $1.56\pm0.05$ &$\dots$ & $\dots$ & $\dots$ & $\dots$ & $0.91$ & $1.00$ & \\				
				& Obs501	&LGP+EX& $38.5_{32.7}^{44.3}$\tnote{a} & $1.04^{+1.57}_{-0.68}$ & $\dots$ & $\dots$ & $\dots$ & $1.50^{+0.16}_{-0.24}$ & $0.06^{+0.26}_{-0.20}$ & $\dots$ & $1.04$ & $0.99/615$ & $0.35$/PL+EX\\
				& Obs601& 	&  	  & 						 		 & $\dots$ & $\dots$ & $\dots$ & $1.41^{+0.12}_{-0.19}$ & $0.16^{+0.21}_{-0.14}$ & $1.00$ & $1.04$ & \\
				& Obs701& 	&  	  & 						 		 & $\dots$ & $\dots$ & $\dots$ & $1.54^{+0.15}_{-0.25}$ & $-0.01^{+0.25}_{-0.18}$ & $0.91$ & $1.00$ & \\

\end{longtable}
\end{ThreePartTable}
\end{landscape}

%% file: Tab_lowz.tex
\begin{table}[!h]
	\scriptsize
	\setlength{\tabcolsep}{8pt}
	\renewcommand{\arraystretch}{1.17}
	\caption{\emph{XMM-Newton} spectral fits for six FSRQ-type blazars and two LBLs of the low-redshift sample, performed in the $0.2-10\,$keV energy range for the pn detector and $0.3-10\,$keV for MOS detectors. The Galactic value is held fixed \citep{Willingale13:GalacticH2} and it is in units of $(10^{20}\,\text{cm}^{-2})$. Errors and upper limits were computed within \texttt{XSPEC} at $90\%$ confidence level. For other columns see the description of Table~\ref{tab:spectral_analysis}.}
	\label{tab:lowz_XMM}
	\centering
	\begin{ThreePartTable}
		\begin{tabular}{ccccccccccc}
			\toprule
			\multicolumn{1}{c}{Source}& 
			\multicolumn{1}{c}{$z$}& 			
			\multicolumn{1}{c}{Model}& 
			\multicolumn{1}{c}{$N_H^{Gal}$}& 
			\multicolumn{1}{c}{$N_H(z)$}&
			\multicolumn{1}{c}{$\Gamma_{low}$}& 
			\multicolumn{1}{c}{$\Gamma$}& 
			\multicolumn{1}{c}{$E_b$}&
			\multicolumn{1}{c}{$a$} &
			\multicolumn{1}{c}{$b$}& 
			\multicolumn{1}{c}{$\chi^2_{\nu}/\nu$}\\
			&
			& 
			&
			&
			&
			&
			& 
			(keV)&
			(@1keV)&
			&
			\\
			\midrule
			TXS 2331+073\tnote{a}    & 0.401 & PL 	& $7.2$ & $\dots$ & $\dots$ & $1.88\pm0.03$ & $\dots$ & $\dots$ & $\dots$ & $1.04/326$\\
			& 		& BKN   & 	 	& $\dots$ & $2.03^{+0.05}_{-0.06}$ & $1.67^{+0.06}_{-0.10}$ & $1.61^{+0.53}_{-0.23}$ & $\dots$ & $\dots$ & $0.91/324$\\
			& 		& LGP    &  	 	& $\dots$ & $\dots$ & $\dots$ & $\dots$ & $1.96\pm0.03$ & $-0.27\pm0.06$ &  $0.92/325$\\
			& 		& LGP+EX &  	 	& $<0.04$ & $\dots$ & $\dots$ & $\dots$ & $1.98^{+0.11}_{-0.02}$ & $-0.29^{+0.06}_{-0.14}$ & $0.92/324$ \\
			4C 31.63    & 0.295 & PL 	& $11.9$ & $\dots$ & $\dots$ & $2.01$ & $\dots$ & $\dots$ & $\dots$ & $2.22/398$\\
			& 		& BKN   & 	 	 & $\dots$ & $2.41^{+0.08}_{-0.06}$ & $1.82\pm0.03$ & $1.05\pm0.10$ & $\dots$ & $\dots$ & $1.18/396$\\
			& 		& LGP    &  	 	 & $\dots$ & $\dots$ & $\dots$ & $\dots$ & $2.16\pm0.02$ & $-0.43\pm0.03$ &  $1.10/397$\\
			& 		& LGP+EX &  	 	 & $<0.003$& $\dots$ & $\dots$ & $\dots$ & $2.16\pm0.02$ & $-0.43\pm0.03$ &  $1.11/396$ \\
			B2 1128+31  & 0.29  & PL 	& $2.0$ & $\dots$ & $\dots$ & $2.07$ & $\dots$ & $\dots$ & $\dots$ & $2.17/369$\\
			&		& BKN   & 	 	& $\dots$ & $2.20\pm0.02$ & $1.66\pm0.04$ & $1.90^{+0.12}_{-0.13}$ & $\dots$ & $\dots$ & $1.25/367$\\
			& 		& LGP    &  	 	& $\dots$ & $\dots$ & $\dots$ & $\dots$ & $1.91\pm0.02$ & $-0.35\pm0.03$ &  $1.22/368$\\
			& 		& LGP+EX &  	 	& $<0.011$& $\dots$ & $\dots$ & $\dots$ & $1.92\pm0.02$ & $<-0.35$\tnote{b} &  $1.22/396$ \\
			PKS 2004-447\tnote{c}   & 0.24  & PL& $3.68$& $\dots$ & $\dots$ & $1.56\pm0.02$ & $\dots$ & $\dots$ & $\dots$ & $1.02/349$\\
			& 		&  	&  		& $\dots$ & $\dots$ & $1.63\pm0.02$ & $\dots$ & $\dots$ & $\dots$ & $1.13/281$\\
			&		&  	&  		& $\dots$ & $\dots$ & $1.63\pm0.02$ & $\dots$ & $\dots$ & $\dots$ & $1.07/698$\\
			& 		& BKN & & $\dots$ & $1.60\pm0.03$ & $1.38^{+0.09}_{-0.25}$ & $2.91^{+1.70}_{-0.71}$ & $\dots$ & $\dots$ & $0.99/347$\\
			& 		&    & 	& $\dots$ & $1.67\pm0.03$ & $1.43^{+0.11}_{-0.12}$ & $3.03^{+0.82}_{-0.89}$ & $\dots$ & $\dots$ & $1.10/279$\\
			& 		&    & 	& $\dots$ & $1.67\pm0.03$ & $1.51^{+0.07}_{-0.08}$ & $2.68^{+0.87}_{-0.89}$ & $\dots$ & $\dots$ & $1.04/694$\\		
			& 		& LGP   & & $\dots$ & $\dots$ & $\dots$ & $\dots$ & $1.61\pm0.03$ & $-0.12\pm0.05$ &  $0.98/348$\\
			& 		&      & & $\dots$ & $\dots$ & $\dots$ & $\dots$ & $1.67\pm0.03$ & $-0.11\pm0.06$ &  $1.11/280$\\
			& 		&      & & $\dots$ & $\dots$ & $\dots$ & $\dots$ & $1.55\pm0.05$\tnote{d} & $-0.08\pm0.04$ &  $1.04/696$\\
			& 		& LGP+EX &  & $<0.016$ & $\dots$ & $\dots$ & $\dots$ & $1.61\pm0.03$ & $-0.12\pm0.05$ & $0.99/347$ \\
			& 		&  &  	   & $<0.017$ & $\dots$ & $\dots$ & $\dots$ & $1.67\pm0.03$ & $-0.11\pm0.06$ & $1.11/279$ \\
			& 		&  &  	   & $<0.007$ & $\dots$ & $\dots$ & $\dots$ & $1.55\pm0.05$\tnote{d} & $-0.08\pm0.04$  & $1.04/695$ \\
			PMN J0623-6436  & 0.129 & PL 	& $4.68$& $\dots$ & $\dots$ & $2.17\pm0.02$ & $\dots$ & $\dots$ & $\dots$ & $1.77/322$\\
			&		& BKN   & 	 	& $\dots$ & $2.35^{+0.03}_{-0.06}$ & $1.91^{+0.04}_{-0.08}$ & $1.36^{+0.24}_{-0.12}$ & $\dots$ & $\dots$ & $1.20/320$\\
			& 		& LGP    &  	 	& $\dots$ & $\dots$ & $\dots$ & $\dots$ & $2.24\pm0.02$ & $-0.36\pm0.04$ &  $1.15/321$\\
			& 		& LGP+EX &  	 	& $<0.012$& $\dots$ & $\dots$ & $\dots$ & $2.24^{+0.06}_{-0.02}$ & $-0.37^{+0.04}_{-0.08}$ &  $1.15/320$ \\
			PKS 0521-365  & 0.055   & PL 	& $4.15$& $\dots$ & $\dots$ & $1.83\pm0.01$ & $\dots$ & $\dots$ & $\dots$ & $1.90/476$\\
			&			& BKN   & 	 	& $\dots$ & $1.94\pm0.02$ & $1.72\pm0.02$ & $1.33^{+0.16}_{-0.17}$ & $\dots$ & $\dots$ & $1.23/474$\\
			& 		& LGP    &  	 	& $\dots$ & $\dots$ & $\dots$ & $\dots$ & $1.88\pm0.01$ & $-0.17\pm0.01$ &  $1.19/475$\\
			& 		& LGP+EX &  	 	& $<0.0004$& $\dots$ & $\dots$ & $\dots$ & $1.88\pm0.01$ & $-0.17\pm0.01$ &  $1.19/474$ \\
			\midrule
			OJ 287\tnote{e}   & 0.306& PL  & $2.78$ & $\dots$ & $\dots$ & $1.79\pm0.01$ & $\dots$ & $\dots$ & $\dots$ & $1.12/832$\\
			& 	 &     &  	    & $\dots$ & $\dots$ & $1.75\pm0.01$ & $\dots$ & $\dots$ & $\dots$ & 		  \\
			& 	 & BKN &		& $\dots$ & $1.91^{+0.16}_{-0.04}$ & $1.73^{+0.02}_{-0.03}$ & $0.98^{+0.26}_{-0.33}$ & $\dots$ & $\dots$ & $1.04/828$\\
			& 	 &    	& 		& $\dots$ & $1.78^{+0.18}_{-0.02}$ & $1.72^{+0.03}_{-0.08}$ & $1.59^{+0.78}_{-0.16}$ & $\dots$ & $\dots$ & 			\\
			& 	 & LGP   & 		& $\dots$ & $\dots$ & $\dots$ & $\dots$ & $1.82\pm0.01$ & $-0.13\pm0.02$ &  $1.03/830$\\
			& 	 &      & 		& $\dots$ & $\dots$ & $\dots$ & $\dots$ & $1.76\pm0.01$ & $-0.05\pm0.03$ &  \\
			& 	 & LGP+EX &  	& $<0.003$ & $\dots$ & $\dots$ & $\dots$ & $1.82\pm0.01$ & $-0.13\pm0.02$ & $1.03/829$ \\
			& 	 &  	&  		& 		   & $\dots$ & $\dots$ & $\dots$ & $1.76\pm0.01$ & $-0.05\pm0.03$ &  \\
			BL Lac.\tnote{f} & 0.069& PL  & $30.3$ & $\dots$ & $\dots$ & $2.01\pm0.01$ & $\dots$ & $\dots$ & $\dots$ & $1.27/1697$\\
			& 	 &      &  	  			& $\dots$ & $\dots$ & $2.00\pm0.01$ & $\dots$ & $\dots$ & $\dots$ & 		  \\
			& 	 &      &  	  			& $\dots$ & $\dots$ & $1.94\pm0.01$ & $\dots$ & $\dots$ & $\dots$ & 		  \\
			& 	 &      &  	  			& $\dots$ & $\dots$ & $1.90\pm0.01$ & $\dots$ & $\dots$ & $\dots$ & 	      \\											
			& 	 & BKN  &	  & $\dots$ & $2.01\pm0.01$ & $1.91^{+0.07}_{-0.08}$ & $4.22^{+0.96}_{-1.31}$ & $\dots$ & $\dots$ & $1.10/1689$\\
			& 	 &    	& 	  & $\dots$ & $2.04^{+0.02}_{-0.01}$ & $1.87^{+0.04}_{-0.05}$ & $2.87\pm0.43$ & $\dots$ & $\dots$ & 			\\
			& 	 &    	& 		& $\dots$ & $1.98^{+0.02}_{-0.01}$ & $1.79^{+0.04}_{-0.05}$ & $2.92\pm0.40$ & $\dots$ & $\dots$ & 			\\
			& 	 &    	& 		& $\dots$ & $2.01^{+0.03}_{-0.02}$ & $1.71^{+0.05}_{-0.04}$ & $2.31^{+0.21}_{-0.33}$ & $\dots$ & $\dots$ & 	\\						    
			& 	 & LGP   & 		& $\dots$ & $\dots$ & $\dots$ & $\dots$ & $2.01\pm0.01$ & $-0.02\pm0.03$ &  $1.11/1693$\\
			& 	 &      & 		& $\dots$ & $\dots$ & $\dots$ & $\dots$ & $2.05\pm0.02$ & $-0.10\pm0.03$ &  \\
			& 	 &      & 		& $\dots$ & $\dots$ & $\dots$ & $\dots$ & $2.00\pm0.02$ & $-0.11\pm0.03$ &  \\
			& 	 &      & 		& $\dots$ & $\dots$ & $\dots$ & $\dots$ & $2.04\pm0.02$ & $-0.25\pm0.03$ &  \\
			&	 & LGP+EX\tnote{b} 	&  	& $0.008^{+0.003}_{-0.006}$ & $\dots$ & $\dots$ & $\dots$ & $2.02\pm0.01$ & $0$ & $1.10/1693$ \\
			& 	 &  	 &  		& 	  		 & $\dots$ & $\dots$ & $\dots$ & $2.08^{+0.01}_{-0.03}$ & $<-0.09$ &  \\
			& 	 &  	 &  		& 		   	& $\dots$ & $\dots$ & $\dots$ & $2.03^{+0.01}_{-0.02}$ & $<-0.11$ &  \\
			& 	 &  	 &  		& 		   	& $\dots$ & $\dots$ & $\dots$ & $2.07^{+0.01}_{-0.02}$ & $<-0.25$ &  \\						    
			\bottomrule
		\end{tabular}
		\begin{tablenotes}
			\item[a] The fit was performed using two \emph{XMM-Newton} archive observations (ObsIDs 0650384501 and 0650384901) with all parameters tied together.
			\item[b] The curvature term was left free to vary between the error boundaries of the value obtained in the LGP fit.
			\item[c] The first and second rows for each model show results related to different \emph{XMM-Newton} observations (ObsIDs 0200360201 and 0790630101, respectively). The third is the $0.2-79\,$keV broadband fit in which the latter \emph{XMM-Newton} observation was tied to the (quasi-simultaneous) \emph{NuSTAR} data, and only fit parameters of the latter are reported.
			\item[d] In this case $a$ is the photon index at 5\,keV.	
			\item[e] The fits were performed using two \emph{XMM-Newton} archive observations (ObsIDs 0401060201 and 0502630201, for the first and second raw, respectively) with only absorption parameters tied together.
			\item[f] The fits were performed using four \emph{XMM-Newton} archive observations (ObsIDs 0501660201, 0501660301, 0501660401 and 0504370401, in raw order) with only absorption parameters tied together.
		\end{tablenotes}
	\end{ThreePartTable}
\end{table}